\documentclass[aps,pre,twocolumn,groupedaddress,showpacs]{revtex4-1}
\usepackage{hyperref}                   %
\usepackage{bm}                         %
\usepackage{amsmath,amsfonts,amssymb}   %
\usepackage{color}                      %
\usepackage{graphicx}                   %
\usepackage{overpic}                    %
\renewcommand{\eqref}{Eq.~\ref}
\newcommand{\eqsref}{Eqs.~\ref}
\newcommand{\figref}{Fig.~\ref}
\newcommand{\figsref}{Figs.~\ref}
\newcommand{\tabref}{Table~\ref}
\newcommand{\secref}{Sec.~\ref}
\newcommand{\secsref}{Secs.~\ref}

\newcommand{\Tr}[1] {\textrm{Tr}[#1]}   %
\newcommand{\SYM}[1] {[#1]^{\rm S}}     %
\newcommand{\ASYM}[1] {[#1]^{\rm A}}    %
\newcommand{\ST}[1] {[#1]^{\rm ST}}     %

\newcommand{\tsr}[1] {\mathbf{#1}}              %
\newcommand{\vtr}[1] {\mathbf{#1}}              %
\newcommand{\uvtr}[1] {\hat{\mathbf{#1}}}       %
\newcommand{\OO}[1] {\mathcal{O}(#1)}           %

\newcommand{\edot}[0]{{\dot{\epsilon}}}           %
\newcommand{\gdot}[0]{{\dot{\gamma}}}           %
\newcommand{\De}[0] {\textrm{De}}               %

\newcommand{\modenumber}{m}
\definecolor{orange}{rgb}{1,0.5,0}

\newcommand{\params}{\textit{Parameters: }}
\newcommand{\ie}{i.e.,~}
\newcommand{\eg}{e.g.,~}
\newcommand{\etal}{\textit{et al.}~}

\newcommand{\smf}[1] {}
\newcommand{\smfchanged}[1] {#1}
\newcommand{\mec}[1] {}
\newcommand{\mecchanged}[1] {#1}
\newcommand{\ejh}[1] {}
\newcommand{\ejhchanged}[1] {#1}
\newcommand{\mecc}[1] {#1}

\begin{document}

\title{Viscoelastic and elastomeric active matter: Linear instability and nonlinear dynamics}

\author{E. J. Hemingway$^1$, M. E. Cates$^{2}$, and S. M. Fielding$^1$}

\affiliation{$^1$Department of Physics, Durham University, Science Laboratories, South Road, Durham DH1 3LE, United Kingdom}
\affiliation{$^2$DAMTP, Centre for Mathematical Sciences, University of Cambridge, Wilberforce Road, Cambridge CB3 0WA, United Kingdom}

\date{\today}

\begin{abstract}
  We consider a continuum model of active viscoelastic matter, whereby an active nematic liquid-crystal is coupled to a minimal model of polymer dynamics with a viscoelastic relaxation time $\tau_C$.  To explore the resulting interplay between active and polymeric dynamics, we first generalise a linear stability analysis (from earlier studies without polymer) to derive criteria for the onset of spontaneous heterogeneous flows (strain rate) and/or deformations (strain). We find two modes of instability. The first is a viscous mode, associated with strain rate perturbations. It dominates for relatively small values of $\tau_C$ and is a simple generalisation of the instability known previously without polymer.  The second is an elastomeric mode, associated with strain perturbations, which dominates at large $\tau_C$ and persists even as $\tau_C\to\infty$.  We explore the dynamical states to which these instabilities lead by means of direct numerical simulations.  These reveal oscillatory shear-banded states in 1D, and activity-driven turbulence in 2D even in the elastomeric limit $\tau_C\to\infty$. Adding polymer can also have calming effects, increasing the net throughput of spontaneous flow along a channel in a type of ``drag-reduction''.  Finally the effect of including strong, antagonistic coupling between nematic and polymer is examined numerically, revealing a rich array of spontaneously flowing states.

\end{abstract}

\pacs{47.57.Lj, 61.30.Jf, 87.16.Ka, 87.19.rh}

\maketitle

\section{Introduction}
\label{sec:intro}

Examples of active matter include bacterial swarms, the cellular cytoskeleton, and \textit{in vitro} `cell extracts' that comprise only polymeric filaments, molecular motors and a fuel supply \citep{Marchetti2013,Dombrowski2004,Nedelec1997,Sanchez2012}. Such materials are not only of direct biophysical significance, but also represent a wider class of systems in statistical physics in which strong deviations from thermal equilibrium emerge at the collective macroscopic scale due to the underlying active dynamics of the system's microscopic subunits (rotating bacterial flagella, marching molecular motors that bridge cytoskeletal filaments, {\it etc.}). Arising at the level of the microscopic subunits, this driving is distinct from the macroscopic boundary driving of, for example, an imposed shear flow.

For an active particle subject to no externally imposed force, the simplest perturbation it can exert on the local fluid environment is that of a force dipole. Depending on the sign of this dipole the activity is classified as extensile (where the two forces act from the centre of mass of the dipole outwards towards the fluid) or contractile (where the opposite applies). Collectively, a fluid of these active particles can exhibit non-equilibrium emergent phenomena on macroscopic lengthscales that greatly exceed the particle size or spacing. These include activity-induced ordering, and bulk fluid flows that arise spontaneously even in the absence of external driving.  Depending on the strength of the activity, these flows may remain steady and laminar at the scale of the system; show oscillatory limit cycles at that scale or below; or exhibit spatiotemporal chaos.  The latter effect closely resembles conventional inertial turbulence in a passive Newtonian fluid and is accordingly often termed ``active turbulence" or ``bacterial turbulence" \citep{Dombrowski2004,Fielding2011,Wensink2012,Giomi2013,Giomi2014a,Thampi2013}.  However its mechanism is distinct from that of inertial turbulence: it
stems from a balance between active stress and orientational relaxation, rather than between inertia and viscosity.

Depending on their symmetry, ordered phases of active fluids can be described by either a polar \citep{Voituriez2005,Julicher2007} or nematic order parameter \citep{AditiSimha2002,Marchetti2013,Ramaswamy2010}. In this work we consider the latter case, and throughout denote the nematic order parameter by $\tsr{Q}$.  Indeed, many theoretical descriptions of active matter are based on simple continuum models for the hydrodynamics of a suspension of rod-like objects, originally developed to describe a passive liquid crystal \citep{beris1994thermodynamics,degennes1995physics}. To describe an active material, such a model is then augmented by the leading-order terms characterising the violations of time-reversal symmetry that arise from activity.  In particular, this gives an additional active contribution $\tsr{\Sigma}_A = -\zeta \tsr{Q}$ to the fluid's stress tensor, where the activity parameter $\zeta$ is positive for extensile systems and negative for contractile ones.

Even without detailed knowledge of the value of $\zeta$, the approach just described is capable of robust predictions. For example, beyond a critical threshold level of activity an initially quiescent fluid is generically predicted to become unstable to the formation of a spontaneously flowing state \cite{AditiSimha2002}. This threshold, for a finite system size, depends on whether the system is extensile or contractile. However in both cases it decreases with increasing system size \citep{Voituriez2005}, tending to zero in the limit of a bulk sample. Any level of activity, however small, can then trigger the formation of spontaneous flows.  Consistent with these analytical predictions, numerical solutions of active nematic continuum models
\citep{Fielding2011,Giomi2013,Giomi2014a,Thampi2013} have indeed revealed a host of spontaneously flowing states resembling experimental observations in bacterial swarms \citep{Dombrowski2004} and microtubule-based cell extracts \citep{Sanchez2012}.  Both of these are extensile nematics, to which we restrict ourselves in our numerical studies below \cite{[{By contrast, actomyosin systems, including the cytoskeleton, are typically contractile and polar; see }] Julicher2007}.

Active nematic fluids are often referred to informally as `active gels' \citep{Voituriez2005,Fielding2011}, particularly in a biological context. But while all liquid crystals are somewhat viscoelastic, for example due to the slow motion of topological defects, most existing models of active matter assume fast local relaxations and so fail to model ``gels" in the conventional sense of the term, as understood by polymer or colloid physicists \citep{Callan2011,Joanny2009a}.  Certainly the standard continuum models of active matter \citep{Ramaswamy2010,Marchetti2013} do not capture the types of slow viscoelastic dynamics that one might expect in the cytoskeleton, which contains long-chain flexible polymers and other cytoplasmic components that are expected to have long intrinsic relaxation times (or even divergent ones in the case of a cross-linked network). This is a major shortcoming, because this slow viscoelastic dynamics would be expected to couple strongly to the active liquid crystalline dynamics and thereby potentially radically modify the effects of activity.

The effect of a viscoelastic polymeric background is also likely to play an important role in modifying active flows and diffusion \citep{Bozorgi2013} at a supra-cellular level. Indeed, long-chain molecules are present in mucus, saliva, and many other viscoelastic fluids both inside and outside the body that are susceptible to colonization by swarms of motile bacteria. Moreover, many bacteria secrete their own polymers \citep{Decho1990}, particularly during biofilm formation (which is however not the topic of this paper). This suggests an evolutionary advantage for bacteria in controlling the viscoelasticity of their surroundings, supporting the view that viscoelasticity and active motion are coupled in a nontrivial way.

With this motivation, the aim of this paper is to study in greater depth the predictions of a model first presented in Ref.~\citep{Hemingway2015}, which addresses at a continuum level the interplay between active liquid-crystalline dynamics and slow polymeric modes of relaxation.  This approach is distinct from, but complements, recent studies of individual swimmers in viscoelastic fluids \citep{Lauga2007,Teran2010,Zhu2012,Spagnolie2013,Riley2014}, which show that swimming speeds can be either enhanced or suppressed relative to those in a Newtonian solvent, depending on details of the swimming mechanism and particle geometry. Such details do not enter our continuum picture, however, which focuses on emergent and potentially universal many-body behavior at larger length scales.

In a biological context, active matter is often found in confined geometries such as the interior of a cell.  Indeed it has been argued that the confinement of subcellular active matter may in part be responsible for cytoplasmic streaming \citep{Woodhouse2012,Goldstein2015}, an important process whereby coherent fluid flows facilitate the circulation of nutrients and organelles within the cell \citep{Serbus2005}.  At a larger scale, a recent study of cell migration in artificial channels observed increases in mean cell velocity and flow coherence as the channel was narrowed \citep{Vedula2012}.  Also, suspensions of \textit{B.~subtilis} were observed to form stable spiral structures when confined in a droplet \citep{Wioland2013}. In view of these observations, it is important that any numerical study should consider carefully the effects of system size. In what follows we choose a rectangular channel geometry of fixed aspect ratio bounded by a pair of parallel hard walls.  The size of this channel is notionally fixed in simulation units, but we study finite size effects by then varying instead the microscopic length scales in the problem.

As we shall explore in detail below, the interplay of active and polymeric dynamics leads to a host of exotic spontaneous flow states.  These include oscillatory shear-banded states in 1D, while in 2D we find activity-driven turbulence even in the limit of infinite viscoelastic relaxation time in which our model describes an active elastomer.  In other regimes we find that adding polymer can have calming effects, increasing the net throughput of spontaneous flow along a channel in a type of ``drag-reduction''.

The paper is structured as follows. In \secref{sec:model} we review the equations of motion that describe the coupled dynamics of an active nematic with a viscoelastic polymer, as derived in Ref.~\cite{Hemingway2015}. In \secref{sec:simulation} we discuss the simulated sample geometry and give details of our numerical methods. In \secref{sec:lsa} we perform a linear stability analysis to derive the threshold of instability to spontaneous flow. The results of this linear calculation then provide a routemap for performing full nonlinear simulations in \secsref{sec:1D_nonlinear} and \ref{sec:2D_nonlinear}. In \secref{sec:1D_nonlinear} we restrict those simulations to one spatial dimension for simplicity, while in \secref{sec:2D_nonlinear} we perform full 2D simulations.  \secref{sec:conclusions} contains a summary of our results, and the outlook for future work.

\section{Model Equations}
\label{sec:model}

In this section we remind the reader of the model equations, as developed originally in Ref.~ \cite{Hemingway2015}.  The state of liquid crystal (nematic) ordering is denoted by a traceless symmetric tensor $\tsr{Q}=q\langle \uvtr{n}\uvtr{n}-\tfrac{1}{3}\tsr{I}\rangle$, where $\uvtr{n}$ is the nematic director and $q$ is the degree of ordering. The polymeric conformation is similarly denoted by the symmetric tensor $\tsr{C} = \langle{\bf rr}\rangle$, where ${\bf r}$ is the end-to-end vector of a chain (or subchain, depending on the level of description used), normalized so that $\tsr{C} = \tsr{I}$ in equilibrium. We adopt a single-fluid description in which the concentration fields of polymer and liquid crystal are assumed to remain uniform (in contrast to a two-fluid approach as taken by, for example, Ref.~  \cite{Ramaswamy2003}).  Accordingly, all three components -- liquid crystal, polymer and solvent -- share the same centre-of-mass velocity, $\vtr{v}$. The symmetric and antisymmetric parts of the velocity-gradient tensor $\left(\nabla \vtr{v}\right)_{ij} \equiv \partial_i v_j$ are denoted $\tsr{D}$ and $\tsr{\Omega}$ respectively.  For any other tensors the symmetric, antisymmetric, and traceless parts carry superscripts \textit{S}, \textit{A} and \textit{T}.

We introduce a free energy density $f = f_Q(\tsr{Q},\nabla\tsr{Q}) + f_C(\tsr{C})+ f_{QC}(\tsr{Q},\tsr{C})$, where $f_{Q}$ and $f_{C}$ are the standard forms for nematics \citep{beris1994thermodynamics} and dumb-bell polymers \citep{Milner1993} respectively.  Accordingly we have
\begin{eqnarray}
  f_Q &=& G_Q \left[ \frac{(1-\gamma/3)}{2} \mbox{Tr}\mathbf{Q}^2 - \frac{\gamma}{3}
  \mbox{Tr}\mathbf{Q}^3 + \frac{\gamma}{4} (\mbox{Tr}\mathbf{Q}^2)^2 \right]\nonumber\\
      & & + \frac{K}{2}
  (\nabla_i Q_{jk})^2\;,
  \label{nem_eng}
\end{eqnarray}
in which $G_Q$ sets the scale of the bulk free energy density, $K$ is
the nematic elastic constant, and $\gamma$ is a control parameter for the
isotropic-nematic transition. Likewise
\begin{equation}
  f_C=\frac{G_C}{2}\left(\mbox{Tr}\mathbf{C} -\ln \det \mathbf{C}\right),
  \label{indep}
\end{equation}
where $G_C$ is the polymer elastic modulus.

The lowest order passive coupling between $\tsr{Q}$ and $\tsr{C}$ is
\begin{equation}
  f_{QC} = \kappa \Tr{\tsr{C}-\tsr{I}}\Tr{\tsr{Q}^2} + 2 \chi \Tr{\tsr{C}\tsr{Q}},
  \label{eq:free_QC_coupling}
\end{equation}
where both terms vanish for undeformed polymers ($\tsr{C} = \tsr{I}$).
Here $\kappa$ controls how the polymer pressure shifts the
isotropic-nematic transition and for simplicity we set $\kappa = 0$
throughout. The second term makes it energetically preferable for
$\tsr{Q}$ and $\tsr{C}$ to align \mecchanged{with major axes parallel} for $\chi < 0$, and
perpendicular for $\chi > 0$.

From the volume-integrated free energy $F = \int f dV$, the nematic
molecular field $\tsr{H} \equiv -\ST{\delta F/\delta \tsr{Q}}$ follows
as
\begin{align}
  \tsr{H} &= -G_Q\left[\left(1 - \frac{\gamma}{3}\right)\tsr{Q} - \gamma \tsr{Q}^2 + \gamma \tsr{Q}^3 \right] - G_Q \gamma \frac{\tsr{I}}{3}\Tr{\tsr{Q}^2} \nonumber\\ &+ K \nabla^2 \tsr{Q}
  -2 \kappa \Tr{\tsr{C}-\tsr{I}}\tsr{Q} - 2\chi\tsr{C}^{\rm T}.
  \label{eq:molec_field_H}
\end{align}
The corresponding molecular field $\tsr{B} \equiv -\SYM{\delta
  F/\delta \tsr{C}}$ for the polymer obeys
\begin{equation}
  \tsr{B} = -G_C(\tsr{I}-\tsr{C}^{-1})/2 - \kappa \tsr I \Tr{\tsr{Q}^2}-2\chi\tsr{Q}.
  \label{eq:molec_field_B}
\end{equation}

Using these molecular fields we follow Ref.~\cite{Hemingway2015} in developing minimally coupled equations of motion for $\tsr{Q}$ and $\tsr{C}$ that respectively reduce to the Beris-Edwards liquid crystal theory and the Johnson-Segalman (JS) polymer model in appropriate limits \citep{beris1994thermodynamics}. We then allow for conformational diffusion in the polymer sector \citep{Olmsted2000}, which adds a gradient term in $\tsr{C}$ of kinetic origin \citep{Hemingway2015}.  (Alternatively one can incorporate a non-local term in the polymer
free energy, though this then produces a more complex form for the polymer stress \citep{PDO_corr}.)  To this is added a minimal set of active terms \citep{AditiSimha2002}, supposing for simplicity that the polymers are not themselves active, but that the origin of the activity resides entirely in $\tsr{Q}$. This is enough to capture, for example, the effect of adding polymer to a cell extract, or the collective dynamics of bacterial suspensions in mucus.  (A contrasting approach is to build a system of polymers directly from active elements \citep{Jayaraman2012}). There remain two active terms linear in $\tsr{Q}$. One of these can be absorbed into $f_Q$, and the other is the familiar active deviatoric stress $\tsr{\Sigma}_A =
-\zeta\tsr{Q}$ \citep{AditiSimha2002}.

The resulting equations of motion for $\tsr{Q}$ and $\tsr{C}$ are:
\begin{align}
  \left(\partial_t + \vtr{v}.\nabla\right)\tsr{Q} &= \tsr{Q}\tsr{\Omega} - \tsr{\Omega}\tsr{Q} + \frac{2\xi}{3} \tsr{D} + 2 \xi \ST{\tsr{Q}\tsr{D}} \nonumber \\
  &\quad - 2\xi \tsr{Q}\Tr{\tsr{Q}\tsr{D}} + \tau_Q^{-1} \tsr{H}/G_Q, \label{eq:eom_Q}\\
  \left(\partial_t + \vtr{v}.\nabla\right)\tsr{C} &= \tsr{C}\tsr{\Omega} - \tsr{\Omega} \tsr{C} + 2 a \SYM{\tsr{C}\tsr{D}} \nonumber \\
  &\quad + \tau_C^{-1}(2\SYM{\tsr{B}\tsr{C}}/G_C+\ell^2_C\nabla^2 \tsr{C}). \label{eq:eom_C}
\end{align}
Here $\xi$ is the flow-alignment parameter of the nematic \citep{Stark2003} and $a$ is the slip parameter of the JS model.  Setting $a=1$ recovers the Oldroyd B model \citep{LarsonConstit1988}.  Each controls the relative tendency of molecules to align with streamlines versus rotating with the local vorticity.  Parameters $\tau_Q, \tau_C$ are intrinsic local relaxation times for nematic and polymer, while $\ell_C$ governs conformational
diffusion in the JS sector \citep{Olmsted2000}.

The velocity field $\vtr{v}$ obeys the Navier Stokes equation for an incompressible fluid
\begin{align}
  \rho \left(\partial_t + v_\beta\partial_\beta\right)v_\alpha &= \partial_\beta \Sigma_{\alpha\beta},\label{eq:NS}\\
  \partial_\alpha v_\alpha &= 0.\label{eq:incomp}
\end{align}
Here the total stress $\tsr{\Sigma} = -P\tsr{I} + 2\eta\tsr{D}+ \tsr{\Sigma}_A+\tsr{\Sigma}_{Q} + \tsr{\Sigma}_C $ combines an isotropic pressure $P$, a contribution from a Newtonian solvent of viscosity $\eta$, two reactive stresses
\begin{align}
  \tsr{\Sigma}_Q &= - K(\nabla \tsr{Q}):(\nabla\tsr{Q}) + 2 \ASYM{\tsr{Q}\tsr{H}}  \nonumber \\
  &\quad - \frac{2 \xi}{3} \tsr{H} - 2 \xi \ST{\tsr{Q}\tsr{H}}+ 2 \xi \tsr{Q} \Tr{\tsr{Q}\tsr{H}}, \label{QReactive}\\
  \tsr{\Sigma}_C &= -2a\SYM{\tsr{C}\tsr{B}} + 2 \ASYM{\tsr{C}\tsr{B}},\label{CReactive}
\end{align}
and the active stress $\tsr{\Sigma}_A = - \zeta \tsr{Q}$.  The colon in \eqref{QReactive} denotes contraction over the second and third Cartesian indices. In what follows we shall assume inertialess (creeping) flow, and set $\rho=0$ in \eqref{eq:NS}.

Having set out this model in which the dynamics of the polymer and liquid crystal are fully coupled, we now consider certain limits in which the coupling between them is diminished. Clearly, setting the constants $\chi$ and $\kappa$ to zero in Eqn.~\ref{eq:free_QC_coupling} eliminates any thermodynamic coupling between $\tsr{Q}$ and $\tsr{C}$.  It is important to understand, however, that even in this case of zero thermodynamic coupling, $\tsr{Q}$ and $\tsr{C}$ remain non-trivially coupled in a purely kinematic way: any changes in $\tsr{C}$ or $\tsr{Q}$ perturb the stress ${\bf\Sigma}$, which in turn perturbs the flow field $\vtr{v}$, which in turn drives both $\tsr{Q}$ and $\tsr{C}$.  Accordingly, even without any thermodynamic coupling we shall find in what follows a rich array of regimes in which the dynamics of $\tsr{Q}$ and $\tsr{C}$ show strong coupling.  We therefore defer the case of true thermodynamic
coupling to ~\secref{sec:2D_nonlinear_expl}.

As already noted, in the limit in which we simply remove the polymeric and active components ($G_C=\kappa=\chi=\zeta = 0$), the model reduces to the Beris-Edwards theory of liquid crystals. Removing instead the liquid crystalline component ($G_Q=K=\kappa=\chi=0$) recovers the Johnson-Segalman model of polymeric fluids. These separate models have been studied comprehensively in the earlier literature and we do not consider them further here.

Note finally, and importantly, that in the limit $\tau_C\to\infty$ our polymer behaves as an elastomer, without any local mechanism for stress relaxation. At zero activity, our model then describes a passive nematic elastomer \citep{warner2003liquid}.

\section{Simulation Details}
\label{sec:simulation}

\subsection{Geometry and boundary conditions}

We consider a two-dimensional (2D) slab of active viscoelastic matter confined between parallel plates separated by a distance $L_y$ in the $y$-direction, and of length $L_x$ in the $x$-direction.  At the plates we choose boundary conditions of no-slip and no-permeation for the fluid velocity $\vtr{v}$:
\begin{align*}
 v_\alpha =  0 \quad &\textrm{at} \quad y = \{0, L_y\} \quad \forall \; \alpha,
\end{align*}
and zero-gradient boundary conditions for $\tsr{Q}$ and $\tsr{C}$:
\begin{align*}
  \partial_y Q_{\alpha\beta} = \partial_y C_{\alpha\beta} = 0 \quad &\textrm{at} \quad y = \{0, L_y\} \quad  \forall \; \alpha, \beta.
\end{align*}
In the direction $x$ parallel to the plates we adopt periodic boundary conditions for all variables.

The adoption of a zero-gradient boundary condition for the polymeric conformation tensor $\tsr{C}$ is standard practice in the literature on shear banding, though it remains to be justified microscopically.  A zero-gradient boundary condition on the nematic order parameter, allowing free rotation of $\tsr{Q}$ at the wall, has been used in some previous studies of active materials \citep{Cates2008,Edwards2009}, whereas other studies have adopted anchoring boundary conditions \citep{Marenduzzo2008}.  Recent work comparing both these boundary conditions has shown the essential physics to be qualitatively unchanged by the choice made \citep{Fielding2011}, particularly at large activities where flowing states develop structure on a length-scale much smaller than the system size.

Despite this, we attach the following note of caution.  In \secref{sec:lsa} below we calculate the critical threshold activity required for the onset spontaneous flow, starting from an initial base state of no flow and a perfectly ordered director field.  Crucially, we will find this threshold to depend strongly on the orientation of the director in that initial state: states where the director is initially aligned in the $x$- and $y$-directions become separately unstable at activities an order of magnitude apart.  That observation in turn has important implications for the boundary condition on $\tsr{Q}$, because a boundary condition that favors anchoring of the director parallel (resp.  perpendicular) to the walls would be likely to favor an initial condition in which the director is aligned in the $x$ (resp. $y$) direction, leading to different threshold activities in each case.  This suggests that, in some instances, the choice of boundary conditions can have an important effect on the system's behavior.  Indeed recent experiments demonstrate that `living liquid-crystals', in which conventional liquid crystals are made active by the addition of swimming bacteria, can be influenced by anchoring boundary conditions \citep{Zhou2014}.

On the other hand, the adoption of a zero gradient boundary condition considerably simplifies our linear stability analysis below, because it permits a spatially homogeneous base state. This in turn means that heterogeneous perturbations to that base state can easily be expressed in a cosine basis. In contrast, anchoring boundary conditions could lead to inhomogeneous base states, complicating the analysis. Accordingly, we work with zero gradient boundary conditions
throughout.

We restrict the fluid velocities to lie in the $x-y$ plane, setting $v_z = 0$, and assume translational invariance of all variables along $z$. In some calculations we further assume translational invariance along $x$, restricting our states to vary only in 1D, along $y$.  However we work throughout with 3D tensors for both $\tsr{Q}$ and $\tsr{C}$, allowing non-zero components $Q_{\alpha\beta}$ and $C_{\alpha\beta}$ for $\alpha,\beta=x,y,z$.  In principle this could allow the principal axis (director) of $\tsr{Q}$ and/or $\tsr{C}$ to point out of the $xy$ plane of the simulation.  Analysing our numerical results, however, we find in practice that in most of our simulations $\tsr{Q}$ and $\tsr{C}$ do remain confined to the $xy$ plane \citep{See1990,Olmsted1994}. We therefore expect the key physics to be robust to the dimensionality of the order parameter tensors \citep{PrashantPreprint2015}.

\subsection{Units and Parameters}
\label{sec:simulation_params}

We work in units of length such that the gap size $L_y = 1$, of time such that the liquid crystal's local relaxation time $ \tau_Q = 1$, and of mass such that its characteristic modulus $G_Q = 1$. The value of the parameter $\eta_Q=G_Q\tau_Q$ that controls the passive nematic viscosity is then also equal to unity. We work in the creeping flow limit of zero Reynolds number, setting $\rho \to 0$. We use a cell aspect ratio $L_x / L_y = 4$.

To enable a direct comparison with previous numerical studies of active matter without polymer present \citep{Fielding2011}, we fix the solvent viscosity $\eta = 0.567$ throughout. In the $\tsr{Q}$ sector of the dynamics we set the isotropic-nematic control parameter $\gamma = 3$ so that the passive liquid crystal lies well within its nematic phase~\footnote{This value in fact corresponds to the spinodal stability limit of the isotropic phase.}. We set the flow-alignment parameter at $\xi = 0.7$, which lies within the flow-aligning regime.

In the polymer sector we set the slip parameter $a = 1$, such that the Johnson-Segalman model reduces to the Oldroyd B model. This eliminates any possibility of a shear-banding instability originating purely from the polymer dynamics.  Theoretical studies suggest that this value, which is widely adopted for flexible polymers, may also be reasonable for dense cross-linked filaments \citep{Head2003,Storm2005}.

With many of the model's parameters having been fixed as just described, there remain only four to be explored numerically: the activity $\zeta$, a diffusivity parameter $\Delta$ (defined below) that governs \mecchanged{relaxation of} spatial gradients in the system, the polymer relaxation time $\tau_C$ and its modulus $G_C$, and (when non zero) the coupling constant $\chi$. In our chosen units, these are adimensionalised as summarised in \tabref{tbl:simulation_params_dimensionless}.  The tildes denote adimensional quantities, but our units have been chosen to make the tildes redundant, and we now drop them.

\begin{table}
  \def\arraystretch{1.5}
  \begin{center}
    \begin{tabular}{| l | l | p{3.6cm} |}
      \hline
      \textbf{Symbol} & \textbf{values explored} & \textbf{parameter}\\
      \hline $\tilde{\zeta} = \frac{\zeta}{G_Q}$ &  0.001$\to$ 10 & extensile activity \\
      $\tilde{\Delta} \equiv \frac{K}{G_Q L_y^2} $ & $10^{-5} \to 6.4 \times10^{-4}$ & diffusion constant \\
      $\tilde{\tau}_C \equiv \frac{\tau_C}{\tau_Q}$ & $10^{-2}\to10^6 \to \infty$ & ratio of relaxation times \\
      $\tilde{G}_C \equiv \frac{G_C}{G_Q}$ & $\eta_C / \tau_C = 10^{-6} \to 10^2$ & polymer modulus\\
      $\tilde{\chi} \equiv \frac{\chi}{G_Q}$ & $0\le\chi \ll G_Q, G_C$ & explicit coupling strength\\
      \hline
    \end{tabular}
    \caption{List of key dimensionless parameters. Hereafter we drop
      the tildes for clarity. }
    \label{tbl:simulation_params_dimensionless}
  \end{center}
\end{table}

\tabref{tbl:simulation_params_dimensionless} also shows the representative ranges that we shall explore numerically for these parameters.  While these are chosen on the basis of order-of-magnitude estimates at best, they might reasonably describe the recent experiments of Sanchez \etal on an extensile kinesin-microtubule mixture akin to a cytoskeletal gel \cite{Sanchez2012}, where the level of activity was varied via the concentration of ATP. In Ref.~\cite{Koenderink2009}, the modulus of a (contractile) actin gel was found to increase by a factor of ten compared to the passive case in the presence of myosin motor activity, suggesting an upper bound to the dimensionless activity parameter of $\zeta / G_Q \approx 10$.

In our model, the polymer could in principle describe a range of viscoelastic behaviors within the cell, including the effects of the cytosol, which comprises entangled protein filaments and organelles \citep{Mofrad2009}. To encompass this diversity, we vary the polymer relaxation time from $\tau_C = 10^{-2}$, for which the polymeric dynamics are rapid and only contribute extra viscosity to the fluid as a whole, to $\tau_C \to \infty$, where the polymer effectively acts as an elastomeric solid.

Throughout most of what follows we assume a constant value of $\eta_C \equiv G_C\tau_C$, which controls the polymer viscosity, setting $\eta_C=1$. To maintain this between runs in which the polymer relaxation time $\tau_C$ changes, the polymer elastic modulus $G_C$ must change in inverse proportion to $\tau_C$.  This assumption of a constant polymer viscosity is however relaxed in those sections where we consider the true elastomeric limit $\tau_C\to\infty$ (at a fixed value of $G_C$) of an active nematic in a crosslinked polymeric background.

The crossover length-scale at which elastic distortions in a passive nematic compete with its bulk free energy is $\ell_Q \equiv \sqrt{K / G_Q}$. Recasting the $\tsr{Q}$ constitutive equation (\eqref{eq:eom_Q}) in this notation, the diffusive term in the molecular field then becomes $\frac{\ell_Q^2}{\tau_Q} \nabla^2 \tsr{Q}$. The diffusive JS model prescribes an analogous term in the polymer sector, with coefficient $\ell_C^2/\tau_C$ \citep{Olmsted2000}.  In most of our numerical runs we assume these to be equal, setting:
\begin{equation}
  \frac{K}{\eta_Q}\equiv \frac{\ell_Q^2}{\tau_Q}  = \frac{\ell_C^2}{\tau_C} \equiv \Delta. \label{eq:simulation_diffusivities}
\end{equation}
Note that taking the elastomeric limit $\tau_C\to\infty$ at a fixed value of the diffusivity parameter $\Delta$ means taking $l^2_C\to\infty$ in tandem.  In some places below we instead consider fixed $l_C=l_Q$. Taking the elastomeric limit in that case then eliminates the diffusive term $\Delta_C=l_C^2/\tau_C$ from the polymer dynamics; setting $l_C = l_Q$ is then the same as setting $l_C = 0$ in this limit.  However many of our 2D simulations then show an unphysical instability that leads to structure on the scale of the numerical grid. The retention of finite $\Delta_C = \Delta$ as $\tau_C\to\infty$ avoids this problem; we discuss the issue further in Section \ref{sec:2D_nonlinear_elast}.

Values for $\Delta$ are then selected so that the associated lengthscale $l_Q$ obeys $\Delta x < \ell_Q \ll L_y$, where $\Delta x$ is the numerical grid size (for which $L_x/1024$ or $L_x/2048$ is a reasonable lower bound, constrained by a maximum feasible number of grid points $N_x=1024$ or $2048$). In this way, the spatial structures predicted by the model are well resolved on the scale of the numerical mesh, and also fit comfortably within the simulation box. The velocity correlation length for kinesin microtubule mixtures has been estimated
at $l_v \approx 100 \mu$m \citep{Sanchez2012}, and if we assume that our lengthscale $\ell_Q$ is an order of magnitude smaller than this \citep{PrashantPreprint2015}, the range of values of $l_Q$ that we explore numerically would imply a channel width in the range $400 \mu$m $\to 3$mm, as might describe a typical microfluidic device.

As noted above, throughout most of what follows we disable explicit thermodynamic coupling between $\tsr{Q}$ and $\tsr{C}$ by setting $\chi = 0$ and $\kappa=0$. The only remaining coupling between $\tsr{Q}$ and $\tsr{C}$ is then purely kinematic, caused by the two components sharing a common fluid velocity $\vtr{v}$. In \secref{sec:2D_nonlinear_expl}, we move finally to address the case of explicit thermodynamic coupling with $\chi \neq 0$.

\subsection{Numerical Details}

In our numerical scheme we evaluate spatially local terms in both constitutive equations using an explicit Euler time-stepping scheme. To solve the Stokes equation we use a stream function formulation, with a hybrid numerical scheme in which gradient terms in the $x$ and $y$ directions are computed using Fourier and Crank-Nicolson methods respectively \citep{NumericalRecipiesBook,pozrikidis2009fluid}.  Spatially diffusive terms in the constitutive equations are likewise solved using this hybrid approach.  Advective terms are computed using third order upwind differencing.

We performed the following checks of our numerics.  For each run we verified the results to be robust against halving either the spatial mesh size or the time step.  Note however that because of the erratic (and quite possibly chaotic) nature of many of the dynamical states explored, converging individual trajectories is not a realistic prospect.  We instead performed convergence tests on time-averaged quantities, and on the overall characteristics of states in the phase diagram. Finally, in our nonlinear runs we have carefully checked the behavior in the early stages of instability for consistency with the growth rates found from our linear stability analysis.

\section{Linear stability analysis}
\label{sec:lsa}

In this section we perform a linear stability analysis of an initially homogeneous, non-flowing base state to determine the regions of parameter space in which such a state is linearly unstable to the formation of a heterogeneous state with spontaneous flows.  Doing so enables us to understand the key instabilities present in the model, and to form a helpful roadmap in selecting parameter values for the much more computationally expensive nonlinear runs in later sections. We set the explicit thermodynamic couplings to zero throughout this section: $\kappa=\chi=0$.

\subsection{General procedure}

We consider an initially homogeneous, quiescent, uniaxial base state with
\begin{subequations}
  \begin{align}
    \overline{Q}_{\alpha\beta} &= q \left(n_\alpha n_\beta - \delta_{\alpha\beta} \right),\label{eq:active_bs_Q}\\
    \overline{C}_{\alpha\beta} &= \delta_{\alpha\beta},\\
\partial_\alpha\overline{v}_\beta &= 0.
  \end{align}
\label{eq:base}
\end{subequations}
Here $q$ is the magnitude of the nematic order parameter, and
$n_\alpha$ its director.  We shall consider two alternative choices
for the initial director field: $\uvtr{n} = \left(1, 0, 0\right)$ and
$\left(0, 1, 0\right)$, oriented parallel and perpendicular to the
walls respectively.  Note that, although we have specified the
$Q_{\alpha\beta}$ of this initial base state to be uniaxial, our
calculations for departures from the base state do not impose that
restriction. \mecchanged{The polymer is initially undeformed, consistent with the
absence of velocity gradients; with our boundary conditions
this absence also implies $\overline{v}_\alpha = 0$ in the base state \ejhchanged{such that the fluid is initially at rest}.}

We now examine the linear stability of this base state to perturbations that are heterogeneous in one spatial dimension (1D), with wavevector in the flow gradient direction, $y$. Within this 1D assumption, the condition of fluid incompressibility further demands that the only element of the velocity gradient tensor that \mecchanged{might} become nonzero is $\gdot \equiv \partial_y v_x$.  To simplify the notation, we collect all the relevant variables into the \mecchanged{quantity} $\bm\phi =\left(\tsr{Q}, \tsr{C}, \gdot\right)$. \mecchanged{To the base state $\overline{\bm\phi}=\left(\overline{\tsr{Q}},\overline{\tsr{C}},\overline{ \gdot}\right)$ prescribed in \eqref{eq:base} above, we then add Fourier-mode perturbations
\begin{align}
  \bm\phi=\overline{\bm\phi}+ \sum_k \bm\phi^k(t) \textrm{cos}\left(k y\right),
  \label{eq:lsa_stab_cos}
\end{align}
with small amplitudes $\bm\phi^k=\left(\tsr{Q}^k,\tsr{C}^k,\gdot^k\right)$.
Our boundary conditions at $y = 0,L_y$ require that the
wavevector $k$ takes discrete values, $k = \pi \modenumber /
L_y$, with integer $\modenumber$.}

Substituting \eqref{eq:lsa_stab_cos} into
the model equations derived in \secref{sec:model} and expanding in
powers of the \mecchanged{mode amplitudes gives}, at first
order, a linearised equation set for the dynamics of the
perturbations:
\begin{equation}
  \partial_t \vtr{p}^k = \tsr{M}^k\cdot\vtr{p}^k,\;
  \label{eq:lsa_stab_matrix}
\end{equation}
where $\vtr{p}^k \equiv \left({Q}^k_{xx}, {Q}^k_{xy},
  {Q}^k_{yy}\right.$, $\left.{C}^k_{xx}, {C}^k_{xy},
  {C}^k_{yy}\right)^T$.

Note that $\vtr{p}^k$ contains only a subset of the full
list of variables originally specified in $\bm{\phi}^k$. \mecchanged{The reasons
for this are twofold.  First, the} Stokes equation requires $\vtr{\nabla}
\cdot \tsr{\Sigma} = 0$, with $\tsr{\Sigma}$ the total stress tensor.
This enables us to express $\gdot^k$ directly in terms of
${\tsr{Q}}^k$ and ${\tsr{C}}^k$ as
\begin{align*}
  {\gdot}^k = \frac{-1}{\eta}\left(\tsr{\Sigma}^k_A + \tsr{\Sigma}^k_Q + \tsr{\Sigma}^k_C\right)_{xy},
\end{align*}
in an obvious notation.  This eliminates $\gdot^k$ as a dynamical \mecchanged{variable: physically,} the shear rate is enslaved to the viscoelastic stress components by the requirement of instantaneous force balance in the limit of creeping flow.

\mecchanged{Secondly, we find that the set of components $Q^k_{\alpha z}$ and $C^k_{\alpha z}$
evolve independently from the
$xx$, $xy$, $yy$ components contained in $\vtr{p}$; this $z$-related set furthermore
always have negative eigenvalues, rendering them stable.
Accordingly, the only dynamical variables relevant to us are those
listed in $\vtr{p}$ above.}

The eigenvalues of the matrix $\tsr{M}^k$ then determine whether the initial base state is stable or unstable to the growth of \mecchanged{perturbations.  Any} eigenvalue with a positive real part corresponds to an unstable mode that grows exponentially (in this linear regime), taking the system away from the initial homogeneous, non-flowing base state, and towards a heterogeneous state with a spontaneous flow.

In what follows we seek a threshold value of the activity at which the
base state first becomes linearly unstable, with the values of the
other model parameters held fixed.  Sweeping the activity level
upwards, the critical threshold $\zeta_c$ at which instability first
sets in is the value at which
\begin{equation}
  Re(\omega^+) = 0,
  \label{eq:lsa_w_zero}
\end{equation}
where $\omega^+$ is the eigenvalue with the largest real part.

It is \ejhchanged{possible} to show that the only non-trivial (potentially
unstable) modes of $\tsr{M}^k$ have eigenvalues $\omega^\pm$ given by

\begin{equation}
2\omega^\pm = -b\pm\sqrt{b^2-4c},
  \label{eq:quad}
\end{equation}
where
\begin{equation}
b=\frac{a^2G_C}{\eta}+\frac{\Lambda}{\eta}(2G_Q\Lambda \ell_Q^2k^2-\zeta)+\frac{1+\ell_C^2k^2}{\tau_C}+\frac{\ell_Q^2k^2}{\tau_Q},
\end{equation}
and
\begin{eqnarray}
\label{eq:c}
c=\frac{\ell_Q^2k^2}{\tau_Q}\frac{a^2G_C}{\eta}&+&\frac{\Lambda}{\eta}\frac{1+\ell_C^2k^2}{\tau_C}(2G_Q\ell_Q^2k^2\Lambda-\zeta) \nonumber \\ &+&\frac{\ell_Q^2k^2}{\tau_Q}\frac{1+\ell_C^2k^2}{\tau_C},
\end{eqnarray}
with
\begin{equation}
\label{eq:Lambda}
  \Lambda = \left\{
    \begin{array}{l l}
      (5\xi - 3) / 12 & \qquad \text{for } \uvtr{n} = \left(1, 0, 0\right),\\
      (5\xi + 3) / 12 & \qquad \text{for } \uvtr{n} = \left(0, 1, 0\right).
    \end{array}
  \right.
\end{equation}
For each value of $\Lambda$, then, two different modes of instability are possible as determined by \eqref{eq:quad}. We term these the viscous and elastomeric instabilities, and explore them in the next two subsections respectively. \mecchanged{For each such mode of instability, the activity threshold depends on $\Lambda$. This in turn depends on whether the initial director obeys $\uvtr{n} = \left(0, 1, 0\right)$, for which the growing mode involves a bend deformation, or obeys $\uvtr{n} = \left(1, 0, 0\right)$, for which it involves splay.}

\subsection{Viscous instability}
\label{sec:viscous_criterion}

For the first mode of instability, the
discriminant $b^2 - 4c$ in \eqref{eq:quad} remains positive and the
eigenvalues of $\tsr{M}^k$ are purely real.  The condition
$Re(\omega^+) > 0$ for the onset of instability then simplifies to the
condition $c<0$. Predividing $c$ by the final term on the right hand
side of \eqref{eq:c} and multiplying by $\eta$ then gives the
conditions for instability as
\begin{equation}
\label{eq:torearrange}
\frac{a^2G_C\tau_C}{1+\ell_C^2k^2}+\frac{\Lambda}{\ell_Q^2k^2/\tau_Q}(2G_Q\ell_Q^2k^2\Lambda-\zeta)+\eta<0.
\end{equation}
We identify the four terms on the left hand side as the
wavevector-dependent zero-shear viscosities associated with the
polymer, the passive nematic stress, the active nematic stress and the
solvent respectively. We can therefore rewrite this criterion as
\begin{equation}
  \eta^k_{\rm total}=\eta^k_{C}+ \eta^k_{Q,{\rm passive}}+\eta^k_{Q,{\rm active}}+\eta^k<0,
  \label{eq:lsa_nematic_zc_visc}
\end{equation}
where in fact $\eta^k_{Q,{\rm passive}}$ and $\eta^k$ are independent of $k$.  Each term in \eqref{eq:lsa_nematic_zc_visc} is positive, apart from the \smfchanged{term $\eta^k_{Q,{\rm active}}$. A simplified analysis, that casts the full tensorial nematic order parameter $\tsr{Q}$ in a simplified form in terms of a director $\uvtr{n}$ with a fixed degree of nematic ordering \cite{*[{The equivalence of $\tsr{Q}$ and $\uvtr{n}$ formulations is discussed further in }] [{}] Marenduzzo2007a}, then suggests that}
\begin{equation}
  \label{eq:viscous}
  \eta^k_{Q,{\rm active}} = -\zeta\times \frac{\Lambda}{\ell_Q^2k^2/\tau_Q}=\frac{\partial \Sigma^k_{Q,\rm active}}{\partial n^{k}}\times\frac{\partial n^{k}}{\partial \gdot^k},
\end{equation}
where the multiplication symbols serve to emphasize the correspondence of terms between the two expressions.
In this equation, an initial base state director orientation along $\uvtr{x}$ is linked to a director perturbation $\delta n$ in the $y$ direction, corresponding to a splay instability; whereas an initial base state director along $\uvtr{y}$ is linked to a perturbation $\delta n$ in the $\uvtr{x}$ direction, corresponding to a bend instability. Recall that the perturbation wavevector $\vtr{k}$ is in the $\uvtr{y}$ direction in all these 1D calculations. (Note that all other analytical and numerical results employ the full $\tsr{Q}$ formulation, with the exception of \eqref{eq:elastic} where we make the analogous argument for the elastomeric instability.) Within \eqsref{eq:lsa_nematic_zc_visc} and \ref{eq:viscous} we recognise the mechanism that drives this instability as follows: a perturbation in the shear rate causes a perturbation in the director orientation, which in turn causes a perturbation in the active contribution to the stress field, which must provide a counterbalancing contribution from the other stress components to maintain force balance. The two shear rate perturbations in this loop are in the same sense, signifying positive feedback.

Rearranging Eq.~\ref{eq:torearrange}, we find the critical activity for the onset of instability to be
\begin{eqnarray}
  \label{eq:visc}
  \zeta_c^{\rm visc}&=& \frac{\ell_Q^2k^2/\tau_Q}{\Lambda}\left[   \frac{a^2G_C\tau_C}{1+\ell_C^2k^2}+ 2G_Q\tau_Q\Lambda^2+\eta\right]\nonumber\\
  &=&\frac{r_Q}{\Lambda}\left[ \eta^k_{C}+\eta^k_{Q,{\rm passive}}+\eta^k \right],
\end{eqnarray}
where we recognise $r_Q\equiv \ell_Q^2k^2/\tau_Q$ as the on-diagonal rate of relaxation of a perturbation to the director field at wavelength $k$.

For values of the activity just above threshold, the perturbations grow very slowly.  They furthermore do so without any oscillatory component because the eigenvalues are real in this case.  Relative to this slow growth, the fluid's intrinsic relaxation timescales $\tau_C$ and $\tau_Q$ become infinitely fast as one approaches the threshold. Accordingly the fluid's resistance to this instability appears only via its zero-frequency viscosities, $\eta_C$ and $\eta_Q$: the relaxation times $\tau_C$ and $\tau_Q$ cannot appear in \eqref{eq:lsa_nematic_zc_visc} independently of these viscosities. The effect of the polymer is therefore simply to increase the fluid's overall viscosity, delaying the onset of instability.  The dependence of the threshold on $\eta_C=G_C\tau_C$ is thus independent of whether $\tau_C$ is varied at fixed $G_C$, or $G_C$ at fixed $\tau_C$. Accordingly, we call this mode the `viscous instability' in what follows.

In view of the relatively trivial role of the polymer in determining the onset of this viscous instability, we expect the result just obtained to match onto the one in the earlier literature for a purely nematic active fluid, without polymer. Indeed, substituting $K \equiv \ell_Q^2 G_Q$ recovers the criterion in the form originally derived in Refs.~\cite{Voituriez2005,Edwards2009}, with an additional term  to account for the extra viscosity contributed by the polymer.

In the limit of infinite system size, $k\to 0$ in \eqref{eq:visc}, we further recover the prediction $\zeta_c^{\rm visc}\to 0$, implying that a bulk active nematic will be unstable for any level of activity, however small \citep{AditiSimha2002,Voituriez2005}.  One important prediction of the present work is that this generic instability of a bulk active nematic persists even with polymer present.

\subsection{Elastomeric instability}

An alternative mode of instability arises when the discriminant $b^2 - 4c$ in \eqref{eq:quad} is negative and the eigenvalues of $\tsr{M}^k$ have an imaginary part, suggesting a Hopf bifurcation with oscillatory dynamics \citep{Strogatz2008Book}.  The criterion $Re(\omega^+) = 0$ for the onset of instability then simplifies to the condition
\begin{equation}
\label{eq:torearrange1}
b=\frac{a^2G_C}{\eta}+\frac{\Lambda}{\eta}(2G_Q\Lambda \ell_Q^2k^2-\zeta)+\frac{1+\ell_C^2k^2}{\tau_C}+\frac{\ell_Q^2k^2}{\tau_Q}<0.
\end{equation}

At a given wavevector $k$, we identity the five terms in $b$ as being respectively proportional to (i) the rate of change of polymer stress during an impulsive strain; (ii) the rate of change of passive nematic stress during an impulsive strain; (iii) the rate of change of active nematic stress during an impulsive strain; (iv) the on-diagonal rate of relaxation $r_c\equiv (1+\ell_C^2k^2)/\tau_C$ for a perturbation in the polymer conformation, and (v) the on-diagonal rate of relaxation $r_Q$ for a perturbation in the director field.

Therefore we write the criterion for the onset of instability as
\begin{equation}
\frac{1}{\eta}\frac{d\Sigma^k_{C}}{d\gamma}+\frac{1}{\eta}\frac{d\Sigma^k_{Q,{\rm passive}}}{d\gamma}+\frac{1}{\eta}\frac{d\Sigma^k_{Q,{\rm active}}}{d\gamma}+r_C+r_Q<0.
\end{equation}
Among the terms on the left hand side, each is always positive apart from the third. \ejhchanged{Following the approach described previously for the viscous instability, in which we rewrite the derivative in terms of the director field $n$, the third term reads}
\begin{equation}
\label{eq:elastic}
\frac{\partial \Sigma^k_{Q,\rm active}}{\partial n^{k}}\times\frac{\partial n^{k}}{\partial \gamma^k}.
\end{equation}
We recognise the mechanism that drives this instability as follows: a perturbation in the shear strain field causes a perturbation in the director orientation, which in turn causes a perturbation in the shear stress field, which in turn must cause a perturbation in the strain field to maintain force balance. The two shear strain perturbations in this loop are in the same sense, signifying positive feedback. Indeed this is essentially the same mechanism as for the viscous instability above, but with the strain rate replaced by the strain. Accordingly, we call this mode the `elastomeric instability'.

Rearranging \eqref{eq:torearrange1} we get an expression for the critical activity for the onset of instability as
\begin{eqnarray}
  \zeta_c^{\rm elast} &=& \frac{\eta}{\Lambda}\left[ \frac{a^2G_C}{\eta}+ 2G_Q\frac{\Lambda^2}{\eta} l_Q^2k^2+\frac{1+l_C^2k^2}{\tau_C}+\frac{l_Q^2k^2}{\tau_Q}\right],\nonumber\\
  &=& \frac{1}{\Lambda}\left[\frac{d\Sigma^k_{C}}{d\gamma}+ \frac{d\Sigma^k_{Q,{\rm passive}}}{d\gamma}+\eta(r_C+r_Q)\right].
  \label{eq:lsa_nematic_zc_elast}
\end{eqnarray}
In contrast to the viscous instability discussed above, at the threshold of this new instability the perturbations grow subject to a superposed oscillatory component
that has a time period prescribed by the imaginary \mecchanged{part} of the
eigenvalue, $\tau_\omega = 2 \pi / Im(\omega^+)$. The presence of this time scale
means that the relative value of the polymer's relaxation time is
relevant to this instability, in contrast to its unimportance to the
viscous one: $\tau_C$ enters \eqref{eq:lsa_nematic_zc_elast} directly
rather than merely via the viscosity $\tau_CG_C$. Put differently, the
polymer's {\em viscoelasticity} is important to this new instability,
in contrast to the viscous instability, which depends only on its
viscosity.

Two \mecchanged{distinct thresholds again arise} for the elastomeric instability
according to whether the deformation corresponds to a bend or a splay
in the director field, which is in turn determined by whether the
initial director lies in the $y$ or $x$ direction \ejhchanged{as set by $\Lambda$ (see \eqref{eq:Lambda})}.

\subsection{Viscous - elastomeric crossover}

\begin{figure}[t]
  \centering
  \includegraphics[width=\columnwidth]{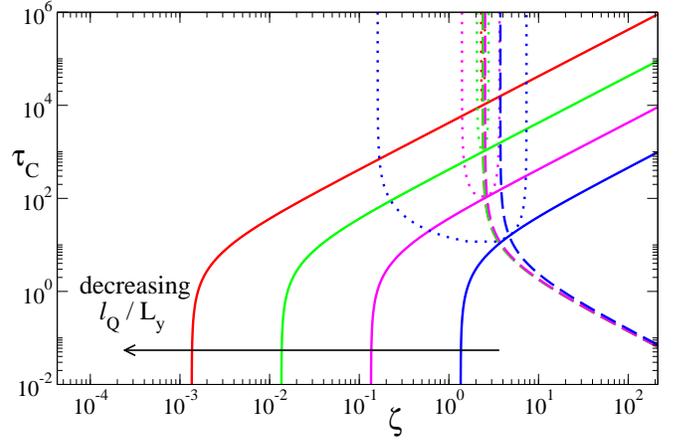}
  \caption{Phase diagram showing the threshold activity for the onset
    of the splay mode of the viscous (solid lines) and elastomeric
    (dashed lines) instabilities, for $k=\pi/L_y$.  The unstable
    region is to the right of the threshold curve in each case.  The
    dotted lines enclose the regime in which the eigenvalues have an
    imaginary contribution, signifying oscillatory dynamics. Sets of
    curves from left to right have $\ell_C^2 = \ell_Q^2 = 10^n$ with
  $n=-5,-4,-3,-2$. For all curves the polymer modulus $G_C=0.1$}
  \label{fig:one}
\end{figure}

\begin{figure}[t]
  \centering
  \includegraphics[width=\columnwidth]{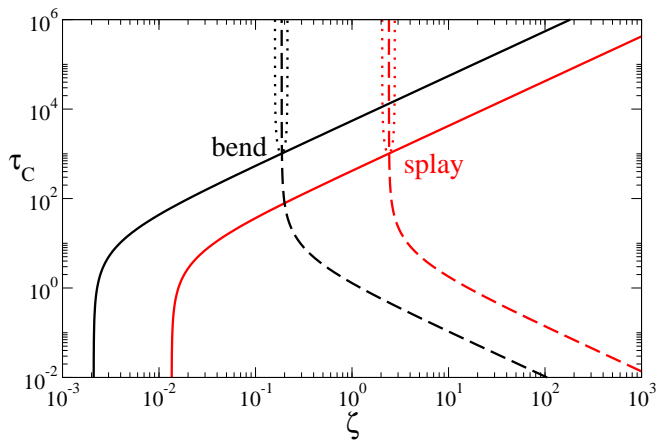}
  \caption{Phase diagram showing the threshold activity for the onset of the viscous (solid lines) and elastomeric (dashed lines) instabilities, for $k=\pi/L_y$. Leftmost and rightmost curves correspond to the bend and splay modes of instability respectively. The unstable region is to the right of the threshold curve in each case.  The dotted lines enclose the regime in which the eigenvalues have an imaginary contribution, signifying oscillatory dynamics.  For all curves $G_C=0.1$ and $l_C^2 = l_Q^2 = 10^{-4}$.}
  \label{fig:two}
\end{figure}

As described in the previous two subsections, viscoelastic active
matter is predicted to exhibit two distinct modes of instability: a
viscous mode and an elastomeric one.  \mecchanged{The activity thresholds for
these two modes are compared} in the
phase diagrams of \figsref{fig:one} and \ref{fig:two}.

As can be seen, the thresholds for the two instabilities display
qualitatively different dependences on the polymer relaxation time
$\tau_C$ and on the system size. Indeed, taking the elastomeric limit
$\tau_C\to\infty$ for any fixed $G_C$ and fixed $\ell_C=\ell_Q$
\mecchanged{raises indefinitely} the threshold for the viscous instability: $\zeta_c^{\rm
  visc}\to\infty$.  \mecchanged{The fluid's
viscosity approaches infinity in this limit, switching off the
viscous instability entirely}. In contrast the threshold for the
onset of the elastomeric instability remains finite even in the
\mecchanged{purely}
elastomeric limit $\tau_C\to\infty$:
\begin{equation}
  \lim_{\substack{\tau_C \to \infty,\\ k \to 0}} \zeta_c^\textrm{elast} = \frac{a^2 G_C}{\Lambda},
  \label{eq:lsa_nematic_zc_elast_lims}
\end{equation}
(where we have also taken the infinite system size limit for clarity
of expression).  Crucially, then, \mecchanged{this second} instability can arise even in the
limit of an elastomeric solid, \mecchanged{because it}
involves perturbations in strain rather than strain rate.

\mecchanged{Now consider instead fixing $\tau_C,G_C$, so that the polymer
viscosity remains finite, and taking the limit
$\ell_C=\ell_Q\to 0$. This is equivalent to taking the limit of
infinite system size $L_y\to\infty$, for which the threshold for the viscous
instability tends to zero}: for a bulk system, any level of activity,
however small, \mecchanged{triggers this} instability.  In
contrast the \mecchanged{activity threshold for} the elastic instability
remains finite even in this bulk limit. \ejhchanged{These two facts can be seen in \figref{fig:one}.}

The physical origin of this important \mecchanged{difference in system-size dependence can} be understood by
comparing the driving terms \mecchanged{for the two modes}.  The term $\partial
n/\partial\gdot$ in the viscous criterion (recall \eqref{eq:viscous})
scales as $1/k^2$, \mecchanged{because an infinitesimal shear rate perturbation
can cause finite rotation of} the director field in the limit of $k\to 0$.
(This \mecchanged{stems from the} Goldstone mode associated with the spontaneously broken
rotational symmetry. \mecchanged{A nematic phase} exerts no restoring
force against global rotations of a state of uniform $\tsr{Q}$, implying
by continuity that the force likewise vanishes in the limit $k \to 0$
\citep{Chaikin2000Book}.)  In contrast, the term $\partial
n/\partial\gamma$ in the elastic criterion (recall \eqref{eq:elastic})
is \mecchanged{finite at low $k$, because small strain perturbations cause
only small changes in director, even in the limit of zero
wavevector.}

With these remarks in mind, \mecchanged{we now see that dominance of the viscous
instability over the elastomeric one (or vice versa) will depend on the ratio $\tau_C/\tau_Q$ of the polymer and
nematic relaxation times, and also on the system size, measured in relation to the
lengths $l_Q,l_C$ (as set by the gradient terms in the equations of motion).} For any fixed system size, below a critical value of the polymer relaxation time $\tau_C = \tau_C^*$ the \mecchanged{viscous instability threshold $\zeta_c^\textrm{visc}$ lies below the elastic one: the viscous instability arises first on increasing the activity, and $\zeta_c^\textrm{visc}$ sets the stability criterion for our base state. In contrast, for $\tau_C > \tau_C^*$ the elastomeric instability arises before the viscous one, and $\zeta_c^\textrm{elast}$ sets the relevant criterion.}

This crossover value $\tau^*_C$ can easily be shown to obey
\begin{equation}
 \tau_C^* \approx (1+\ell_C^2k^2) \frac{\tau_Q}{\ell_Q^2k^2},
  \label{eq:lsa_nematic_crossover}
\end{equation}
which for large system sizes simplifies to $\tau_C^* \approx
\tau_Q/\ell_Q^2k^2$.  This form can be understood as follows. The
quantity $\tau_Q/\ell_Q^2k^2$ is the orientational relaxation time for
long-wavelength distortions in the nematic phase, which diverges as
$k\to 0$.  (Recall that this is a Goldstone mode, with no penalty for
global rotations in the director field.) For polymer relaxation times
faster than this, $\tau_C < \tau_C^*$, the polymer behaves in a
viscous way relative to the nematic mode, and the viscous instability
dominates.  In the opposite case the polymer behaves \mecchanged{elastically}, and the elastomeric instability dominates.

In the limit of infinite system size the crossover value \mecchanged{diverges:}
$\tau_C^*\to\infty$.  In this bulk limit the viscous instability \mecchanged{is triggered first (at} zero activity threshold) for all
values of $\tau_C$, \mecchanged{whereas} the threshold of the elastomeric instability
remains finite even for a bulk sample. As noted in the introduction,
however, many biological environments are strongly confined.  This
\mecchanged{motivates} the use of a finite box size in our simulations, and a
correspondingly careful exploration of both the viscous and
elastomeric instabilities.  For our typical choice of parameter values
given in \tabref{tbl:simulation_params_dimensionless}, the crossover
between the two instabilities occurs at $\tau_C^* \approx
\eta_Q/(Kk^2) = \mathcal{O}(10^3\tau_Q)$.

\section{Nonlinear dynamics (1D)}
\label{sec:1D_nonlinear}

\begin{figure}[t]
  \centering
  \includegraphics[width=\columnwidth]{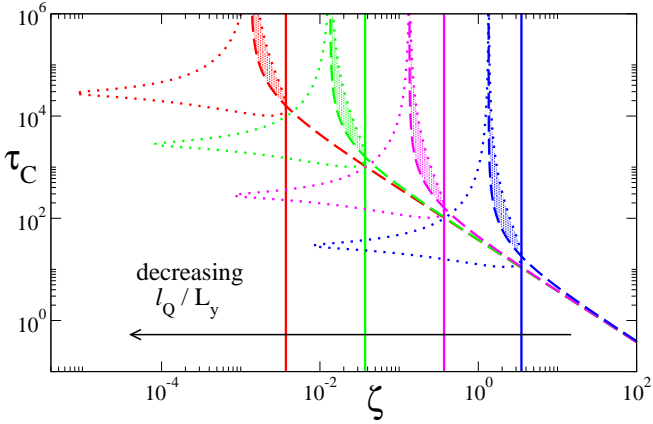}
  \caption{Phase diagram showing the threshold activity for the onset of the viscous (solid lines) and elastomeric (dashed lines) instabilities in the splay mode, for $k=\pi/L_y$.  The unstable region is to the right of the threshold curve in each case.  The dotted curves enclose the regime in which the eigenvalues have an imaginary contribution, signifying oscillatory dynamics (shaded regions are both oscillatory and unstable for $k = \pi / L_y$). Sets of curves from left to right have $l_C^2 = l_Q^2 = 10^{-5, -4, -3, -2}$.  For all curves the polymer viscosity $\eta_C = G_C\tau_C = 1$. }
  \label{fig:1D_nonlinear_analytical_fixed_eta}
\end{figure}

In the previous section we derived the activity threshold for linear instability to a state of spontaneous flow in an active nematic with added polymer.  We now use those results as a roadmap to explore the model's full nonlinear \mecchanged{dynamics}.  We shall restrict ourselves in this section to 1D calculations, with heterogeneity in the $y$ direction only, deferring a study of 2D effects to \secref{sec:2D_nonlinear} below.  In each run we initialise the system in the homogeneous base state given by \eqref{eq:base}, with the director $\uvtr{n}$ aligned in the $x$-direction.  Accordingly, any instabilities that arise below originate from splay deformations. We subject this base state to small added perturbations at the start of the run of the form $\delta \tilde{\gdot} = \sum_{\modenumber=1}^{N} \delta A_{\modenumber} \cos(\pi \modenumber y / L_y)$, where $\delta \sim \mathcal{O}(10^{-10})$, $A_\modenumber$ is a random number drawn from a uniform distribution on $[-1, 1]$ and $N = 32$.  In this way, the early-time regime of these nonlinear runs acts as an independent cross-check of the linear stability calculations in the previous section.

Our model has three competing local relaxation times: the liquid-crystal relaxation time $\tau_Q$, the polymer relaxation time $\tau_C$, and the active forcing timescale $\tau_a = \eta / \zeta$, which was identified in Ref.~\cite{Giomi2012} by balancing the active modulus $\zeta$ against the solvent viscosity $\eta$. As discussed previously, the first of these, \mecchanged{$\tau_Q$, is} our unit of time. We now examine the competition between the second and third timescales by producing phase diagrams in the $(\zeta, \tau_C)$ plane.  Related to $\tau_a$ is the active timescale $\bar{\tau}_a = \eta / (\zeta - \zeta_c)$, with the denominator shifted to account for the fact that in a finite system the critical threshold for the onset of activity is non-zero.  (Note that $\bar{\tau}_a \to \tau_a$ as $L_y \to \infty$ or $\zeta \to \infty$.)

Whenever the active timescale is shorter than the two relaxation times, we can expect oscillatory dynamics \citep{Giomi2012}. We will confirm this expectation: first (in \secref{sec:1D_nonlinear_fixed_eta}) for a material of fixed polymer viscosity $\eta_C = 1$, where we find the dominant period of oscillation $\tau_{osc}$ to be set by $\tau_C$, and then (in \secref{sec:1D_nonlinear_spont_flow_solid}) for an elastomeric material with $\tau_C\to\infty$, where instead we find $\tau_{osc} \propto \bar{\tau}_a$.  In a biological context, oscillatory states have been observed in fibroblast cells \citep{Salbreux2007} with a period of oscillation proportional to the myosin motor activity. \mecc{Similar states have also been studied in \textit{Drosophila} embryos \cite{Dierkes2014}, where oscillations are thought to play a key role in cell shape formation (\textit{morphogenesis}).} %

\subsection{Phase diagram: fixed viscosity}
\label{sec:1D_nonlinear_fixed_eta}

\begin{figure}[t]
  \centering
  \includegraphics[width=\columnwidth]{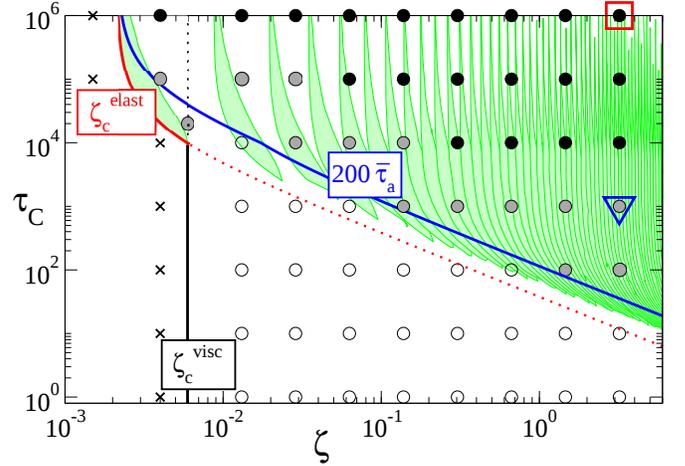}
  \caption{Phase diagram showing the competing effects of the activity $\zeta$ and the polymer's viscoelastic relaxation timescale $\tau_C$, for fully nonlinear runs in 1D with fixed polymer viscosity $\eta_C = 1$. Marked are the critical activities $\zeta_c^\textrm{visc}$ (\eqref{eq:lsa_nematic_zc_visc}, black line) and $\zeta_c^\textrm{elast}$ (\eqref{eq:lsa_nematic_zc_elast}, red line) predicted by the linear stability analysis. Nonlinear states are denoted by symbols: quiescent (crosses), static bands (empty circles), oscillating bands (shaded circles) or oscillating and flipping bands (filled circles) where the flow switches direction. In the green shaded regions our linear stability analysis predicts unstable oscillatory growth, at some wavevector $k$. Examples of the states highlighted with the triangle and square are given in \figref{fig:1D_nonlinear_stp_PiC} (left and right respectively).  Oscillations are observed for $\tau_C / \bar{\tau}_a > 200$ (blue curve). \params $\eta_C = 1$, $\ell_Q = \ell_C = 0.004$.}
  \label{fig:1D_nonlinear_tz_PD_fixed_eta}
\end{figure}

\begin{figure*}[t]
  \centering
  \includegraphics[width=\textwidth]{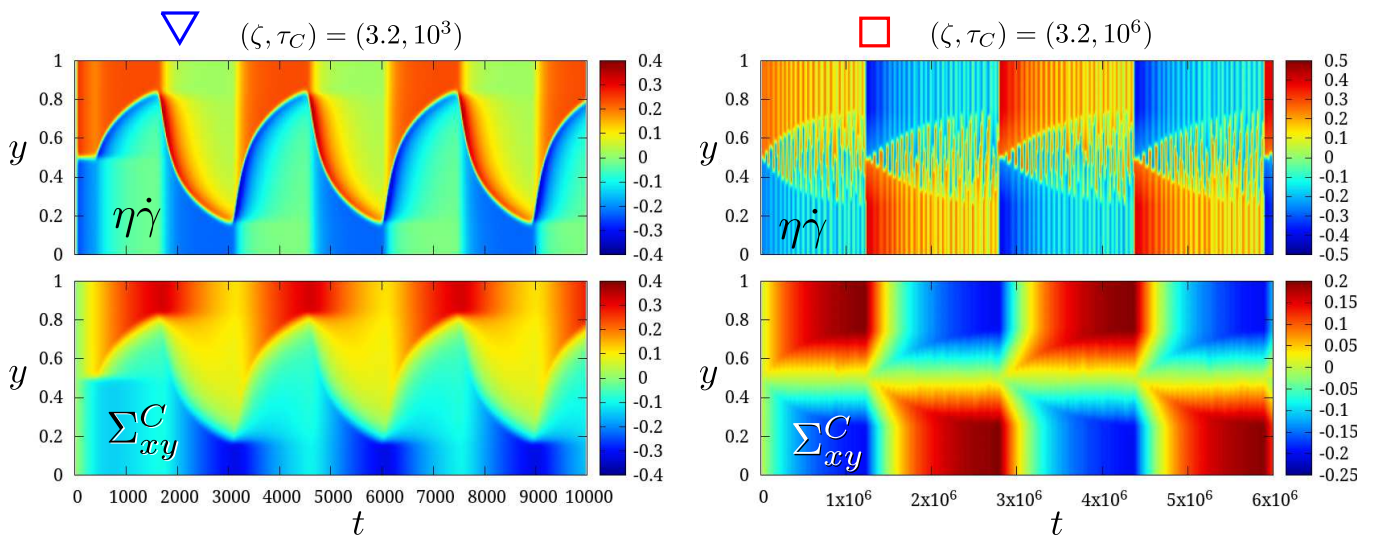}
  \caption{Spacetime plots corresponding to the blue triangle and red square in the phase diagram of \figref{fig:1D_nonlinear_tz_PD_fixed_eta}. The color scale represents: $\eta\gdot$ (top row), $\Sigma_{xy}^C$ (bottom row).  \textit{Left column:} State with oscillating interface ($\zeta = 3.2$, $\tau_C = 10^3$), \textit{Right column:} State with oscillating interface and switching flow direction ($\zeta = 3.2$, $\tau_C = 10^6$). }
  \label{fig:1D_nonlinear_stp_PiC}
\end{figure*}

We begin by considering the case of a fixed polymer viscosity $\eta_C=G_C\tau_C=1$. (Recall that the passive nematic viscosity $\eta_Q=1$ by our choice of units, and the solvent viscosity $\eta=0.567$ for consistency with Ref.~\cite{Fielding2011}.)  Between runs we vary the polymer relaxation time $\tau_C$, and in tandem therefore also vary $G_C = 1/\tau_C$.  This will allow us to confirm the direct role of the polymer relaxation time $\tau_C$ in the dynamics, while removing any potential complications arising from variations in the overall viscosity.  In varying $\tau_C$ in this section we choose to keep $\ell_C=\ell_Q$ fixed. Note that this effectively eliminates the diffusivity $\Delta_C=\ell_C^2/\tau_C$ from the polymer dynamics in the elastomeric limit $\tau_C\to\infty$.  The predictions of our linear stability analysis in this case are shown in the $(\zeta, \tau_C)$ plane in \figref{fig:1D_nonlinear_analytical_fixed_eta}, with four sets of curves corresponding to different values of $\ell_C=\ell_Q$.

The results of our nonlinear simulations at fixed $\ell_Q = \ell_C = 0.004$ are shown in \figref{fig:1D_nonlinear_tz_PD_fixed_eta}, with the linear stability thresholds also repeated on the same axes for comparison. The green shaded region indicates when oscillatory instabilities are expected from the linear stability analysis at any point in the discrete spectrum of modes $k = \modenumber \pi / L_y$ (cf.~\figref{fig:1D_nonlinear_analytical_fixed_eta} where oscillatory instabilities are also marked with shaded regions, but where we instead fix $k = \pi / L_y$ and plot multiple values of $\ell_Q = \ell_C$).  As can be seen, the nonlinear results are consistent with those of the linear stability analysis.  For values of the activity below $\zeta_c = \textrm{min}\left(\zeta_c^\textrm{visc}, \zeta_c^\textrm{elast}\right)$ we observe quiescent states (black crosses) in our nonlinear simulations, consistent with the analytical prediction of a stable base state in this regime.  For values of the activity above $\zeta_c$ we observe spontaneously flowing shear banded states (black circles).

These spontaneously flowing states may be categorised according to whether they approach a steady state at long times, or instead oscillate.  In particular, for activity values exceeding the threshold for the viscous instability, but below that for the elastomeric instability, we observe steady shear banded states (open circles), as seen in earlier studies of the model in the absence of polymer \citep{Fielding2011}. In contrast, for values of the activity that also exceed the threshold for the elastomeric instability, and roughly correspond to $\tau_C / \bar{\tau}_a > 200$ (blue line in \figref{fig:1D_nonlinear_tz_PD_fixed_eta}), we find time-dependent states in which the shear bands oscillate with a period set by the viscoelastic relaxation timescale $\tau_C$.  No counterpart to these oscillatory states was observed in the earlier study without polymer \citep{Fielding2011}. This confirms our view that the polymer relaxation time is integral to their dynamics.

The separation of timescales (large $\tau_C/\tau_a$) required to see these oscillations is reminiscent of earlier work addressing rheological chaos \citep{Aradian2004}.  Intuitively, if the active timescale $\tau_a=\eta/\zeta$ is sufficiently long, then both polymer and liquid crystal can relax any activity-driven deformation and the resulting shear-banded steady state is time-independent.  Conversely, if the polymer cannot relax the activity-induced stress quickly enough, the polymer dynamics lag behind, resulting in oscillations.  This is also somewhat analogous to the mechanism described in Ref.~\cite{Giomi2012}, although in that case the coupling was to a slowly diffusing concentration field rather than a slowly relaxing polymer.

To gain further insight into these oscillatory shear banded states, we now explore in detail two particular examples denoted respectively by the blue triangle and red square in the phase diagram of \figref{fig:1D_nonlinear_tz_PD_fixed_eta}.

\textit{Oscillatory interface, fixed flow direction.}
The first example, denoted by the blue triangle in \figref{fig:1D_nonlinear_tz_PD_fixed_eta}, corresponds to a time-dependent state in which the position of the interface between two shear bands oscillates as a function of time.  See \figref{fig:1D_nonlinear_stp_PiC} for a space-time plot of the shear rate $\gdot$ (top left) and of the polymer shear stress, $\Sigma^C_{xy}$ (bottom left).  The time period is set by the polymer relaxation time $\tau_C$, making it clear that these oscillations are governed by the viscoelastic dynamics of the polymer.  As the interface position deviates from the centre, large polymeric stresses are seen to develop in the narrower band until eventually the interface turns round and returns in the opposite direction, relaxing the polymer somewhat.  The total shear stress and integrated throughput, $\Psi = \int_0^{L_y} v_x dy$ also have time-dependent signals (not shown) with a period set by $\tau_C$.

In this example the sign of the throughput --- {\it i.e.,} the overall
direction of the flow --- remains constant in time. This can be seen by
noting that the region of negative shear rate remains always below the
one of positive shear rate in \figref{fig:1D_nonlinear_stp_PiC} (top
left), meaning that the velocity profile is always overall nose-shaped
and pointing to the left. (\mecchanged{Note that a} state with sustained throughput to the
right could equally well have developed, \mecchanged{had we made} small changes in
the initial conditions.)

\textit{Oscillatory interface, switching flow direction.}
The second example is denoted by the red square in \figref{fig:1D_nonlinear_tz_PD_fixed_eta}, with space-time plots of $\gdot$ and $\Sigma_{xy}^C$ in \figref{fig:1D_nonlinear_stp_PiC} (right). Here two distinct kinds of oscillation are seen in the course of any run. On a slow timescale $\approx 2\tau_C$ the flow periodically reverses direction, as evidenced by the up-down switching between the bands of positive and negative shear rate. Another, more rapid, oscillation is evident: the throughput direction remains the same but the position of the interface fluctuates with a mechanism resembling the one discussed in the previous example above.

As already noted above, in this section we have kept the values of $\ell_C = \ell_Q$ fixed as the polymer relaxation time changes. Had we instead chosen to keep $ \ell_Q^2 / \tau_Q = \ell_C^2 / \tau_C \equiv \Delta$ fixed as $\tau_C$ varied, the initial linear instability would not be oscillatory. While a oscillatory linear instability doesn't necessarily imply oscillatory nonlinear dynamics, our results in this section demonstrate a reasonable correlation between the two (compare green region in \figref{fig:1D_nonlinear_tz_PD_fixed_eta} with oscillatory banded states (shaded and filled circles)).

\subsection{Spontaneous flow in elastomeric solids}
\label{sec:1D_nonlinear_spont_flow_solid}

\mecc{The case of an active elastomeric solid can be approached within our formalism by more than one route, yielding slightly different limiting descriptions that may be appropriate to different physical cases. One choice concerns the way in which the infinite viscosity limit is taken; a second and independent choice concerns the behavior of the polymer stress diffusivity $\Delta_C$. } In the previous section we studied the dependence of the model's behavior on the polymer relaxation \mecchanged{time $\tau_C$ at fixed polymer viscosity} $\eta_C=G_C\tau_C=1$. We also fixed the values of the microscopic lengthscales $\ell_C=\ell_Q$, in relation to the macroscopic system size. In the regime of large polymer relaxation times ($\tau_C\simeq 10^6$) we found sustained oscillations in the system's long time response. Such large values of $\tau_C$ correspond for all practical purposes to an elastomeric solid, but with a tiny modulus $G_C$ to maintain the overall constraint of fixed polymer viscosity $\eta_C=G_C\tau_C=1$.

In this section we consider instead taking the ``true" limit of an elastomeric solid $\tau_C \to \infty$ at a fixed value of the polymer modulus $G_C$, so that the polymer viscosity $\eta_C=G_C\tau_C$ now also diverges as $\tau_C\to\infty$. We also fix the value of the diffusivity parameter $\ell_Q^2 / \tau_Q = \ell_C^2 / \tau_C = \Delta$. In this way, taking the limit $\tau_C\to\infty$ means taking the limit $l_C\to\infty$ also.

\mecchanged{Our choice of nonzero polymer stress diffusivity $\Delta_C=\Delta>0$ in the elastomeric limit is not obvious, since physically there is no reason for $l_C$ to diverge in that limit.  However, we have found that in our 2D nonlinear work setting $\Delta_C = 0$ generally leads to a short-scale numerical instability \ejhchanged{unless prohibitively small time- and spatial- stepsizes are used.} We explore this issue further in Sec.~\ref{sec:2D_nonlinear_elast}, but even there set $\Delta_C = \Delta$ to retain direct comparability of parameters between the 1D and 2D numerics.}

\begin{figure}[t]
  \centering
  \includegraphics[width=\columnwidth]{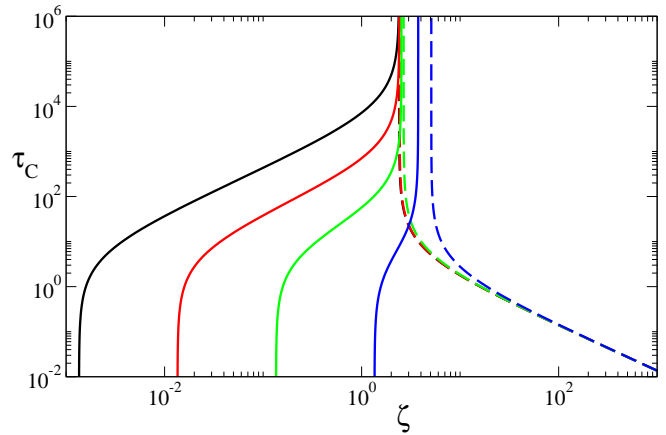}
  \caption{Phase diagram showing the threshold activity for the onset of the viscous (solid lines) and elastomeric (dashed lines) instabilities in the splay mode, for $k=\pi/L_y$.  The unstable region is to the right of the threshold curve in each case.  Sets of curves from left to right have $\Delta_C = \Delta_Q = 10^{-5,-4,-3,-2}$. For all curves the polymer modulus $G_C=0.1$. }
  \label{fig:six}
\end{figure}

\begin{figure}[t]
  \centering
  \includegraphics[width=\columnwidth]{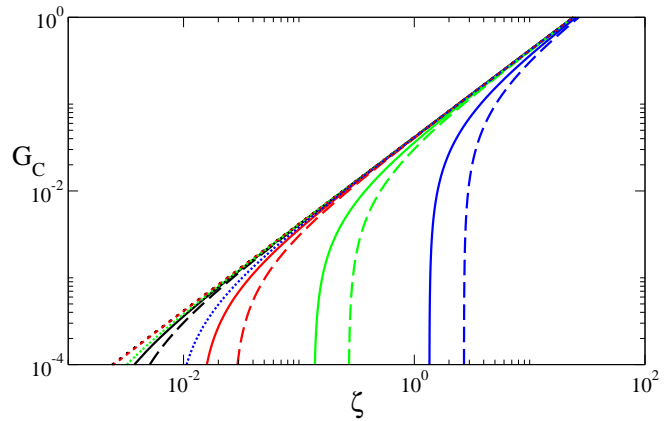}
  \caption{Phase diagram showing the threshold activity for the onset of the viscous (solid lines) and elastomeric (dashed lines) instabilities in the splay mode, for $k=\pi/L_y$.  The unstable region is to the right of the threshold curve in each case. The dotted line shows the locus where $b^2 - 4c = 0$; when $\ell_Q^2 / \tau_Q = \ell_C^2 / \tau_C = \Delta$ this quantity never becomes negative and therefore the linear instability is never oscillatory.  Sets of curves from left to right have $\Delta_C = \Delta_Q = 10^{-5,4,3,2}$. For all curves the polymer relaxation time $\tau_C=\infty$.}
  \label{fig:fifteen}
\end{figure}

The predictions of our linear stability analysis in this case are shown in Fig.~\ref{fig:six} in the $(\zeta,\tau_C)$ plane, for fixed $G_C=0.1$ and various fixed values of $\Delta$.  In this figure, $\tau_C\to\infty$ is approached by taking the ordinate to infinity.  Note how \figref{fig:six}, which has fixed $G_C$ and $\Delta_C$, differs from \figref{fig:one}, which had fixed $G_C$ and $\ell_C$.  In particular, the viscous instability is not eliminated when the elastomeric limit is taken at fixed $\Delta_C$ in \figref{fig:six}, apparently because the polymer stress can still redistribute itself spatially by diffusion despite having a divergent local viscoelastic relaxation time $\tau_C$. (The latter ensures that the volume-averaged polymer stress cannot decay; see Sec.~\ref{sec:2D_nonlinear_elast} for further discussion.)  Furthermore, the regime in which an oscillatory instability is predicted has been eliminated in \figref{fig:six} compared to its strong presence in \figref{fig:one}.  Linear stability results calculated in the limit $\tau_C=\infty$, again at various fixed $\Delta_Q = \Delta_C \equiv \Delta$, are shown in the plane of polymer modulus and activity $(\zeta, G_C)$ in Fig.~\ref{fig:fifteen}.

\begin{figure}
  \centering
  \includegraphics[width=\columnwidth]{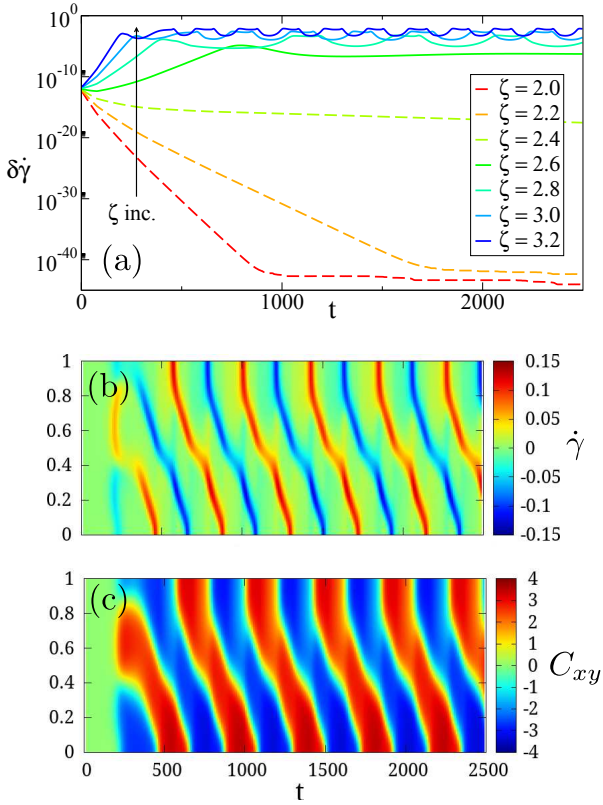}
  \caption{1D simulations with infinite polymer relaxation time. (a) Evolution of shear-rate perturbations for a range of activities $\zeta$, showing growth for $\zeta > \zeta_c$ (solid lines) and decay for $\zeta<\zeta_c$ (dashed lines). (b,c) Space time plots for $\zeta = 3.2 > \zeta_c$. Colourscale is (b) shear-rate $\partial_y v_x = \gdot$, (c) $C_{xy}$. \params $\tau_C \to \infty$, $G_C = 0.1$, $\Delta = 10^{-4}$.}
  \label{fig:1D_nonlinear_elast_numerics}
\end{figure}

\begin{figure}
  \includegraphics[width=\columnwidth]{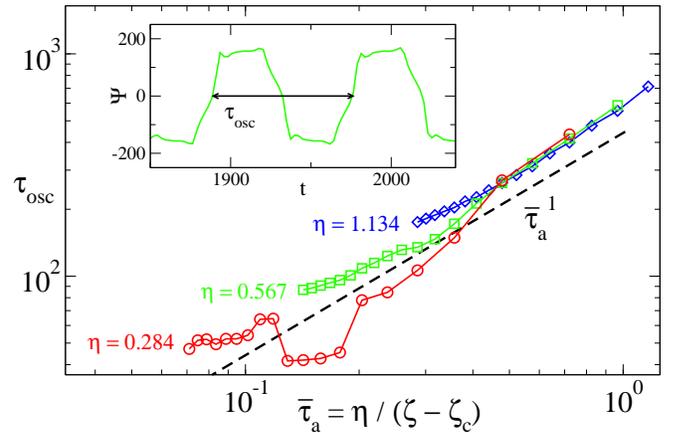}
  \caption{Dominant period of oscillation $\tau_{osc}$ versus the active timescale $\bar{\tau_a}$ (\eqref{eq:1D_nonlinear_t_osc}). Inset: Throughput-time series for $\zeta = 6.4$, $\eta = 0.567$, with $\tau_{osc}$ marked. \params $\eta = 0.284$ (red circles), $0.567$ (green squares), $1.134$ (blue diamonds), $\tau_C \to \infty$, $G_C = 0.1$, $\Delta = 10^{-4}$. }
  \label{fig:1D_nonlinear_elast_t_osc}
\end{figure}

With these linear stability results in mind, we now perform nonlinear runs in the elastomeric limit $\tau_C=\infty$ for the particular case of $G_C=0.1$ and $\Delta=10^{-4}$. For these parameters the critical activity threshold is $\zeta_c=2.41$.  Accordingly, we simulate a sequence of activity values $\zeta = 2 \to 3.2$ straddling this threshold. The results are shown in \figref{fig:1D_nonlinear_elast_numerics}. The upper panel shows the time evolution of the (largest mode of the) perturbation to the base state, clearly showing exponential decay (linear stability) for activity values $\zeta<\zeta_c$ and early time exponential growth (linear instability) for $\zeta>\zeta_c$, giving way to saturation at long times due to nonlinear effects. We have checked that the slopes of the lines in \figref{fig:1D_nonlinear_elast_numerics}a) correspond to the eigenvalues predicted by the linear stability
analysis.

For the particular run $\zeta = 3.2$ we now examine the system's behavior at long times, once it has settled to the ultimate nonlinear attractor of the dynamics. As before we find states that are shear banded.  Note, however, that conventional static shear bands are forbidden by the elastomeric nature of the polymer: a static band with a fixed non-zero shear rate would indefinitely load the polymer, leading to divergent polymer stresses. Indeed, inspecting the shear-rate profiles $\gdot(y,t)$ in \figref{fig:1D_nonlinear_elast_numerics}b, we see instead travelling bands of local shear-rate $\gdot = \pm 0.15$, such that the fluid at any location in the flow cell \mecchanged{is alternately} sheared forward then backward. Corresponding to this, the shear component $C_{xy}$ of the polymer stress (\figref{fig:1D_nonlinear_elast_numerics}b) also \mecchanged{alternates in} sign. These results are reminiscent of traveling density bands seen experimentally in cytoskeletal extracts \citep{Schaller2010}.  Note that the initial {\em linear} instability is not oscillatory, consistent with the prediction in our linear stability analysis that eigenvalues are purely real for $\ell_Q^2 / \tau_Q = \ell_C^2 / \tau_C = \Delta$; recall \figsref{fig:six} and \ref{fig:fifteen}.

Because the polymer relaxation timescale $\tau_C$ is infinite in these runs, the period of oscillation $\tau_{osc}$ must now be set by a separate timescale.  To identify this timescale we performed simulations for a range of values of the activity $\zeta$ and solvent viscosity $\eta$.  In each, we measured $\tau_{osc}$ via the largest peak in the Fourier series of the throughput as a function of time. See \figref{fig:1D_nonlinear_elast_t_osc} inset. Collecting the data from all these runs in \figref{fig:1D_nonlinear_elast_t_osc} (main), we find reasonable evidence for the scaling
\begin{equation}
  \tau_{osc}\propto \bar{\tau}_a \equiv \eta / (\zeta - \zeta_c).
  \label{eq:1D_nonlinear_t_osc}
\end{equation}
Physically $\bar{\tau}_a$ is related to the timescale of active forcing identified in Ref.~\cite{Giomi2012}.  As discussed in that study, if the active forcing is faster than the relaxation timescales, $\bar{\tau}_a < \tau_Q \ll \tau_C$, both nematic and polymer lag behind, resulting in oscillatory behavior.  Note, however, that for large values of the activity the flow becomes increasingly aperiodic and $\tau_{osc}$ becomes less clearly defined, resulting in minor deviations from the suggested scaling law.

To summarise, we have shown both by linear stability analysis and also by full nonlinear simulations (so far in one spatial dimension) that an elastomeric active nematic generically undergoes instability towards a spontaneously flowing state which must be oscillatory, otherwise the polymers would suffer infinite loading.

\section{Nonlinear dynamics (2D)}
\label{sec:2D_nonlinear}

In the previous section we saw a rich array of dynamical states with heterogeneity in one spatial dimension (1D). These would clearly have been forbidden in any description that constrained the system to remain homogeneous in all directions (0D). We now increase the dimensionality further, to consider spatio-temporal dynamics in 2D. Two new features are immediately anticipated that were forbidden in 1D. One is the presence of $\pm \frac{1}{2}$ point defects in the director field \citep{Thampi2013,Giomi2014a,PrashantPreprint2015}. Another is the possibility of extensional flow, which was forbidden by the constraint of fluid incompressibility in a 1D flow field of the form $\vtr{v}=v(y)\hat{\vtr{x}}$.

\begin{figure}
  \centering
  \includegraphics[width=0.48\textwidth]{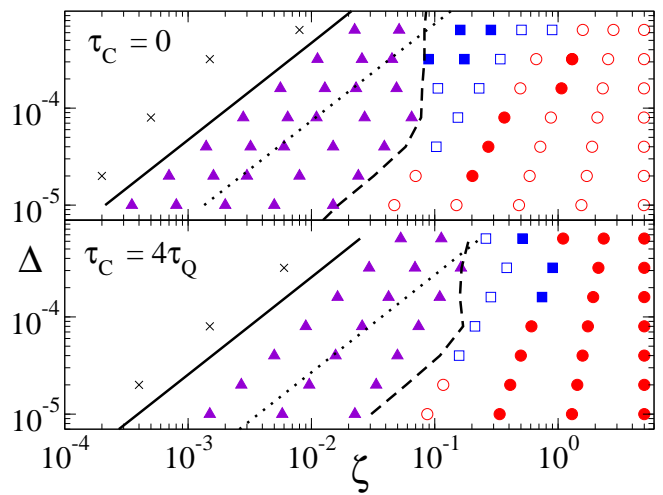}
  \caption{Phase diagrams in the plane of diffusivity $\Delta$ and
    activity $\zeta$, without polymer (top panel) and with polymer of
    relaxation time $\tau_C = 4$ and viscosity $\eta_C=1$ (lower
    panel).  Solid line: threshold for the onset of a 1D bending
    instability starting from a homogeneous initial state with the
    director $\uvtr{n}$ along $y$.  Dotted line: counterpart for a 1D
    splay instability given an initial director along $x$.  Symbols
    show the results of 2D numerical runs \ejhchanged{with director $\uvtr{n}$
    initially aligned along $y$ (subject to perturbations as described in the text)}. Triangles: 1D banded states; squares: oscillatory
    states; and circles: unsteady or chaotic.
    Dashed line: numerically observed crossover line $\zeta_c^{2D}$
    beyond which 1D states always destabilise to the formation of 2D
    flows.  Open and filled symbols denote states with zero and
    non-zero net throughput respectively.}
  \label{fig:2D_nonlinear_PD_tauC_4}
\end{figure}

Consistent with these expectations, an earlier study of active nematics (without polymer) indeed found a much richer spectrum of phase behavior in 2D than in 1D~\citep{Fielding2011}. We now briefly remind the reader of those earlier findings, as a starting point from which to understand the effects of polymer below.  Accordingly, the model's phase behaviour without polymer is summarised in \figref{fig:2D_nonlinear_PD_tauC_4} (top) in the plane of $(\zeta, \Delta)$, where $\Delta \equiv \ell_Q^2/\tau_Q$. The solid line shows the linear instability threshold (\eqref{eq:lsa_nematic_zc_visc}) for the onset of the 1D viscous bending instability, given a homogeneous initial state with the director $\uvtr{n}$ along $y$. The dotted line is the counterpart for the 1D viscous splay mode, given an initial director along $x$.  As expected the instability manifests itself at high $\zeta$ and low $\Delta$, consistent with the fact that smaller system sizes (large $\Delta$) tend to be stabilising.

In the same diagram, the results of numerical runs performed \mecchanged{without polymer} in 2D are shown by symbols. \ejhchanged{Runs are initialised with $\uvtr{n}$ along $y$, with a perturbation $\delta \tilde{Q}_{\alpha\beta} = \sum_{n}^{32} \sum_{m}^{32} \delta A_{\alpha\beta}^{mn} \cos{(\pi m y / L_y)} \cos{(2 \pi n x / L_x + \theta_{\alpha\beta}^{n})}$, where $\delta = 10^{-10}$, $A_{\alpha\beta}^{mn}$ is a randomly selected magnitude drawn from a uniform distribution on $[-1, 1]$ and $\theta_{\alpha\beta}^n \in [0, 2\pi)$ is a randomly selected phase (analogous perturbations are also added to $C_{\alpha\beta}$ when present).} \mecchanged{In} the region where no linear instability is predicted the system remains homogeneous and quiescent (no spontaneous flow), even in these 2D runs. In contrast, beyond the threshold of the (bend) instability we find spontaneously flowing states, as expected. For activities only just above threshold, the ultimate solution corresponds to a steady 1D shear-banded state (triangular symbols), even though these runs in principle allow heterogeneity in 2D. In contrast, further into the unstable regime we find fully 2D spontaneous flow states. \ejhchanged{At intermediate activities we observe a range of oscillatory states, including states in which defect pairs undulate along the channel; these we denote by squares in \figref{fig:2D_nonlinear_PD_tauC_4}. Spatially, these states exhibit repeating structures at a lengthscale roughly set by the channel width $L_y$. As the activity is increased (or $\Delta$ decreased), these oscillatory states gain additional frequency components when plotting a given scalar observable (\eg throughput) against time. When there is no longer any discernible periodicity in this signal, we term the state aperiodic. In this limit the characteristic lengthscale of the resulting nematic texture decreases \citep{Thampi2014b,PrashantPreprint2015} and the order parameter fields lose any obvious spatial periodicity.} The numerically observed crossover line $\zeta_c^{2D}$ separating 1D from 2D states is shown as a dashed line.

In each run, once the system has attained its ultimate attractor we
measure the net throughput of fluid along the channel in the main flow
direction $x$.  The criterion that we adopt to represent significant
throughput is discussed in the Appendix. States meeting
this \mecchanged{throughput criterion are} represented in
\figref{fig:2D_nonlinear_PD_tauC_4} by solid symbols, and those
without significant throughput are shown by open symbols.  As might be
expected, the 1D laminar banded states show significant throughput. In
contrast most of the 2D states, with just a few exceptions, lack any
coherent throughput: the more complicated velocity rolls associated
with them have no overall flow direction.

With \mecchanged{the above} discussion in mind, we now consider the
effects of adding polymer. We start in
\secref{sec:2D_nonlinear_visc_PD} by considering a viscoelastic
polymer with a fixed viscosity $\eta_C=G_C\tau_C$.  We then turn to
the elastomeric limit $\tau_C\to\infty$ in
\secref{sec:2D_nonlinear_elast}. In
\secsref{sec:2D_nonlinear_visc_PD} and \ref{sec:2D_nonlinear_elast} we
restrict ourselves for simplicity to the case of zero thermodynamic
coupling between the nematic and polymer, $\chi=\kappa=0$. (As
discussed \mecchanged{previously, a strong kinematic coupling is however still present}.)
Finally
in \secref{sec:2D_nonlinear_expl} we consider the effects of including
explicit thermodynamic coupling.

\subsection{Viscoelastic active matter}
\label{sec:2D_nonlinear_visc_PD}

In this section we consider a polymer with a finite viscoelastic relaxation timescale $\tau_C$.  To avoid any complications associated with changing the \mecchanged{overall viscosity}, we consider fixed $\eta_C=G_C\tau_C$. (Accordingly, between runs in which $\tau_C$ progressively increases, $G_C$ is progressively decreased in tandem.)  \ejhchanged{We first explore the effects of polymer on the 2D nonlinear dynamics by revisiting the $(\zeta, \Delta)$ phase diagram discussed above.}

\subsubsection{Long-time behavior: active drag reduction}

\begin{figure}
  \centering
  \includegraphics[width=0.48\textwidth]{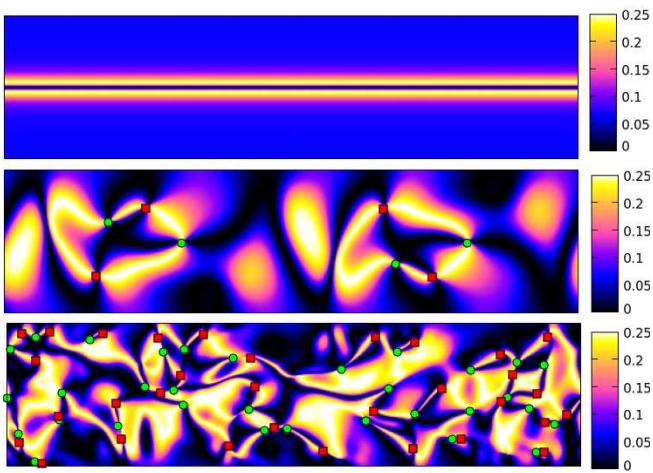}
  \caption{Representative snapshot states with net throughput, from the $\tau_C = 4$ phase diagram (\figref{fig:2D_nonlinear_PD_tauC_4} bottom panel): banded ($\zeta=0.025$, $\Delta = 10^{-5}$), oscillatory ($\zeta = 0.6$, $\Delta = 6.4 \times 10^{-4}$) and chaotic ($\zeta = 1.75$, $\Delta = 8 \times 10^{-5}$); colour scale indicates $\left(n_x n_y\right)^2$. Defects of topological charge $\pm 1/2$ are identified by green dots (+) and red squares (-). }
  \label{fig:2D_nonlinear_PD_tauC_4b}
\end{figure}

The lower panel of \figref{fig:2D_nonlinear_PD_tauC_4} shows the phase diagram of the model, with polymer now included, for a polymer relaxation time $\tau_C=4.0$.  Compared to the polymer-free case (top panel) a dramatic difference is immediately apparent. Whereas without polymer the regime of significant throughput is confined predominantly to that of laminar 1D banded states, with added polymer even the majority of the more complicated 2D time-dependent flow states show a significant throughput.  Three representative snapshot states with a significant throughput are shown in \figref{fig:2D_nonlinear_PD_tauC_4b}.

To explore this further, we pick a single set of parameters $\zeta = 3.2$, $\Delta = 2\times10^{-5}$ representative of the regime of fully developed active turbulence in \figref{fig:2D_nonlinear_PD_tauC_4} (for both $\tau_C=0.0$ and $\tau_C=4.0$) and perform a separate series of runs across which we vary the value of the polymeric relaxation time $\tau_C$, exploring the full range from the Newtonian limit (for the polymer) $\tau_C \to 0$ to the elastomeric limit $\tau_C \to \infty$.  (Recall that the polymer modulus $G_C$ must decrease in inverse proportion to $\tau_C$, given our fixed value of $\eta_C=G_C\tau_C=1$; likewise $\ell_C^2$ must increase in proportion to $\tau_C$ to keep a fixed value of $\Delta$.)

\begin{figure}
  \includegraphics[width=0.48\textwidth]{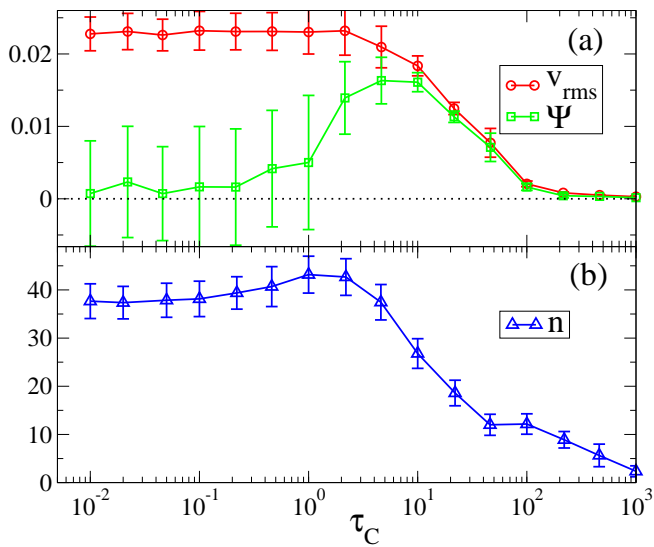}
  \caption{
    (a) Root mean square velocity $v_\textrm{rms} =
    \sqrt{\langle|\vtr{v}|^2\rangle_\vtr{x}}$ and mean throughput
    $\Psi$. (b) Mean defect density $n$. Each quantity is
    time-averaged across a run, after discarding the run's initial
    transient, with the vertical bars denoting the standard deviation
    of the time-series. \params $\zeta = 3.2$, $\Delta =
    2\times10^{-5}$, $\eta_C = 1$. }
  \label{fig:2D_nonlinear_tauC_sweep}
\end{figure}

For this series of runs, we plot in
\figref{fig:2D_nonlinear_tauC_sweep} the root-mean-square fluid velocity
\mecchanged{$v_\textrm{rms} \equiv
\sqrt{\langle|\vtr{v}|^2\rangle_\vtr{x}}$},
the mean throughput, and the mean defect density \cite{*[{We count defects in our numerics using an algorithm adapted an algorithm adapted from }] [{}] Huterer2005}. (Each such quantity is time-averaged across a run, after discarding the initial transient.) As expected, in the regime of small $\tau_C$ where the polymer acts simply as an additional solvent, none of these quantities change with
$\tau_C$.

In contrast, once $\tau_C > \tau_Q$ the active stress has to work against an increasingly elastic fluid, and \mecchanged{$v_\textrm{rms}$} decreases with increasing polymer relaxation time. Indeed, for $\tau_C\to\infty$ the polymer effectively arrests the flow altogether. \mecchanged{(This holds only for the final behavior, after discarding the initial transient; we show below that this transient can in fact be long-lived, and have its own rich dynamics.)}

Despite this monotonic decrease of \mecchanged{the root-mean-square fluid velocity, the mean throughput has a non-monotonic}
dependence on $\tau_C$. For small $\tau_C$, the throughput is
essentially zero, consistent with the model's limiting behavior in
the absence of polymer discussed above.  As $\tau_C$ increases to
become comparable with $\tau_Q$ the throughput increases dramatically,
suggesting a more coherent flow state capable of sustaining a net flow
in one direction.  (Indeed, not only does the mean throughput increase
but the fluctuations of the throughput about this mean decrease.)
Finally, for large
$\tau_C$ the throughput again decreases as the overall flow arrests.

This \mecchanged{gradual} transition with increasing $\tau_C$ to a state of significant
throughput, \mecchanged{followed by arrest, suggests a progressive
increase in the spatial coherence of the flow} with
increasing $\tau_C$.  Consistent with this, we see in
\figref{fig:2D_nonlinear_tauC_sweep}b that the defect density $n$ in
the nematic's director field markedly falls with increasing $\tau_C$,
with the \mecchanged{onset of this decline} roughly
coinciding with the peak in throughput.  A strong correlation
between $v_\textrm{rms}$ and defect density has been noted previously
in Refs.~\cite{Thampi2014b,Giomi2015,PrashantPreprint2015}.

We have demonstrated, then, that adding polymer to a fluid showing fully developed active turbulence calms the short scale structure of the spontaneous active flow, decreasing the nematic defect density, and thereby increasing the flow correlation length to give a more organized flow state, \mecchanged{often with} a sustained net throughput~\citep{Bozorgi2014,Hemingway2015}

For a passive Newtonian fluid displaying conventional inertial turbulence in a pressure-driven pipe flow at high Reynolds number \citep{Reynolds1883}, the addition of a small amount of long-chain polymer is known to calm the flow and reduce the ratio of imposed pressure gradient to throughput \citep{White2008}, in an effect known as `polymer drag reduction'.  The phenomenon just described for our active fluid, whereby (zero Reynolds number) active turbulence is calmed by the addition of polymer is strongly reminiscent of that effect, and we term it `active drag reduction' accordingly. The persistent, coherent motion to which it leads may have a biophysical analogue in cytoplasmic streaming, where it is believed that active cytoskeletal materials such as actomyosin play a role in generating coherent flows, facilitating the transport of nutrients and organelles within the cell \citep{Woodhouse2012,Goldstein2015}.

\subsubsection{Transient dynamics: extensional catastrophe}

\begin{figure}
  \centering
  \includegraphics[width=\columnwidth]{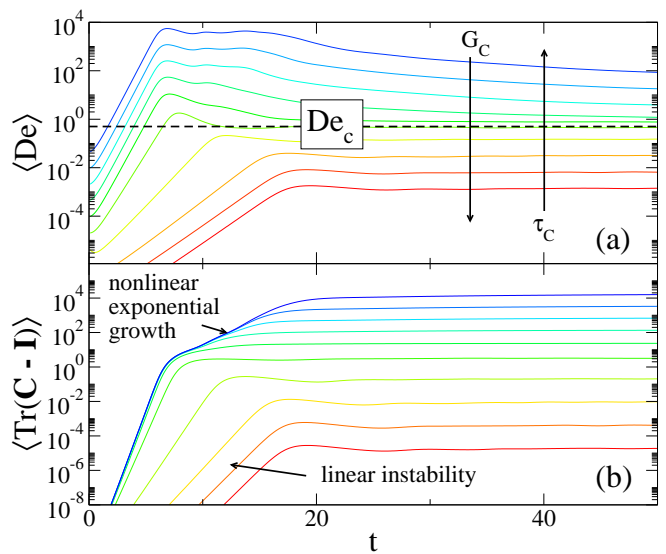}
  \caption{(a) Spatially averaged Deborah number. Marked as a dashed line is the threshold $\De_\textrm{c} = \frac{1}{2}$ above which we might expect nonlinear exponential growth in the polymer strain. (b) Spatial average of $\Tr{\tsr{C} - \tsr{I}}$, quantifying the mean polymer strain. When $\langle \De \rangle > \De_c$ in (a), we indeed observe nonlinear, exponential loading of the polymer in (b).  Curves upwards have $\tau_C$ from $10^{-2}$ (red line) $\to$ $10^4$ (blue line) with logarithmic spacing.  $\zeta = 2.2$, $\Delta = 4\times10^{-5}$, $\eta_C = 1$. }
  \label{fig:2D_nonlinear_exp_growth}
\end{figure}

In the previous subsection we discussed the long-time behavior of viscoelastic active matter in 2D, \mecchanged{ignoring} the initial transient en route to the ultimate attractor of the dynamics.  We now turn to consider that initial transient.

\mecchanged{Again} we pick a single set of parameters ($\zeta = 2.2$, $\Delta = 4\times10^{-5}$), representative of the regime of fully developed active turbulence in the phase diagram of
\figref{fig:2D_nonlinear_PD_tauC_4}. \mecchanged{We then} perform a series of runs, \mecchanged{varying} the polymer relaxation time $\tau_C$ \mecchanged{across} the full range from $\tau_C \to 0$ to the elastomeric limit $\tau_C \to \infty$. \mecchanged{We again fix both the overall polymer viscosity ($\eta_C = 1$) and the diffusivity $\Delta$.}

In each run we initialise the system as usual in a state that is homogeneous and \mecchanged{quiescent, with the polymer undeformed, $\tsr{C} = \tsr{I}$. For the given parameters we expect perturbations to grow exponentially until nonlinear effects become important; this is confirmed in the early} time growth dynamics in \figref{fig:2D_nonlinear_exp_growth}. \ejhchanged{(Note that the time signals in \figref{fig:2D_nonlinear_exp_growth} don't appear chaotic until the turbulent state becomes fully developed, \ie at later times in the run.)}

The degree to which the polymer becomes deformed by this instability,
as measured by the growth in $\mathrm{Tr}(\tsr{C} - \tsr{I})$,
increases with increasing $\tau_C$. This is to be expected: polymeric
strains build up at a rate set by the velocity gradients in the fluid,
and relax at a rate set by $1/\tau_C$.  Indeed a polymer obeying the
Oldroyd B equation will, if subject to an indefinitely sustained
simple shear flow of rate $\gdot$, acquire a shear strain
$\gdot\tau_C$; and, if subject to an indefinitely sustained
extensional flow of rate $\edot$, acquire an extensional strain
$\edot\tau_C/(1-2\edot\tau_C)$ \citep{Rallison1988}. The latter is an
infamous result, predicting the \mecchanged{polymers to suffer unbounded}
deformation for any sustained extensional flow of rate $\edot>
1/2\tau_C$. (In reality this growth would ultimately be cut off by
finite chain extensibility or chain rupture, neither of which is
captured by the Oldroyd-B model.)

Although the actively turbulent flow states that develop here do
not subject any given fluid element to an indefinitely sustained
extensional flow, they do nonetheless include regions with
significant extension rates.  To characterise these, we define a local
extensional Deborah number $\De=\tau_C \sqrt{\frac{1}{2}
  \left(\tsr{D}: \tsr{D}\right)}$,
and take its spatial average $\langle \De \rangle$. In view of our above
discussion of the Oldroyd B model's extensional catastrophe, we might
then reasonably expect a dramatic growth of polymer strain in any
regime where $\langle \De \rangle > \De_c \approx \frac{1}{2}$.

\mecchanged{In confirmation of this, the} time-evolution of $\langle
\De \rangle$ is plotted in \figref{fig:2D_nonlinear_exp_growth}a for
each in our series of runs. For small $\tau_C$ (warm
colors), $\langle \De \rangle$ remains small and the spatially average
polymer strain $\langle\Tr{\tsr{C} - \tsr{I}}\rangle$ likewise remains
modest.  In contrast, for larger $\tau_C$ (cold colors)
$\langle\De\rangle$ exceeds $\De_c \approx \frac{1}{2}$ leading to a
dramatic effect on the polymer conformation: following the initial
linear instability, we observe a second period of exponential chain
stretching that generates huge polymer strains.  See
\figref{fig:2D_nonlinear_exp_growth}b.

Any growth in the polymer strain gives a corresponding growth in the
polymeric contribution $\tsr{\Sigma}_C = G_C \left(\tsr{C} -
  \tsr{I}\right)$ to the stress. Recalling that $G_C=1/\tau_C$ for
this series of runs, the large values of $\tau_C$ for which dramatic
chain stretching occurs have correspondingly small values of $G_C$.
Accordingly, as long as the polymer strain remains $O(1)$, the
polymeric contribution to the stress will remain small. In contrast,
in the regime of dramatic chain stretch the polymer strains become
huge and the polymer stress does then become comparable with the
nematic and Newtonian ones, despite the small polymer modulus $G_C$.
This in turn feeds back on the flow field, mitigating the extensional
flows and halting the divergence of the polymer stress.

\subsubsection{Limit cycle}

In the previous two subsections we considered state points $(\zeta,\Delta)$ deep in the active-turbulent region of the phase diagram in \figref{fig:2D_nonlinear_PD_tauC_4} (which has $\tau_C=4.0$). \ejhchanged{While the transient dynamics of corresponding runs at much larger $\tau_C$ revealed that polymer deformation proliferates due to the appreciable extensional component of the turbulent activity-driven flow field, it is unclear how this phenomenon manifests in the long time dynamics.} Therefore we consider a point $(\zeta,\Delta)$ in \figref{fig:2D_nonlinear_PD_tauC_4} with a more moderate activity (which is less deeply turbulent) and again perform a corresponding run at a much larger $\tau_C=1000$ (once again at fixed $\eta_C=1$), \ejhchanged{where we now run the simulation until long times $t_{max} \sim \OO{10\tau_C}$.}

\begin{figure}
  \centering
  \includegraphics[width=\columnwidth]{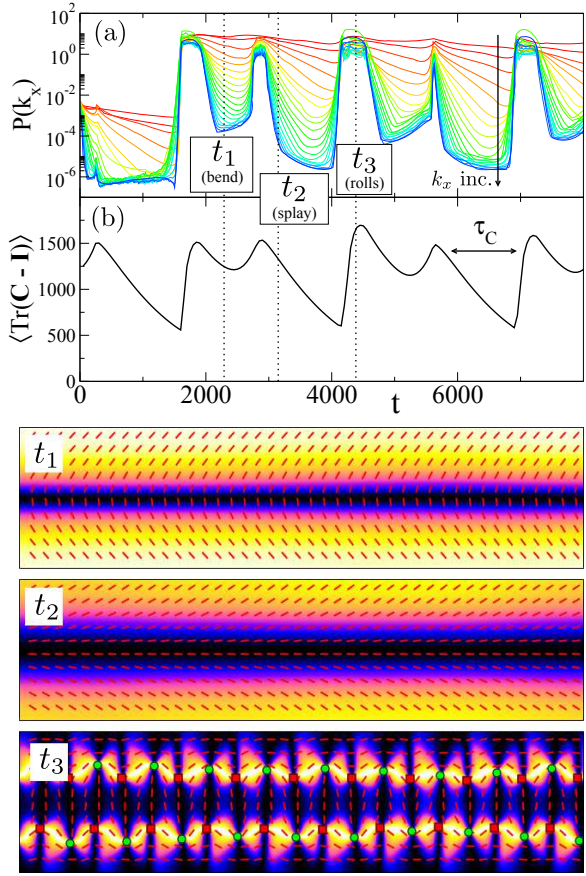}
  \caption{Representative state that continuously oscillates between bend and splay quasi-1D shear banded states via an intermediate roll state. Plotted are (a) the power spectrum $P(k_x)$ for the first 20 Fourier modes $k_x = 1 \textrm{ (red)} \to k_x = 20 \textrm{ (blue)}$ and (b) spatially averaged $\Tr{\tsr{C} - \tsr{I}}$ against time. The transitions between states are 2D in nature, as shown by the periodic spikes in $P(k_x)$.  \textit{Bottom:} Snapshots for the times marked in (a/b) show $(n_x n_y)^2$ (colourmap), director $\uvtr{n}$ (red lines), and defects (symbols). \params $\zeta = 1.24$, $\Delta = 1.6 \times10^{-4}$, $\tau_C = 1000$, $\eta_C = 1$.}
  \label{fig:2D_nonlinear_limit_cycle}
\end{figure}

The results are shown in \figref{fig:2D_nonlinear_limit_cycle}. In this case the system \mecchanged{ultimately settles to an oscillatory state whose period is} set by the \mecchanged{polymer relaxation time} $\tau_C$.  Over the course of \mecchanged{one cycle}, the system switches from a quasi-1D banded state dominated by bending in the director field (see the snapshot at time $t_1$), to a different quasi-1D banded state dominated by splay (at time $t_2$), via a fully 2D intermediate state with velocity rolls. A representative snapshot of such a 2D intermediate state is shown at time $t_3$.

Closer study of the full run reveals that the transition from bend to splay initiates via the director, at a single value of $x$ midway between the plates, rotating by $\pi/2$. This disturbance then propagates along the interface until the whole system is splayed.  (Destabilisation of the bend state to form the splay state has been seen previously in Ref.~ \cite{Thampi2014}.) The splay-banded state then apparently remains stable for a time of order $\tau_C$, before developing a roll-like instability around $t_3 = 4400$ characterised by alternative pairs of $\pm \frac{1}{2}$ topological defects \citep{Chaikin2000Book} in the nematic director field.

The degree of inhomogeneity during these transitions can be monitored quantitatively through the power spectrum $P(k_x, t)$ of the spatial Fourier transform of
the nonzero order parameters ${\phi} = \left(Q_{xx}, Q_{xy}, Q_{yy}, C_{xx}, C_{xy},
  C_{yy}\right)$:
\begin{equation}
  P(k_x) = \sum_{i=1}^6 \int_0^{L_y} dy |\phi_i (k_x, y)|^2.
  \label{eq:chap_active_2D_apx_power_spec}
\end{equation}
Any purely 1D state would have $P(k_x, t) = 0$ for $k_x > 0$. As can
be seen, the quasi-1D bend and splay states indeed have small $P$. The
2D roll-like states that arise during the transition between these
quasi-1D states have, in contrast, a much larger $P=O(1)$. See
\figref{fig:2D_nonlinear_limit_cycle}a.

\mecchanged{This oscillating effective dimensionality of the flow state directly influences the
polymeric stretching, which we monitor as before with a scalar
extension measure $\langle \Tr{\tsr{C}-\tsr{I}}\rangle$ (see
\figref{fig:2D_nonlinear_limit_cycle}b).}  The constraint of fluid
incompressibility means that an extensional component to the flow
field can only arise in the 2D states.  Because of this, during times
when the system occupies a quasi-1D bend or splay state the polymer
\mecchanged{extension relaxes} on a timescale $\tau_C$. In contrast, the
intervening 2D roll state contains regions of extensional flow leading
to significant polymer stretching.

\ejhchanged{It is worth noting that the nonlinear oscillatory state reported here would not have been predicted from our 1D linear stability analysis, for which the corresponding linear instability is not oscillatory. However a 2D linear stability calculation that linearises about one of the inhomogeneous banded states (shown in \figref{fig:2D_nonlinear_limit_cycle}) may well predict this behavior.} \mecchanged{Note that oscillatory dynamics in active systems} is widely seen in a biophysical context.  For example, shape oscillations in developing cells are believed to be driven by actomyosin networks \citep{He2010}, which have been theoretically described using an elastic model \citep{Dierkes2014}.

\subsection{Elastomeric active matter}
\label{sec:2D_nonlinear_elast}

\begin{figure}
  \centering
  \includegraphics[width=\columnwidth]{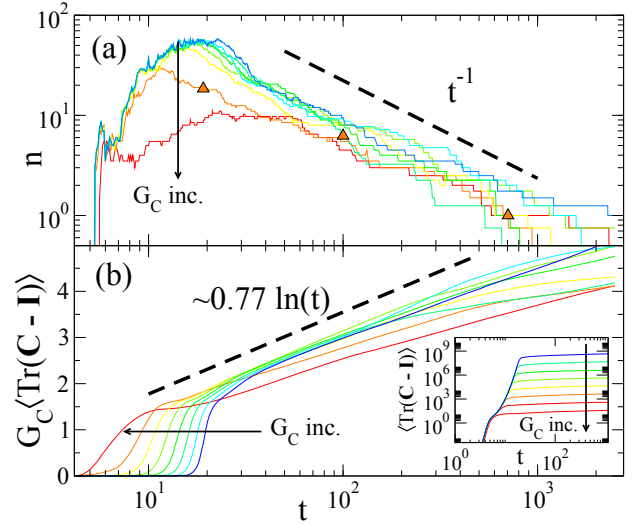}
  \caption{Results from 2D runs with $\tau_C \to \infty$. We vary $G_C$ from $10^{-8}$ (blue lines) $\to 10^{-1}$ (red lines) with logarithmic spacing. Snapshots for $G_C = 10^{-2}$ given in \figref{fig:2D_nonlinear_inf_coarsening_snapshots}. (a) Areal defect density $n$ against time. Steps arise because $n$ is discrete. (b) Scalar measure of polymer stress $G_C \langle \Tr{\tsr{C} - \tsr{I}}\rangle$ (inset shows the same data but not scaled by $G_C$, \ie we plot $\langle \Tr{\tsr{C} - \tsr{I}}\rangle$). \params $\zeta = 3.2$, $\Delta = 8 \times10^{-5}$, $\tau_C \to \infty$. }
  \label{fig:2D_nonlinear_inf_coarsening}
\end{figure}

\begin{figure}
  \centering
  \includegraphics[width=\columnwidth]{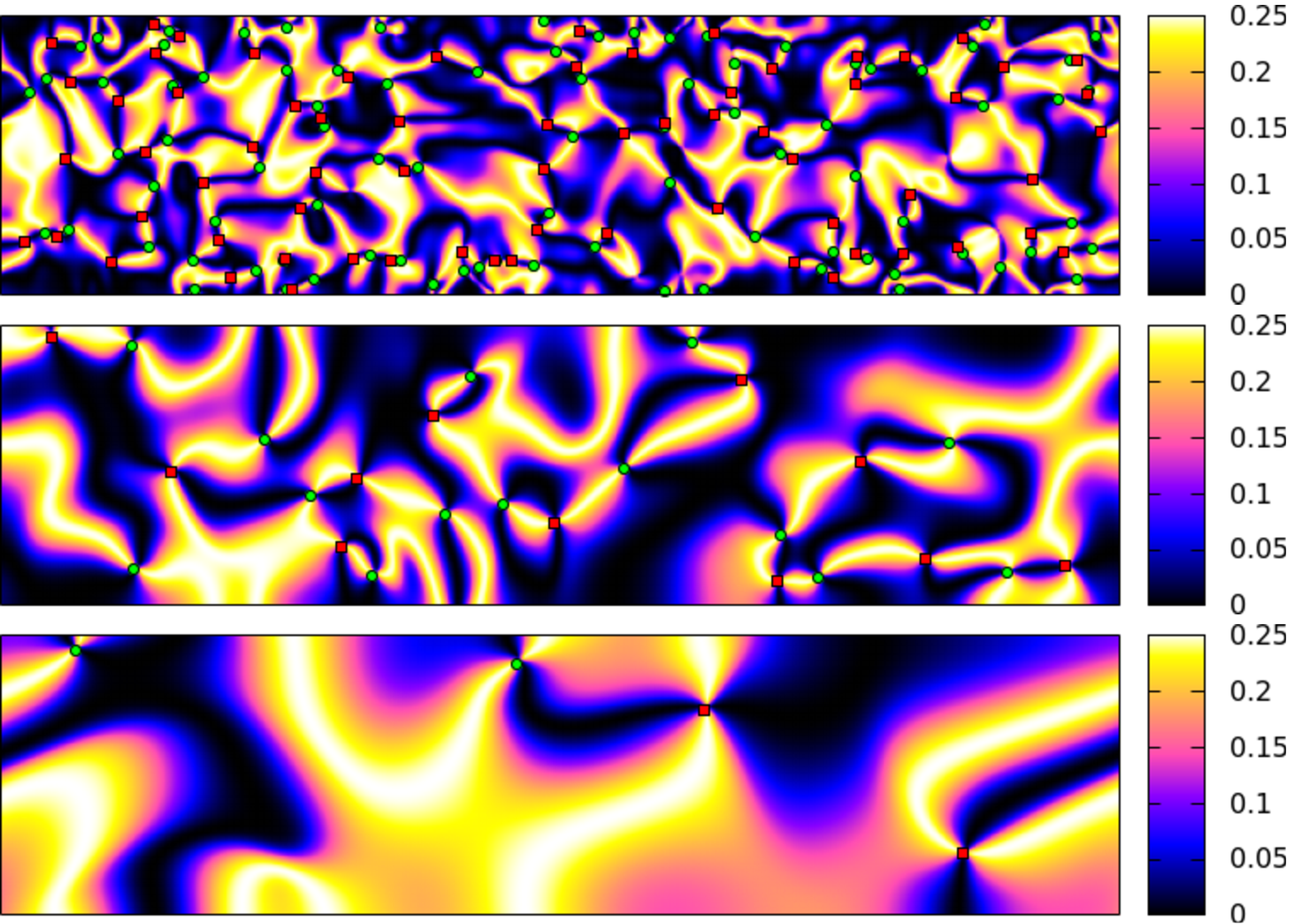}
  \caption{Snapshots for the points marked by triangles in \figref{fig:2D_nonlinear_inf_coarsening}a at times $t = 20, 100, 700$. Images show $(n_x n_y)^2$ (colourmap) and defects (symbols). \params $\zeta = 3.2$, $\Delta = 8 \times10^{-5}$, $\tau_C \to \infty$, $G_C = 10^{-2}$.}
  \label{fig:2D_nonlinear_inf_coarsening_snapshots}
\end{figure}

In the previous subsection we saw that adding a polymer of even modest relaxation time has a dramatic effect on the phase behavior of confined active nematics. For example, it gives rise to an unusual drag reduction effect with strong enhancements in the flow field's coherence and net throughput, for a fixed level of activity.  We also explored the effect of changing the polymeric relaxation time $\tau_C$, moving towards the elastomeric limit of large $\tau_C$.  We did so at a fixed overall polymer viscosity $\eta_C = 1$, with $G_C$ decreasing in inverse proportion to $\tau_C$.

In this section we consider the \mecchanged{``true" elastomeric limit, in which
$\tau_C\to\infty$ is taken at finite $G_C$ so that the polymeric
viscosity diverges}.  Physically, this could be realised by
increasing the crosslink density and/or chain length in an entangled
polymer of fixed concentration: \mecchanged{our model might then describe
an actomyosin} cell
extract in a background of lightly cross-linked polymer gel. To implement
this limit in our simulations we simply remove the local relaxation
term prefactored by $1/\tau_C$ in the polymer equation of motion. \mecchanged{As mentioned
already for the 1D case in \secref{sec:1D_nonlinear_spont_flow_solid}, we do however
retain finite stress diffusivity, $\Delta_C=\ell_C^2/\tau_C$, meaning that $\ell_C\to\infty$ as the elastomeric limit is taken. We discuss this choice further
at the end of this subsection.}

The predictions of our linear instability \mecchanged{analysis in} this limit were shown in \figref{fig:fifteen}.  We now explore the instability more fully by performing nonlinear 2D
simulations, at infinite $\tau_C$, for a range of polymer moduli
$G_C = 10^{-8} \to 10^{-1}$. \mecchanged{We choose values of the activity ($\zeta=3.2$)
and diffusivity ($\Delta=8\times 10^{-5}$) for which the initially
homogeneous state is predicted to be unstable.}

In each run, we find that the initial homogeneous state indeed
destabilises to form a \mecchanged{heterogeneous, complex liquid
crystalline texture} with a high density $n$ of defects in the director
field; see \figref{fig:2D_nonlinear_inf_coarsening}a and \figref{fig:2D_nonlinear_inf_coarsening_snapshots} (top).  Associated with this buildup of defects is
a complex `elastically turbulent' deformation field with regions of
(transient) extensional flow, which result in exponential stretching
of the polymer; see \figref{fig:2D_nonlinear_inf_coarsening}b.
Because in these runs the polymer modulus obeys $G_C<G_Q$, appreciable
polymer strain can develop (inset of \figref{fig:2D_nonlinear_inf_coarsening}b) during this early time regime before the resulting polymeric contribution to the stress becomes comparable
with solvent and nematic stresses, which are $O(1)$ in our units. (A comparable phenomenon is seen in some models of polymeric glasses \citep{Fielding2012}.)

Once the polymer stress does become comparable with the \mecchanged{other} stresses, the `elastic turbulence' \mecchanged{arrests into} a complex
but almost frozen defect pattern. Thereafter the defect density shown in \figref{fig:2D_nonlinear_inf_coarsening}a slowly coarsens over time, roughly as $t^{-1}$, which is the same as the classical result for coarsening in a passive nematic \citep{Bray1994}. \ejhchanged{See \figref{fig:2D_nonlinear_inf_coarsening_snapshots} for snapshots of the nematic texture during this process.} Although a defect-free state $n\to 0$ \mecchanged{might arise} in the true steady state limit $t\to\infty$, the coarsening process that leads to it is
sufficiently slow that the strain pattern created by the arrested
active turbulence might easily be mistaken for a final steady state.
Note that the mean polymer extension continues to grow in this
coarsening regime, albeit \mecchanged{slowly (roughly logarithmically at intermediate times) with some suggestion of eventual saturation seen in the numerics.}
This elastomeric arrest of the active turbulent flow field, where strong polymer stretching in extensional flow regions creates strong stresses that oppose the flow, is mechanistically reminiscent of the drag reduction effects reported earlier. Note that the quasi-homogeneous state that develops at long times is drastically different from the base state in our earlier linear stability calculation (which assumes that the polymer is undeformed), and is  therefore not subject to the same linear instability as found there.

\mecchanged{We now return to an issue raised previously in this section and earlier in \secref{sec:1D_nonlinear_spont_flow_solid}, which is our choice to retain finite polymer stress diffusivity $\Delta_C = \Delta > 0$ even in the elastomeric limit of $\tau_C\to\infty$. Since $\Delta_C  = l_C^2/\tau_C$, this requires matching divergence of the length $l_C$. (The physical meaning of this length is ambiguous, but it is often assumed to be of order the polymer coil size.)}

\mecchanged{Our reason for this choice is ultimately pragmatic: we have found that our 2D numerical simulations become highly unstable, at the discretization scale, if performed with $\Delta_C = 0$. A credible explanation of this behavior is as follows. We have seen that polymers encounter regions of strong elongation in 2D active flows. These convert `globular' initial stress fluctuations into `threadlike' ones, creating fine structure in the transverse direction as set by the local compression axis. Once elongation is strong, numerical instability is inevitable unless there is a restoring term to iron out these short-scale fluctuations. However, in the absence of thermodynamic couplings (see next section), the only term in the polymer dynamics that can achieve this is $\Delta_C$. Thus we retain finite $\Delta_C$ in order to capture the short lengthscale physics of local stress redistribution. This mechanism, whatever its details, should not disappear for lightly cross-linked elastomers, since their physics essentially merges with that of free chains at short lengthscales (below the cross-link separation).}

\mecchanged{Note that in an elastomer of finite $\Delta_C$, although the mean stress caused by an applied deformation can never relax (which is the defining feature of a solid), any inhomogeneous polymer stress will relax to a uniform one eventually, via the stress diffusion term. This may seem a bit artificial, although a similar effect can arise in many elastomeric systems by other pathways. One of these, although absent from the present model, is solvent permeation flow \citep{Milner1993}. Interestingly, one might expect that thermodynamic coupling to the $\tsr{Q}$ field, whose stress diffusivity does not vanish as $\tau_C \to \infty$, could also provide enough stress redistribution to stabilize the numerics. Indeed we find this to be the case, as we now discuss. }

\subsection{Explicit thermodynamic coupling}
\label{sec:2D_nonlinear_expl}

Throughout the preceding results sections we set $\kappa=\chi = 0$ in
\eqref{eq:free_QC_coupling}, so that there is no direct thermodynamic
coupling between $\tsr{Q}$ and $\tsr{C}$. In that way, the only
coupling between $\tsr{Q}$ and $\tsr{C}$ was purely kinematic, arising
\mecchanged{because} they share a common fluid velocity $\vtr{v}$
which is influenced by both sources of stress. In this subsection, we
study the effects of a direct coupling.

\mecchanged{The $\kappa$ term in \eqref{eq:free_QC_coupling} controls how the polymer pressure shifts the isotropic-nematic transition. Our interest in this paper lies far from the transition region, deep within the nematic phase, so for simplicity we set $\kappa = 0$. The remaining coupling parameter, $\chi$,}
governs the energetics of relative orientations of $\tsr{Q}$ and
$\tsr{C}$. For $\chi < 0$, it is favorable for $\tsr{Q}$ and $\tsr{C}$
to align.  Indeed experiments (in the passive limit) suggest that
single semi-flexible polymers can couple to the nematic director field
in this fashion \citep{Dogic2004}.  However, such co-alignment tends to
arise kinematically even when $\chi = 0$, since $\tsr{Q}$ and
$\tsr{C}$ both have similar alignment tendencies in relation to the
spontaneous velocity field (at least for our choices of the
flow-alignment and slip parameters $\xi$ and $a$).  We therefore do
not expect substantially new physics to arise in this case.  Therefore
we focus here on antagonistic coupling $\chi > 0$, where $\tsr{Q}$ and
$\tsr{C}$ now prefer to orient with their major axes perpendicular. We
do not perform an exhaustive search of parameter space but instead
present a selection of the intriguing states observed.

\subsubsection{Persistent elastomeric turbulence}

\begin{figure}[t]
  \centering
  \includegraphics[width=\columnwidth]{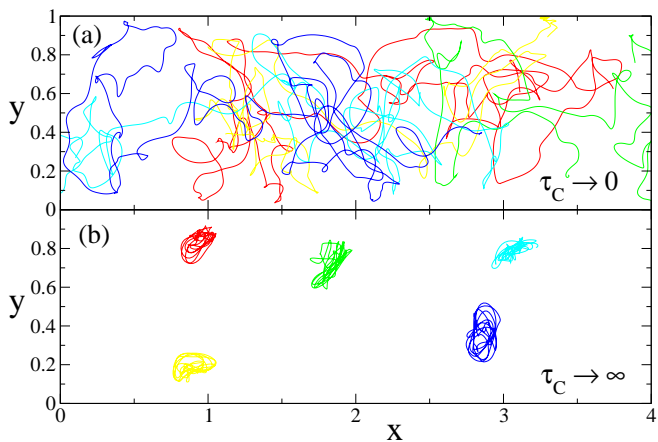}
  \caption{Plot of tracer trajectories for parameters characteristic of the turbulent phase for $t = 0 \to 200 \tau_Q$. Polymer diffusion has been disabled by setting $\Delta_C = \ell_C^2 / \tau_C = 0$. Trajectories are shown in (a) the no polymer limit $\tau_C \to 0$ and in (b) the elastomeric limit $\tau_C \to \infty$. Colours are used to distinguish the separate tracer paths. \params $\zeta = 3.2$, $\Delta_Q = 10^{-4}$, $\Delta_C = 0$, $G_C = 0.1$, $\chi = +0.032$. }
  \label{fig:2D_nonlinear_elastomer_tracers}
\end{figure}

We have shown previously (in 1D) that persistent oscillatory states can develop in the elastomeric limit $\tau_C \to \infty$, provided that we keep $\ell_C$ fixed. Our 2D simulations produce significantly more complex flow fields characterised by strong extensional components which, as previously discussed, can be numerically problematic. Without any polymer diffusion we found that our numerics required impractically severe convergence criteria; with polymer diffusion retained (by imposing $\Delta_C = \ell_C^2 / \tau_C = \Delta_Q$) the simulations are stabilised but the initial ``turbulent solid" state slowly coarsens away.

\begin{figure}[t]
  \centering
  \includegraphics[width=\columnwidth]{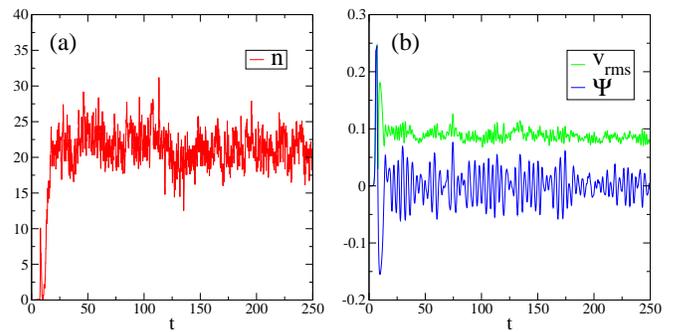}
  \caption{Plot of (a) defect density $n$, (b) throughput $\Psi$ (lower blue line) and $v_{rms}$ (upper green line) in the elastomeric limit $\tau_C \to \infty$. Polymer diffusion has been disabled by setting $\Delta_C = \ell_C^2 / \tau_C = 0$. \params $\zeta = 3.2$, $\Delta_Q = 10^{-4}$, $\Delta_C = 0$, $G_C = 0.1$, $\chi = +0.032$, $\tau_C \to \infty$. }
  \label{fig:2D_nonlinear_elastomer_flow}
\end{figure}

However we might anticipate a similar stabilisation to result also from a small thermodynamic coupling between $\tsr{C}$ and $\tsr{Q}$, the latter of which retains spatial gradients even in the limit $\tau_C \to \infty$. Indeed, with this coupling we have successfully performed runs at zero polymer diffusivity, \ie $\Delta_C = 0$, albeit using a significantly smaller space- and time-steps~\footnote{If $\Delta_C = 0$, the timestep $\Delta t$ needs to be an order of magnitude smaller than for $\Delta_C \neq 0$, and the spatial-stepsize $\Delta x$ must also be at least halved.}. Intriguingly, we find that the transient elastomeric turbulence discussed previously now continues indefinitely, producing an unusual oscillatory state. This can be readily visualised by plotting the paths of massless tracer particles as they are advected by the flow, as shown in \figref{fig:2D_nonlinear_elastomer_tracers}. Without polymer present, these tracers perform a random walk as they are advected by the turbulent flow. However in the elastomeric limit, the tracer trajectories remain confined to periodic orbits for the duration of the simulation. In this limit, the activity-driven, turbulent flow field stretches and distorts the elastic polymer background which, at large enough strains, reacts by producing a restoring force that ``undoes'' any displacement, eventually returning displaced material points to their initial location (or nearby). We also plot the defect density $n$, throughput $\Psi$ and rms velocity in \figref{fig:2D_nonlinear_elastomer_flow}. These all apparently reach quasi-steady state values after a short time, and show no signs of the coarsening discussed previously for $\Delta_C \neq 0$ in \figref{fig:2D_nonlinear_inf_coarsening}. Also note that the throughput oscillates about zero, consistent with our expectation that an elastic solid should undergo no net displacement.  The severe numerical requirements when $\Delta_C = 0$ restrict our discussion to the qualitative aspects of the state; a more sophisticated numerical implementation would be required to fully quantify the elastomeric turbulence seen here.

\subsubsection{Shear Bands with interfacial defects}

\begin{figure}
  \includegraphics[width=\columnwidth]{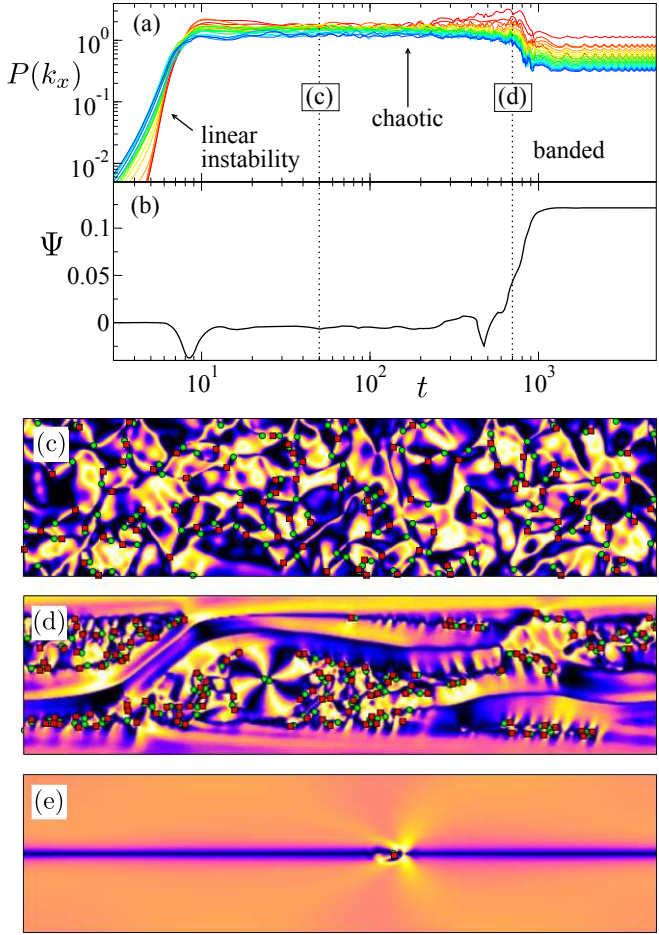}
  \caption{Example where an initially chaotic state organises into a coherent shear-banded state with defects, which advect with the flow, embedded in the interface. Time evolution of (a) the power spectrum $P(k_x, t)$, averaged over $y$ (for $k_x = 1 \textrm{ (red lines)} \to 20 \textrm{ (blue lines)}$), and (b) throughput $\Psi$. (c,d,e) Snapshots of $(n_x n_y)^2$ for the states marked left at times (c) $t = 50$, (d) $t = 700$ and (e) $t = 3800$. \params $\zeta = 3.2$, $\Delta = 4 \times 10^{-5}$, $\tau_C = 10$, $\chi = 0.002$.}
\label{fig:2D_nonlinear_expl_defect_band}
\end{figure}

For the remaining examples we restore $\Delta_C = \Delta_Q = \Delta$ and fix a finite value of $\tau_C$ comparable to $\tau_Q$. The first of these (which has $\chi = 0.002$) demonstrates an intriguing, polymer-driven disorder-order transition. Choosing parameters characteristic of the chaotic state (when $\chi = 0$), at early stages of the run we find the defect-rich disordered state observed without polymer (\figref{fig:2D_nonlinear_expl_defect_band}c). However as the simulation progresses, ordered regions of nearly uniform director $\uvtr{n}$ spread in from the walls towards the centre of the channel, forming an increasingly shear-banded like state (\figref{fig:2D_nonlinear_expl_defect_band}d). Eventually, the only remaining evidence of the earlier chaotic state are pairs of defects embedded in the interface (\figref{fig:2D_nonlinear_expl_defect_band}e).

This transition can be seen quantitatively by examining the power spectrum $P(k_x, t)$ in \figref{fig:2D_nonlinear_expl_defect_band}a. During the initial chaotic phase, the first 20 wavevectors (plotted) contribute significantly to $P(k_x, t)$, indicating significant spatial structure in the $x$-direction, with dynamics aperiodic in time. At long times, once the banded state forms, all amplitudes $P(k_x, t)$ are attenuated, particularly at large $k_x$. This (admittedly extreme) example is consistent with our drag reduction argument, whereby polymer calms short-scale structure. Snapshots of the evolution of this state in \figref{fig:2D_nonlinear_expl_defect_band}c/d/e demonstrate a clear transition from a disordered to an almost ordered state.

Correlated with this suppression of short scale structure is a dramatic increase in the throughput (\figref{fig:2D_nonlinear_expl_defect_band}b). While the chaotic state at early times has zero mean throughput, the latter banded state develops a strong net flow in a spontaneously chosen direction, in this example towards the right.

Interestingly, during the intermediate phase between chaotic and banded states, we occasionally find transient, rotating spiral structures, which when viewed macroscopically~\footnote{Microscopically, at the heart of the structure, we find a pair of $+\frac{1}{2}$ defects though these are close enough that the effective director field forms a spiral pattern of topological charge $+1$.} possess an integer topological charge $+1$. (See \figref{fig:2D_nonlinear_expl_defect_band}d.) Such structures are commonly found in polar active materials \citep{Schaller2010,Kruse2004,Elgeti2011,Yang2014a}, but not in apolar nematics which much more commonly display $\pm \frac{1}{2}$ defects \citep{Sanchez2012}, as were found above. In general, integer defects in passive nematics tend to dissociate \citep{Chaikin2000Book}, a tendency maintained in the active case without polymer.  (One exception is in highly confined cylindrical geometries, as studied in Ref.~\cite{Woodhouse2012} where the authors found a single $+1$ defect at low activities, which split into a pair of $+\frac{1}{2}$ defects only at higher activity values.) It appears that antagonistic coupling can help promote integer defects, but the detailed mechanism for this remains unclear.

\subsubsection{Shuffling state}

\begin{figure}[t]
  \includegraphics[width=\columnwidth]{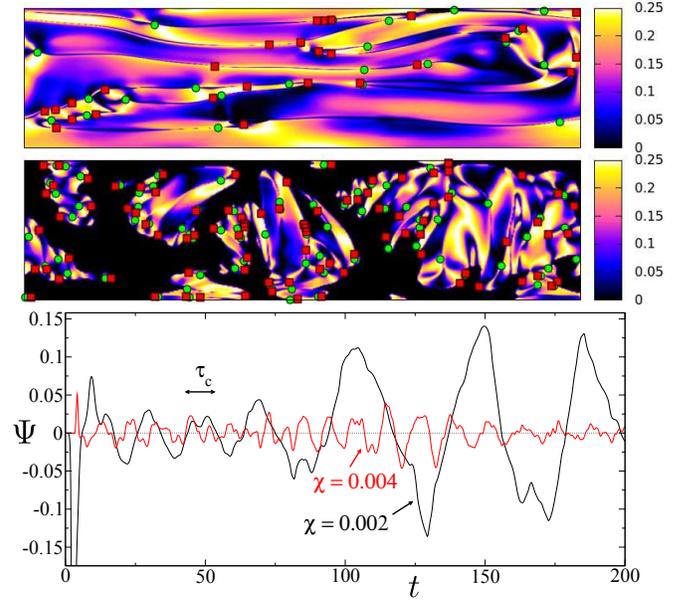}
  \caption{\textit{Top:} An exotic oscillatory state that coherently shuffles alternatively left and right on a timescale $\mathcal{O}(\tau_C)$. \textit{Middle:} Coexistence of `bubbling' active domains and regions where the director is out of plane (black). \textit{Bottom:} Throughput-time series for both states (black line $\chi = 0.002$, red line $\chi = 0.004$). \params $\zeta = 6$, $\Delta = 10^{-4}$, $\tau_C = 10$, $\chi = 0.002$ (upper snapshot), $0.004$ (lower snapshot). }
  \label{fig:expl_examples}
\end{figure}

By increasing the activity with other parameters fixed, we can disrupt the above shear-banded state. The result, \figref{fig:expl_examples} (upper snapshot), is a state that shuffles back and forth as a whole, with defects travelling along regions in which $q$, the largest eigenvalue of $\tsr{Q}$, is small. This mechanism is similar to that reported in Ref.~\cite{Thampi2014} (without polymer), where defect motion in `walls' (regions of high local distortion) was observed. The throughput time series in \figref{fig:expl_examples} reveals that the flow switches direction periodically on a timescale of order the viscoelastic relaxation timescale $\tau_C$, once again confirming the direct influence of polymer on the dynamics.

\subsubsection{Order-disorder coexistence}

With a larger coupling constant ($\chi = 0.004$), even more exotic states can develop. In \figref{fig:expl_examples} (lower snapshot) we show a state exhibiting coexistence between chaotic/oscillatory regions (where the director lies in the $xy$ plane shown) and pseudo-quiescent regions (where the director points along $z$). The active regions travel back and forth, trapped between the bounding walls.  Repeating the simulations a number of times with different seeds for the 2D perturbation, we find that these pseudo-quiescent regions sometimes grow to envelop the whole system, which then remains quiescent indefinitely. Such quiescent regions are likely made possible by our assumption of translational invariance along the vorticity axis $z$. (Recall that in our simulations we consider 3D order parameter tensors $\tsr{Q}$ and $\tsr{C}$ residing in a 2D space). It may be that fully 3D simulations are required to fully resolve the dynamical behavior of this state, which could easily then show structure also in the $z$ direction. We leave such 3D investigations to future work.

\section{Conclusions}
\label{sec:conclusions}

We have studied in depth the linear stability and nonlinear dynamics of the minimally coupled model for viscoelastic active matter derived in Ref.~\cite{Hemingway2015}. While treating 3D structural tensors for both the active nematic and the polymer sectors, our studies assumed either 2D flow states (translationally invariant along the vorticity axis $z$) or 1D states (invariant also along the periodic axis $x$).

For appreciable activities, the phase diagram without polymer is dominated by chaotic states with no net throughput \citep{Fielding2011}. Adding polymer results in a `active drag-reduction effect' whereby the short scale flow structure is calmed, resulting in a reduced density of nematic point defects. The increased flow correlation length is associated with net material transport around the periodic boundary conditions of our simulation. In fully confined geometries this would likely also help promote long-lived steady circulation as opposed to a mixing flow \citep{Goldstein2015}.

While adding polymer can delay the onset of spontaneous flow, our results predict a critical activity which does not diverge in the elastomeric limit $\tau_C \to \infty$, and indeed vanishes for large enough systems just as it does without polymer \citep{Voituriez2005}. Thus activity-driven flows can occur within active materials that are ultimately solid: these might be called ``true active gels". Numerical simulation of such materials in 1D indeed reveals oscillatory shear-banded states where the flow direction switches periodically as the polymer stress is loaded. A range of complex states involving nontrivial interplay of viscoelasticity and activity were seen in systems of large but finite polymer relaxation time $\tau_C$.

As the relaxation time of the polymer was increased, the effect of activity-driven extensional flows was shown to be important. When the spatially averaged Deborah number (which describes the ratio of polymer and extensional timescales) exceeds a critical value, we found that the initial linear instability is followed by a period of rapid, exponential deformation of the polymer. For sufficiently large $\tau_C$, this resulted in oscillatory states which cycled between rapid extensional deformation of the polymer and slow stress relaxation on a timescale $\tau_C$.

In the elastomeric limit, $\tau_C\to\infty$ the ultimate fate of the oscillatory and/or turbulent states that we discovered remains sensitive to numerical details of the model. Retention of finite stress diffusivity (which greatly improve numerical stability) in this limit causes ultimate relaxation to a state of uniform stress; this precludes any permanent state of oscillation or elastic turbulence. However, numerically we were able to switch off the stress diffusivity once thermodynamic coupling between nematic and polymeric degrees of freedom was included. In this case we found that the system could be truly a ``turbulent solid" even in steady state. That is, the coupled system shows chaotic, activity-induced velocity fields that persist indefinitely but that reverse sign often enough locally that the elastic strain remains bounded, as a solid requires.

More generally, we found a wide range of complex and interesting spontaneous flow states when $\tsr{Q}$ and $\tsr{C}$ are antagonistically coupled at the free energy level. Depending on the strength of the coupling, our simulations show that polymer can drive a transition from active turbulence to near-laminar banded flow, result in states with oscillatory dynamics on a timescale $\tau_C$, and create regions where director is oriented out of the plane. Fully 3D simulations should provide useful insight, particularly for the last of these.

Our prediction of active turbulence in soft solid materials ($\tau_Q\to\infty$) arises when $G_C/G_Q \lesssim 0.1$. This looks experimentally feasible for subcellular active matter (though probably not swarms of bacteria) within a lightly cross-linked polymer gel. The symmetry breaking characterised by net throughputs, which we observe at moderate polymer relaxation times ($\tau_C / \tau_Q \approx 10$), could have particular relevance for studies of cytoplasmic streaming, where coherent flow is crucial for material transport within the cell. The same physics may also apply, albeit at larger scales, to cell migration in confined geometries \citep{Vedula2012}. Our results with explicit coupling that show oscillatory, 'shuffling' states may also be relevant in describing the cell shape oscillations of Ref.~\cite{Salbreux2007}.

We hope our work will promote experiments on these and other forms of active viscoelastic matter; while the effects of viscoelasticity on individual swimmers have been studied previously \citep{Shen2011,Gagnon2014}, we are not aware of an equivalent study for bulk, orientationally ordered phases. One related study did however consider \ejhchanged{the linear stability of the bulk \textit{isotropic} phase in an Oldroyd-B fluid \citep{Bozorgi2013}. While the specifics of that model differ from the one presented here~\footnote{There are several differences between our study and Ref.~\cite{Bozorgi2013}. In that work (i) the stability analysis is about an isotropic base state, (ii) liquid-crystalline stresses are not included, (iii) the active particles are self-propelled, and (iv) the concentration field can vary in space (we assume a homogeneous concentration).}, both studies find (i) that increasing $\eta_C$ at fixed $\tau_C$ will always eventually suppress the spontaneous flow instability, (ii) that at small $\tau_C$ the polymer simply renormalises the solvent viscosity (this is our viscous limit) and} (iii) a region at large $\tau_C$ where unstable oscillatory behaviour is predicted (this is our elastomeric limit).

\begin{acknowledgments}
  We thank Ananyo Maitra, Cristina Marchetti, Peter Olmsted, and Sriram Ramaswamy for discussions.  EJH thanks EPSRC for a Studentship.  MEC thanks EPSRC Grant EP/J007404 and the Royal Society for funding. SMF's and EJH's research leading to these results has received funding from the European Research Council under the European Union's Seventh Framework Programme (FP7/2007-2013) / ERC grant agreement number 279365.
\end{acknowledgments}

\appendix*
\section{Throughput criterion}
\label{apx:throughput}

\begin{figure}[t]
  \centering
  \includegraphics[width=\columnwidth]{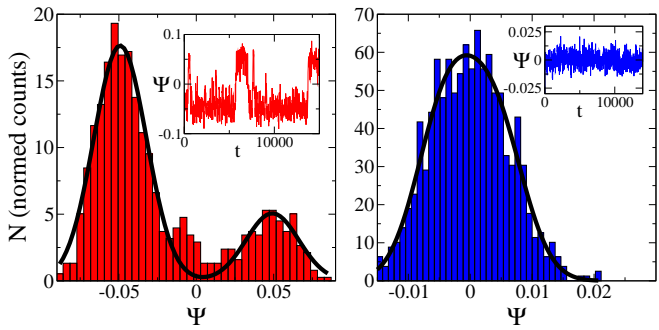}
  \caption{Method for determining throughput, $\Psi$. \textit{Left:} A state with net throughput, in which the throughput direction intermittently switches between right and left. The red bins show the normalized histogram of $\Psi(t)$, the solid black line is a fit using two Gaussian functions at $\pm \mu_\Psi$. In this example, the positive throughput state lasted for a shorter time, hence the difference in heights (means and standard deviations are the same). Both peaks will tend to the same height in the limit $t\to\infty$. Here $\zeta = 5$, $\Delta = 3.2 \times 10^{-4}$, $\tau_C = 1$. \textit{Right:} A state with no net throughput for comparison, with $\zeta = 5$, $\Delta = 10^{-5}$, $\tau_C = 1$. \textit{Insets:} Examples of throughput-time series for each run.}
  \label{fig:throughput}
\end{figure}

At time $t$, the throughput of the system is defined as
\begin{equation}
  \Psi(t) = \frac{1}{L_y} \int_0^{L_y} v_x(t) dy = \langle v_x(t) \rangle_y\;,
\end{equation}
which is independent of $x$ because of fluid
incompressibility. Because this quantity generally exhibits
significant fluctuations in time, particularly in the chaotic regime,
we define our criterion for significant or `net' throughput as being
when the mean of the throughput time-series $\mu_\Psi$ exceeds the
standard deviation $\sigma_\Psi$:
\begin{equation}
 \mu_\Psi>\sigma_\Psi,
  \label{eq:active_2D_apx_throughput}
\end{equation}
where
\begin{equation}
  \mu_\Psi = \frac{1}{t} \int_0^t \Psi(t') dt',
  \label{eq:apx_throughput_mean}
\end{equation}
and
\begin{equation}
  \sigma_\Psi = \sqrt{\frac{1}{t} \int_0^t \left(\Psi(t') - \mu_\Psi\right)^2 dt'},
  \label{eq:apx_throughput_sd}
\end{equation}
which converge to constant values as $t \to \infty$. Note that under
this definition, states that have non-zero mean throughput will
fail the criterion if this mean is less than the standard deviation of
the time-series.

In practice, we find that the flow direction can intermittently
switch.  See, for example, \figref{fig:throughput} left. If naively
averaged as above, this would clearly produce zero mean throughput, at
least in the limit $t \to \infty$. Therefore we instead perform a
least-squares fit, fitting the throughput histogram with two Gaussians
of width $\sigma_\Psi$, centred at $\pm \mu_\Psi$. We have explicitly
checked that our results are robust to the number of histogram bins
used. Examples of both throughput and non-throughput states are shown
in \figref{fig:throughput}. We mark states satisfying criterion
\eqref{eq:active_2D_apx_throughput} in the phase diagrams with solid
symbols, and those failing it with empty symbols.

\input{refs.bbl}


%Merlin.mbs v4.21 2009-07-09.
\begin{thebibliography}{10}%
\makeatletter
\providecommand \@ifxundefined [1]{%
 \ifx #1\undefined \expandafter \@firstoftwo
 \else \expandafter \@secondoftwo
\fi
}%
\providecommand \@ifnum [1]{%
 \ifnum #1\expandafter \@firstoftwo
 \else \expandafter \@secondoftwo
\fi
}%
\providecommand \enquote [1]{``#1''}%
\providecommand \bibnamefont  [1]{#1}%
\providecommand \bibfnamefont [1]{#1}%
\providecommand \citenamefont [1]{#1}%
\providecommand\href[0]{\@sanitize\@href}%
\providecommand\@href[1]{\endgroup\@@startlink{#1}\endgroup\@@href}%
\providecommand\@@href[1]{#1\@@endlink}%
\providecommand \@sanitize [0]{\begingroup\catcode`\&12\catcode`\#12\relax}%
\@ifxundefined \pdfoutput {\@firstoftwo}{%
 \@ifnum{\z@=\pdfoutput}{\@firstoftwo}{\@secondoftwo}%
}{%
 \providecommand\@@startlink[1]{\leavevmode\special{html:<a href="#1">}}%
 \providecommand\@@endlink[0]{\special{html:</a>}}%
}{%
 \providecommand\@@startlink[1]{%
  \leavevmode
  \pdfstartlink
   attr{/Border[0 0 1 ]/H/I/C[0 1 1]}%
   user{/Subtype/Link/A<</Type/Action/S/URI/URI(#1)>>}%
  \relax
 }%
 \providecommand\@@endlink[0]{\pdfendlink}%
}%
\providecommand \url  [0]{\begingroup\@sanitize \@url }%
\providecommand \@url [1]{\endgroup\@href {#1}{\urlprefix}}%
\providecommand \urlprefix [0]{URL }%
\providecommand \Eprint[0]{\href }%
\@ifxundefined \urlstyle {%
  \providecommand \doi [1]{doi:\discretionary{}{}{}#1}%
}{%
  \providecommand \doi [0]{doi:\discretionary{}{}{}\begingroup
  \urlstyle{rm}\Url }%
}%
\providecommand \doibase [0]{http://dx.doi.org/}%
\providecommand \Doi[1]{\href{\doibase#1}}%
\providecommand \bibAnnote [3]{%
  \BibitemShut{#1}%
  \begin{quotation}\noindent
    \textsc{Key:}\ #2\\\textsc{Annotation:}\ #3%
  \end{quotation}%
}%
\providecommand \bibAnnoteFile [2]{%
  \IfFileExists{#2}{\bibAnnote {#1} {#2} {\input{#2}}}{}%
}%
\providecommand \typeout [0]{\immediate \write \m@ne }%
\providecommand \selectlanguage [0]{\@gobble}%
\providecommand \bibinfo [0]{\@secondoftwo}%
\providecommand \bibfield [0]{\@secondoftwo}%
\providecommand \translation [1]{[#1]}%
\providecommand \BibitemOpen[0]{}%
\providecommand \bibitemStop [0]{}%
\providecommand \bibitemNoStop [0]{.\EOS\space}%
\providecommand \EOS [0]{\spacefactor3000\relax}%
\providecommand \BibitemShut [1]{\csname bibitem#1\endcsname}%
%</preamble>
\bibitem{Marchetti2013}%
  \BibitemOpen
  \bibfield{author}{%
  \bibinfo {author} {\bibfnamefont{M.~C.}\ \bibnamefont{Marchetti}}, \bibinfo
  {author} {\bibfnamefont{J.-F.}\ \bibnamefont{Joanny}}, \bibinfo {author}
  {\bibfnamefont{S.}~\bibnamefont{Ramaswamy}}, \bibinfo {author}
  {\bibfnamefont{T.~B.}\ \bibnamefont{Liverpool}}, \bibinfo {author}
  {\bibfnamefont{J.}~\bibnamefont{Prost}}, \bibinfo {author}
  {\bibfnamefont{M.}~\bibnamefont{Rao}},\ and\ \bibinfo {author}
  {\bibfnamefont{R.~A.}\ \bibnamefont{Simha}},\ }%
  \bibfield{journal}{%
  \Doi{10.1103/RevModPhys.85.1143}{\bibinfo {journal} {Rev. Mod. Phys.}}\ }%
  \textbf{\bibinfo {volume} {85}},\ \bibinfo {pages} {1143} (\bibinfo {year}
  {2013})%
  \bibAnnoteFile{NoStop}{Marchetti2013}%
\bibitem{Dombrowski2004}%
  \BibitemOpen
  \bibfield{author}{%
  \bibinfo {author} {\bibfnamefont{C.}~\bibnamefont{Dombrowski}}, \bibinfo
  {author} {\bibfnamefont{L.}~\bibnamefont{Cisneros}}, \bibinfo {author}
  {\bibfnamefont{S.}~\bibnamefont{Chatkaew}}, \bibinfo {author}
  {\bibfnamefont{R.~E.}\ \bibnamefont{Goldstein}},\ and\ \bibinfo {author}
  {\bibfnamefont{J.~O.}\ \bibnamefont{Kessler}},\ }%
  \bibfield{journal}{%
  \Doi{10.1103/PhysRevLett.93.098103}{\bibinfo {journal} {Phys. Rev. Lett.}}\
  }%
  \textbf{\bibinfo {volume} {93}},\ \bibinfo {pages} {098103} (\bibinfo {year}
  {2004})%
  \bibAnnoteFile{NoStop}{Dombrowski2004}%
\bibitem{Nedelec1997}%
  \BibitemOpen
  \bibfield{author}{%
  \bibinfo {author} {\bibfnamefont{F.}~\bibnamefont{N\'{e}d\'{e}lec}}, \bibinfo
  {author} {\bibfnamefont{T.}~\bibnamefont{Surrey}}, \bibinfo {author}
  {\bibfnamefont{A.~C.}\ \bibnamefont{Maggs}},\ and\ \bibinfo {author}
  {\bibfnamefont{S.}~\bibnamefont{Leibler}},\ }%
  \bibfield{journal}{%
  \Doi{10.1038/38532}{\bibinfo {journal} {Nature}}\ }%
  \textbf{\bibinfo {volume} {389}},\ \bibinfo {pages} {305} (\bibinfo {year}
  {1997})%
  \bibAnnoteFile{NoStop}{Nedelec1997}%
\bibitem{Sanchez2012}%
  \BibitemOpen
  \bibfield{author}{%
  \bibinfo {author} {\bibfnamefont{T.}~\bibnamefont{Sanchez}}, \bibinfo
  {author} {\bibfnamefont{D.~T.~N.}\ \bibnamefont{Chen}}, \bibinfo {author}
  {\bibfnamefont{S.~J.}\ \bibnamefont{DeCamp}}, \bibinfo {author}
  {\bibfnamefont{M.}~\bibnamefont{Heymann}},\ and\ \bibinfo {author}
  {\bibfnamefont{Z.}~\bibnamefont{Dogic}},\ }%
  \bibfield{journal}{%
  \Doi{10.1038/nature11591}{\bibinfo {journal} {Nature}}\ }%
  \textbf{\bibinfo {volume} {491}},\ \bibinfo {pages} {431} (\bibinfo {year}
  {2012})%
  \bibAnnoteFile{NoStop}{Sanchez2012}%
\bibitem{Fielding2011}%
  \BibitemOpen
  \bibfield{author}{%
  \bibinfo {author} {\bibfnamefont{S.~M.}\ \bibnamefont{Fielding}}, \bibinfo
  {author} {\bibfnamefont{D.}~\bibnamefont{Marenduzzo}},\ and\ \bibinfo
  {author} {\bibfnamefont{M.~E.}\ \bibnamefont{Cates}},\ }%
  \bibfield{journal}{%
  \Doi{10.1103/PhysRevE.83.041910}{\bibinfo {journal} {Phys. Rev. E}}\ }%
  \textbf{\bibinfo {volume} {83}},\ \bibinfo {pages} {041910} (\bibinfo {year}
  {2011})%
  \bibAnnoteFile{NoStop}{Fielding2011}%
\bibitem{Wensink2012}%
  \BibitemOpen
  \bibfield{author}{%
  \bibinfo {author} {\bibfnamefont{H.~H.}\ \bibnamefont{Wensink}}, \bibinfo
  {author} {\bibfnamefont{J.}~\bibnamefont{Dunkel}}, \bibinfo {author}
  {\bibfnamefont{S.}~\bibnamefont{Heidenreich}}, \bibinfo {author}
  {\bibfnamefont{K.}~\bibnamefont{Drescher}}, \bibinfo {author}
  {\bibfnamefont{R.~E.}\ \bibnamefont{Goldstein}}, \bibinfo {author}
  {\bibfnamefont{H.}~\bibnamefont{L\"{o}wen}},\ and\ \bibinfo {author}
  {\bibfnamefont{J.~M.}\ \bibnamefont{Yeomans}},\ }%
  \bibfield{journal}{%
  \Doi{10.1073/pnas.1202032109}{\bibinfo {journal} {Proc. Natl. Acad. Sci. U.
  S. A.}}\ }%
  \textbf{\bibinfo {volume} {109}},\ \bibinfo {pages} {14308} (\bibinfo {year}
  {2012})%
  \bibAnnoteFile{NoStop}{Wensink2012}%
\bibitem{Giomi2013}%
  \BibitemOpen
  \bibfield{author}{%
  \bibinfo {author} {\bibfnamefont{L.}~\bibnamefont{Giomi}}, \bibinfo {author}
  {\bibfnamefont{M.~J.}\ \bibnamefont{Bowick}}, \bibinfo {author}
  {\bibfnamefont{X.}~\bibnamefont{Ma}},\ and\ \bibinfo {author}
  {\bibfnamefont{M.~C.}\ \bibnamefont{Marchetti}},\ }%
  \bibfield{journal}{%
  \Doi{10.1103/PhysRevLett.110.228101}{\bibinfo {journal} {Phys. Rev. Lett.}}\
  }%
  \textbf{\bibinfo {volume} {110}},\ \bibinfo {pages} {228101} (\bibinfo {year}
  {2013})%
  \bibAnnoteFile{NoStop}{Giomi2013}%
\bibitem{Giomi2014a}%
  \BibitemOpen
  \bibfield{author}{%
  \bibinfo {author} {\bibfnamefont{L.}~\bibnamefont{Giomi}}, \bibinfo {author}
  {\bibfnamefont{M.~J.}\ \bibnamefont{Bowick}}, \bibinfo {author}
  {\bibfnamefont{P.}~\bibnamefont{Mishra}}, \bibinfo {author}
  {\bibfnamefont{R.}~\bibnamefont{Sknepnek}},\ and\ \bibinfo {author}
  {\bibfnamefont{M.~C.}\ \bibnamefont{Marchetti}},\ }%
  \bibfield{journal}{%
  \Doi{10.1098/rsta.2013.0365}{\bibinfo {journal} {Philos. Trans. A. Math.
  Phys. Eng. Sci.}}\ }%
  \textbf{\bibinfo {volume} {372}},\ \bibinfo {pages} {16} (\bibinfo {year}
  {2014})%
  \bibAnnoteFile{NoStop}{Giomi2014a}%
\bibitem{Thampi2013}%
  \BibitemOpen
  \bibfield{author}{%
  \bibinfo {author} {\bibfnamefont{S.~P.}\ \bibnamefont{Thampi}}, \bibinfo
  {author} {\bibfnamefont{R.}~\bibnamefont{Golestanian}},\ and\ \bibinfo
  {author} {\bibfnamefont{J.~M.}\ \bibnamefont{Yeomans}},\ }%
  \bibfield{journal}{%
  \Doi{10.1103/PhysRevLett.111.118101}{\bibinfo {journal} {Phys. Rev. Lett.}}\
  }%
  \textbf{\bibinfo {volume} {111}},\ \bibinfo {pages} {118101} (\bibinfo {year}
  {2013})%
  \bibAnnoteFile{NoStop}{Thampi2013}%
\bibitem{Voituriez2005}%
  \BibitemOpen
  \bibfield{author}{%
  \bibinfo {author} {\bibfnamefont{R.}~\bibnamefont{Voituriez}}, \bibinfo
  {author} {\bibfnamefont{J.-F.}\ \bibnamefont{Joanny}},\ and\ \bibinfo
  {author} {\bibfnamefont{J.}~\bibnamefont{Prost}},\ }%
  \bibfield{journal}{%
  \Doi{10.1209/epl/i2004-10501-2}{\bibinfo {journal} {Europhys. Lett.}}\ }%
  \textbf{\bibinfo {volume} {70}},\ \bibinfo {pages} {404} (\bibinfo {year}
  {2005})%
  \bibAnnoteFile{NoStop}{Voituriez2005}%
\bibitem{Julicher2007}%
  \BibitemOpen
  \bibfield{author}{%
  \bibinfo {author} {\bibfnamefont{F.}~\bibnamefont{J\"{u}licher}}, \bibinfo
  {author} {\bibfnamefont{K.}~\bibnamefont{Kruse}}, \bibinfo {author}
  {\bibfnamefont{J.}~\bibnamefont{Prost}},\ and\ \bibinfo {author}
  {\bibfnamefont{J.-F.}\ \bibnamefont{Joanny}},\ }%
  \bibfield{journal}{%
  \Doi{10.1016/j.physrep.2007.02.018}{\bibinfo {journal} {Phys. Rep.}}\ }%
  \textbf{\bibinfo {volume} {449}},\ \bibinfo {pages} {3} (\bibinfo {year}
  {2007})%
  \bibAnnoteFile{NoStop}{Julicher2007}%
\bibitem{AditiSimha2002}%
  \BibitemOpen
  \bibfield{author}{%
  \bibinfo {author} {\bibfnamefont{R.}~\bibnamefont{{Aditi Simha}}}\ and\
  \bibinfo {author} {\bibfnamefont{S.}~\bibnamefont{Ramaswamy}},\ }%
  \bibfield{journal}{%
  \Doi{10.1103/PhysRevLett.89.058101}{\bibinfo {journal} {Phys. Rev. Lett.}}\
  }%
  \textbf{\bibinfo {volume} {89}},\ \bibinfo {pages} {058101} (\bibinfo {year}
  {2002})%
  \bibAnnoteFile{NoStop}{AditiSimha2002}%
\bibitem{Ramaswamy2010}%
  \BibitemOpen
  \bibfield{author}{%
  \bibinfo {author} {\bibfnamefont{S.}~\bibnamefont{Ramaswamy}},\ }%
  \bibfield{journal}{%
  \Doi{10.1146/annurev-conmatphys-070909-104101}{\bibinfo {journal} {Annu. Rev.
  Condens. Matter Phys.}}\ }%
  \textbf{\bibinfo {volume} {1}},\ \bibinfo {pages} {323} (\bibinfo {year}
  {2010})%
  \bibAnnoteFile{NoStop}{Ramaswamy2010}%
\bibitem{beris1994thermodynamics}%
  \BibitemOpen
  \bibfield{author}{%
  \bibinfo {author} {\bibfnamefont{A.~N.}\ \bibnamefont{Beris}}\ and\ \bibinfo
  {author} {\bibfnamefont{B.~J.}\ \bibnamefont{Edwards}},\ }%
  \emph{\bibinfo {title} {{Thermodynamics of flowing systems: with internal
  microstructure}}},\ Vol.~\bibinfo {volume} {36}\ (\bibinfo {publisher}
  {Oxford University Press, USA},\ \bibinfo {year} {1994})%
  \bibAnnoteFile{NoStop}{beris1994thermodynamics}%
\bibitem{degennes1995physics}%
  \BibitemOpen
  \bibfield{author}{%
  \bibinfo {author} {\bibfnamefont{P.~G.}\ \bibnamefont{de~Gennes}}\ and\
  \bibinfo {author} {\bibfnamefont{J.}~\bibnamefont{Prost}},\ }%
  \emph{\bibinfo {title} {{The Physics of Liquid Crystals}}},\ International
  Series of Monographs on Physics\ (\bibinfo {publisher} {Clarendon Press},\
  \bibinfo {year} {1995})\ ISBN \bibinfo {isbn} {9780198517856}%
  \bibAnnoteFile{NoStop}{degennes1995physics}%
\bibitem{Callan2011}%
  \BibitemOpen
  \bibfield{author}{%
  \bibinfo {author} {\bibfnamefont{A.~C.}\ \bibnamefont{Callan-Jones}}\ and\
  \bibinfo {author} {\bibfnamefont{F.}~\bibnamefont{J\"{u}licher}},\ }%
  \bibfield{journal}{%
  \Doi{10.1088/1367-2630/13/9/093027}{\bibinfo {journal} {New J. Phys.}}\ }%
  \textbf{\bibinfo {volume} {13}},\ \bibinfo {pages} {093027} (\bibinfo {year}
  {2011})%
  \bibAnnoteFile{NoStop}{Callan2011}%
\bibitem{Joanny2009a}%
  \BibitemOpen
  \bibfield{author}{%
  \bibinfo {author} {\bibfnamefont{J.-F.}\ \bibnamefont{Joanny}}\ and\ \bibinfo
  {author} {\bibfnamefont{J.}~\bibnamefont{Prost}},\ }%
  in\ \emph{\bibinfo {booktitle} {S\'{e}minaire Poincar\'{e} XII}}\ (\bibinfo
  {year} {2009})\ p.~\bibinfo {pages} {1}%
  \bibAnnoteFile{NoStop}{Joanny2009a}%
\bibitem{Bozorgi2013}%
  \BibitemOpen
  \bibfield{author}{%
  \bibinfo {author} {\bibfnamefont{Y.}~\bibnamefont{Bozorgi}}\ and\ \bibinfo
  {author} {\bibfnamefont{P.~T.}\ \bibnamefont{Underhill}},\ }%
  \bibfield{journal}{%
  \Doi{10.1122/1.4778578}{\bibinfo {journal} {J. Rheol.}}\ }%
  \textbf{\bibinfo {volume} {57}},\ \bibinfo {pages} {511} (\bibinfo {year}
  {2013})%
  \bibAnnoteFile{NoStop}{Bozorgi2013}%
\bibitem{Decho1990}%
  \BibitemOpen
  \bibfield{author}{%
  \bibinfo {author} {\bibfnamefont{A.~W.}\ \bibnamefont{Decho}},\ }%
  \bibfield{journal}{%
  \bibinfo {journal} {Ocean. Mar. Biol. Annu. Rev}\ }%
  \textbf{\bibinfo {volume} {28}},\ \bibinfo {pages} {73} (\bibinfo {year}
  {1990})%
  \bibAnnoteFile{NoStop}{Decho1990}%
\bibitem{Hemingway2015}%
  \BibitemOpen
  \bibfield{author}{%
  \bibinfo {author} {\bibfnamefont{E.~J.}\ \bibnamefont{Hemingway}}, \bibinfo
  {author} {\bibfnamefont{A.}~\bibnamefont{Maitra}}, \bibinfo {author}
  {\bibfnamefont{S.}~\bibnamefont{Banerjee}}, \bibinfo {author}
  {\bibfnamefont{M.~C.}\ \bibnamefont{Marchetti}}, \bibinfo {author}
  {\bibfnamefont{S.}~\bibnamefont{Ramaswamy}}, \bibinfo {author}
  {\bibfnamefont{S.~M.}\ \bibnamefont{Fielding}},\ and\ \bibinfo {author}
  {\bibfnamefont{M.~E.}\ \bibnamefont{Cates}},\ }%
  \bibfield{journal}{%
  \Doi{10.1103/PhysRevLett.114.098302}{\bibinfo {journal} {Phys. Rev. Lett.}}\
  }%
  \textbf{\bibinfo {volume} {114}},\ \bibinfo {pages} {098302} (\bibinfo {year}
  {2015})%
  \bibAnnoteFile{NoStop}{Hemingway2015}%
\bibitem{Lauga2007}%
  \BibitemOpen
  \bibfield{author}{%
  \bibinfo {author} {\bibfnamefont{E.}~\bibnamefont{Lauga}},\ }%
  \bibfield{journal}{%
  \Doi{10.1063/1.2751388}{\bibinfo {journal} {Phys. Fluids}}\ }%
  \textbf{\bibinfo {volume} {19}},\ \bibinfo {pages} {083104} (\bibinfo {year}
  {2007})%
  \bibAnnoteFile{NoStop}{Lauga2007}%
\bibitem{Teran2010}%
  \BibitemOpen
  \bibfield{author}{%
  \bibinfo {author} {\bibfnamefont{J.~M.}\ \bibnamefont{Teran}}, \bibinfo
  {author} {\bibfnamefont{L.}~\bibnamefont{Fauci}},\ and\ \bibinfo {author}
  {\bibfnamefont{M.~J.}\ \bibnamefont{Shelley}},\ }%
  \bibfield{journal}{%
  \Doi{10.1103/PhysRevLett.104.038101}{\bibinfo {journal} {Phys. Rev. Lett.}}\
  }%
  \textbf{\bibinfo {volume} {104}},\ \bibinfo {pages} {038101} (\bibinfo {year}
  {2010})%
  \bibAnnoteFile{NoStop}{Teran2010}%
\bibitem{Zhu2012}%
  \BibitemOpen
  \bibfield{author}{%
  \bibinfo {author} {\bibfnamefont{L.}~\bibnamefont{Zhu}}, \bibinfo {author}
  {\bibfnamefont{E.}~\bibnamefont{Lauga}},\ and\ \bibinfo {author}
  {\bibfnamefont{L.}~\bibnamefont{Brandt}},\ }%
  \bibfield{journal}{%
  \Doi{10.1063/1.4718446}{\bibinfo {journal} {Phys. Fluids}}\ }%
  \textbf{\bibinfo {volume} {24}},\ \bibinfo {pages} {051902} (\bibinfo {year}
  {2012})%
  \bibAnnoteFile{NoStop}{Zhu2012}%
\bibitem{Spagnolie2013}%
  \BibitemOpen
  \bibfield{author}{%
  \bibinfo {author} {\bibfnamefont{S.~E.}\ \bibnamefont{Spagnolie}}, \bibinfo
  {author} {\bibfnamefont{B.}~\bibnamefont{Liu}},\ and\ \bibinfo {author}
  {\bibfnamefont{T.~R.}\ \bibnamefont{Powers}},\ }%
  \bibfield{journal}{%
  \Doi{10.1103/PhysRevLett.111.068101}{\bibinfo {journal} {Phys. Rev. Lett.}}\
  }%
  \textbf{\bibinfo {volume} {111}},\ \bibinfo {pages} {068101} (\bibinfo {year}
  {2013})%
  \bibAnnoteFile{NoStop}{Spagnolie2013}%
\bibitem{Riley2014}%
  \BibitemOpen
  \bibfield{author}{%
  \bibinfo {author} {\bibfnamefont{E.~E.}\ \bibnamefont{Riley}}\ and\ \bibinfo
  {author} {\bibfnamefont{E.}~\bibnamefont{Lauga}},\ }%
  \bibfield{journal}{%
  \Doi{10.1209/0295-5075/108/34003}{\bibinfo {journal} {Europhys. Lett.}}\ }%
  \textbf{\bibinfo {volume} {108}},\ \bibinfo {pages} {34003} (\bibinfo {year}
  {2014})%
  \bibAnnoteFile{NoStop}{Riley2014}%
\bibitem{Woodhouse2012}%
  \BibitemOpen
  \bibfield{author}{%
  \bibinfo {author} {\bibfnamefont{F.~G.}\ \bibnamefont{Woodhouse}}\ and\
  \bibinfo {author} {\bibfnamefont{R.~E.}\ \bibnamefont{Goldstein}},\ }%
  \bibfield{journal}{%
  \Doi{10.1103/PhysRevLett.109.168105}{\bibinfo {journal} {Phys. Rev. Lett.}}\
  }%
  \textbf{\bibinfo {volume} {109}},\ \bibinfo {pages} {168105} (\bibinfo {year}
  {2012})%
  \bibAnnoteFile{NoStop}{Woodhouse2012}%
\bibitem{Goldstein2015}%
  \BibitemOpen
  \bibfield{author}{%
  \bibinfo {author} {\bibfnamefont{R.~E.}\ \bibnamefont{Goldstein}}\ and\
  \bibinfo {author} {\bibfnamefont{J.-W.}\ \bibnamefont{van~de Meent}},\ }%
  \bibfield{journal}{%
  \Doi{10.1098/rsfs.2015.0030}{\bibinfo {journal} {Interface Focus}}\ }%
  \textbf{\bibinfo {volume} {5}},\ \bibinfo {pages} {20150030} (\bibinfo {year}
  {2015})%
  \bibAnnoteFile{NoStop}{Goldstein2015}%
\bibitem{Serbus2005}%
  \BibitemOpen
  \bibfield{author}{%
  \bibinfo {author} {\bibfnamefont{L.~R.}\ \bibnamefont{Serbus}}, \bibinfo
  {author} {\bibfnamefont{B.-J.}\ \bibnamefont{Cha}}, \bibinfo {author}
  {\bibfnamefont{W.~E.}\ \bibnamefont{Theurkauf}},\ and\ \bibinfo {author}
  {\bibfnamefont{W.~M.}\ \bibnamefont{Saxton}},\ }%
  \bibfield{journal}{%
  \Doi{10.1242/dev.01956}{\bibinfo {journal} {Development}}\ }%
  \textbf{\bibinfo {volume} {132}},\ \bibinfo {pages} {3743} (\bibinfo {year}
  {2005})%
  \bibAnnoteFile{NoStop}{Serbus2005}%
\bibitem{Vedula2012}%
  \BibitemOpen
  \bibfield{author}{%
  \bibinfo {author} {\bibfnamefont{S.~R.~K.}\ \bibnamefont{Vedula}}, \bibinfo
  {author} {\bibfnamefont{M.~C.}\ \bibnamefont{Leong}}, \bibinfo {author}
  {\bibfnamefont{T.~L.}\ \bibnamefont{Lai}}, \bibinfo {author}
  {\bibfnamefont{P.}~\bibnamefont{Hersen}}, \bibinfo {author}
  {\bibfnamefont{A.~J.}\ \bibnamefont{Kabla}}, \bibinfo {author}
  {\bibfnamefont{C.~T.}\ \bibnamefont{Lim}},\ and\ \bibinfo {author}
  {\bibfnamefont{B.}~\bibnamefont{Ladoux}},\ }%
  \bibfield{journal}{%
  \Doi{10.1073/pnas.1119313109}{\bibinfo {journal} {Proc. Natl. Acad. Sci. U.
  S. A.}}\ }%
  \textbf{\bibinfo {volume} {109}},\ \bibinfo {pages} {12974} (\bibinfo {year}
  {2012})%
  \bibAnnoteFile{NoStop}{Vedula2012}%
\bibitem{Wioland2013}%
  \BibitemOpen
  \bibfield{author}{%
  \bibinfo {author} {\bibfnamefont{H.}~\bibnamefont{Wioland}}, \bibinfo
  {author} {\bibfnamefont{F.~G.}\ \bibnamefont{Woodhouse}}, \bibinfo {author}
  {\bibfnamefont{J.}~\bibnamefont{Dunkel}}, \bibinfo {author}
  {\bibfnamefont{J.~O.}\ \bibnamefont{Kessler}},\ and\ \bibinfo {author}
  {\bibfnamefont{R.~E.}\ \bibnamefont{Goldstein}},\ }%
  \bibfield{journal}{%
  \Doi{10.1103/PhysRevLett.110.268102}{\bibinfo {journal} {Phys. Rev. Lett.}}\
  }%
  \textbf{\bibinfo {volume} {110}},\ \bibinfo {pages} {5} (\bibinfo {year}
  {2013})%
  \bibAnnoteFile{NoStop}{Wioland2013}%
\bibitem{Ramaswamy2003}%
  \BibitemOpen
  \bibfield{author}{%
  \bibinfo {author} {\bibfnamefont{S.}~\bibnamefont{Ramaswamy}}, \bibinfo
  {author} {\bibfnamefont{R.~A.}\ \bibnamefont{Simha}},\ and\ \bibinfo {author}
  {\bibfnamefont{J.}~\bibnamefont{Toner}},\ }%
  \bibfield{journal}{%
  \Doi{10.1209/epl/i2003-00346-7}{\bibinfo {journal} {Europhys. Lett.}}\ }%
  \textbf{\bibinfo {volume} {62}},\ \bibinfo {pages} {196} (\bibinfo {year}
  {2002})%
  \bibAnnoteFile{NoStop}{Ramaswamy2003}%
\bibitem{Milner1993}%
  \BibitemOpen
  \bibfield{author}{%
  \bibinfo {author} {\bibfnamefont{S.~T.}\ \bibnamefont{Milner}},\ }%
  \bibfield{journal}{%
  \Doi{10.1103/PhysRevE.48.3674}{\bibinfo {journal} {Phys. Rev. E}}\ }%
  \textbf{\bibinfo {volume} {48}},\ \bibinfo {pages} {3674} (\bibinfo {year}
  {1993})%
  \bibAnnoteFile{NoStop}{Milner1993}%
\bibitem{Olmsted2000}%
  \BibitemOpen
  \bibfield{author}{%
  \bibinfo {author} {\bibfnamefont{P.~D.}\ \bibnamefont{Olmsted}}, \bibinfo
  {author} {\bibfnamefont{O.}~\bibnamefont{Radulescu}},\ and\ \bibinfo {author}
  {\bibfnamefont{C.-Y.~D.}\ \bibnamefont{Lu}},\ }%
  \bibfield{journal}{%
  \Doi{10.1122/1.551085}{\bibinfo {journal} {J. Rheol.}}\ }%
  \textbf{\bibinfo {volume} {44}},\ \bibinfo {pages} {257} (\bibinfo {year}
  {2000})%
  \bibAnnoteFile{NoStop}{Olmsted2000}%
\bibitem{PDO_corr}%
  \BibitemOpen
  \bibfield{author}{%
  \bibinfo {author} {\bibfnamefont{P.}~\bibnamefont{Olmsted}},\ }%
  \bibinfo {howpublished} {Private communication} (\bibinfo {year} {2013})%
  \bibAnnoteFile{NoStop}{PDO_corr}%
\bibitem{Jayaraman2012}%
  \BibitemOpen
  \bibfield{author}{%
  \bibinfo {author} {\bibfnamefont{G.}~\bibnamefont{Jayaraman}}, \bibinfo
  {author} {\bibfnamefont{S.}~\bibnamefont{Ramachandran}}, \bibinfo {author}
  {\bibfnamefont{S.}~\bibnamefont{Ghose}}, \bibinfo {author}
  {\bibfnamefont{A.}~\bibnamefont{Laskar}}, \bibinfo {author}
  {\bibfnamefont{M.~S.}\ \bibnamefont{Bhamla}}, \bibinfo {author}
  {\bibfnamefont{P.~B.~S.}\ \bibnamefont{Kumar}},\ and\ \bibinfo {author}
  {\bibfnamefont{R.}~\bibnamefont{Adhikari}},\ }%
  \bibfield{journal}{%
  \Doi{10.1103/PhysRevLett.109.158302}{\bibinfo {journal} {Phys. Rev. Lett.}}\
  }%
  \textbf{\bibinfo {volume} {109}},\ \bibinfo {pages} {158302} (\bibinfo {year}
  {2012})%
  \bibAnnoteFile{NoStop}{Jayaraman2012}%
\bibitem{Stark2003}%
  \BibitemOpen
  \bibfield{author}{%
  \bibinfo {author} {\bibfnamefont{H.}~\bibnamefont{Stark}}\ and\ \bibinfo
  {author} {\bibfnamefont{T.~C.}\ \bibnamefont{Lubensky}},\ }%
  \bibfield{journal}{%
  \Doi{10.1103/PhysRevE.67.061709}{\bibinfo {journal} {Phys. Rev. E. Stat.
  Nonlin. Soft Matter Phys.}}\ }%
  \textbf{\bibinfo {volume} {67}},\ \bibinfo {pages} {061709} (\bibinfo {year}
  {2003})%
  \bibAnnoteFile{NoStop}{Stark2003}%
\bibitem{LarsonConstit1988}%
  \BibitemOpen
  \bibfield{author}{%
  \bibinfo {author} {\bibfnamefont{R.~G.}\ \bibnamefont{Larson}},\ }%
  \emph{\bibinfo {title} {{Constitutive Equations for Polymer Melts and
  Solutions}}}\ (\bibinfo {publisher} {Butterworths},\ \bibinfo {year} {1988})%
  \bibAnnoteFile{NoStop}{LarsonConstit1988}%
\bibitem{warner2003liquid}%
  \BibitemOpen
  \bibfield{author}{%
  \bibinfo {author} {\bibfnamefont{M.}~\bibnamefont{Warner}}\ and\ \bibinfo
  {author} {\bibfnamefont{E.~M.}\ \bibnamefont{Terentjev}},\ }%
  \emph{\bibinfo {title} {{Liquid Crystal Elastomers}}},\ International Series
  of Monographs on Physics\ (\bibinfo {publisher} {OUP Oxford},\ \bibinfo
  {year} {2003})\ ISBN \bibinfo {isbn} {9780191523632}%
  \bibAnnoteFile{NoStop}{warner2003liquid}%
\bibitem{Cates2008}%
  \BibitemOpen
  \bibfield{author}{%
  \bibinfo {author} {\bibfnamefont{M.~E.}\ \bibnamefont{Cates}}, \bibinfo
  {author} {\bibfnamefont{S.~M.}\ \bibnamefont{Fielding}}, \bibinfo {author}
  {\bibfnamefont{D.}~\bibnamefont{Marenduzzo}}, \bibinfo {author}
  {\bibfnamefont{E.}~\bibnamefont{Orlandini}},\ and\ \bibinfo {author}
  {\bibfnamefont{J.~M.}\ \bibnamefont{Yeomans}},\ }%
  \bibfield{journal}{%
  \Doi{10.1103/PhysRevLett.101.068102}{\bibinfo {journal} {Phys. Rev. Lett.}}\
  }%
  \textbf{\bibinfo {volume} {101}},\ \bibinfo {pages} {068102} (\bibinfo {year}
  {2008})%
  \bibAnnoteFile{NoStop}{Cates2008}%
\bibitem{Edwards2009}%
  \BibitemOpen
  \bibfield{author}{%
  \bibinfo {author} {\bibfnamefont{S.~A.}\ \bibnamefont{Edwards}}\ and\
  \bibinfo {author} {\bibfnamefont{J.~M.}\ \bibnamefont{Yeomans}},\ }%
  \bibfield{journal}{%
  \Doi{10.1209/0295-5075/85/18008}{\bibinfo {journal} {Europhys. Lett.}}\ }%
  \textbf{\bibinfo {volume} {85}},\ \bibinfo {pages} {18008} (\bibinfo {year}
  {2009})%
  \bibAnnoteFile{NoStop}{Edwards2009}%
\bibitem{Marenduzzo2008}%
  \BibitemOpen
  \bibfield{author}{%
  \bibinfo {author} {\bibfnamefont{D.}~\bibnamefont{Marenduzzo}}, \bibinfo
  {author} {\bibfnamefont{E.}~\bibnamefont{Orlandini}}, \bibinfo {author}
  {\bibfnamefont{M.~E.}\ \bibnamefont{Cates}},\ and\ \bibinfo {author}
  {\bibfnamefont{J.~M.}\ \bibnamefont{Yeomans}},\ }%
  \bibfield{journal}{%
  \Doi{10.1016/j.jnnfm.2007.02.005}{\bibinfo {journal} {J. Non-Newtonian Fluid
  Mech.}}\ }%
  \textbf{\bibinfo {volume} {149}},\ \bibinfo {pages} {56} (\bibinfo {year}
  {2008})%
  \bibAnnoteFile{NoStop}{Marenduzzo2008}%
\bibitem{Zhou2014}%
  \BibitemOpen
  \bibfield{author}{%
  \bibinfo {author} {\bibfnamefont{S.}~\bibnamefont{Zhou}}, \bibinfo {author}
  {\bibfnamefont{A.}~\bibnamefont{Sokolov}}, \bibinfo {author}
  {\bibfnamefont{O.~D.}\ \bibnamefont{Lavrentovich}},\ and\ \bibinfo {author}
  {\bibfnamefont{I.~S.}\ \bibnamefont{Aranson}},\ }%
  \bibfield{journal}{%
  \Doi{10.1073/pnas.1321926111}{\bibinfo {journal} {Proc. Natl. Acad. Sci. U.
  S. A.}}\ }%
  \textbf{\bibinfo {volume} {111}},\ \bibinfo {pages} {1265} (\bibinfo {year}
  {2014})%
  \bibAnnoteFile{NoStop}{Zhou2014}%
\bibitem{See1990}%
  \BibitemOpen
  \bibfield{author}{%
  \bibinfo {author} {\bibfnamefont{H.}~\bibnamefont{See}}, \bibinfo {author}
  {\bibfnamefont{M.}~\bibnamefont{Doi}},\ and\ \bibinfo {author}
  {\bibfnamefont{R.~G.}\ \bibnamefont{Larson}},\ }%
  \bibfield{journal}{%
  \Doi{10.1063/1.458598}{\bibinfo {journal} {J. Chem. Phys.}}\ }%
  \textbf{\bibinfo {volume} {92}},\ \bibinfo {pages} {792} (\bibinfo {year}
  {1990})%
  \bibAnnoteFile{NoStop}{See1990}%
\bibitem{Olmsted1994}%
  \BibitemOpen
  \bibfield{author}{%
  \bibinfo {author} {\bibfnamefont{P.~D.}\ \bibnamefont{Olmsted}}\ and\
  \bibinfo {author} {\bibfnamefont{S.~T.}\ \bibnamefont{Milner}},\ }%
  \bibfield{journal}{%
  \Doi{10.1021/ma00100a059}{\bibinfo {journal} {Macromolecules}}\ }%
  \textbf{\bibinfo {volume} {27}},\ \bibinfo {pages} {6648} (\bibinfo {year}
  {1994})%
  \bibAnnoteFile{NoStop}{Olmsted1994}%
\bibitem{PrashantPreprint2015}%
  \BibitemOpen
  \bibfield{author}{%
   \bibinfo {author} {\bibfnamefont{E.~J.}\ \bibnamefont{Hemingway}},
   \bibinfo {author} {\bibfnamefont{P.~M.}\ \bibnamefont{Mishra}},
   \bibinfo {author}
  {\bibfnamefont{S.~M.}\ \bibnamefont{Fielding}},\ and\ \bibinfo {author}
  {\bibfnamefont{M.~C.}\ \bibnamefont{Marchetti}},\ }%
  \bibinfo {howpublished} {Correlation lengths in active nematics (in
  preparation)} (\bibinfo {year} {2015})%
  \bibAnnoteFile{NoStop}{PrashantPreprint2015}%
\bibitem{Note1}%
  \BibitemOpen
  \bibinfo {note} {This value in fact corresponds to the spinodal stability
  limit of the isotropic phase.}%
  \bibAnnoteFile{Stop}{Note1}%
\bibitem{Head2003}%
  \BibitemOpen
  \bibfield{author}{%
  \bibinfo {author} {\bibfnamefont{D.~A.}\ \bibnamefont{Head}}, \bibinfo
  {author} {\bibfnamefont{A.~J.}\ \bibnamefont{Levine}},\ and\ \bibinfo
  {author} {\bibfnamefont{F.~C.}\ \bibnamefont{MacKintosh}},\ }%
  \bibfield{journal}{%
  \Doi{10.1103/PhysRevLett.91.108102}{\bibinfo {journal} {Phys. Rev. Lett.}}\
  }%
  \textbf{\bibinfo {volume} {91}},\ \bibinfo {pages} {108102} (\bibinfo {year}
  {2003})%
  \bibAnnoteFile{NoStop}{Head2003}%
\bibitem{Storm2005}%
  \BibitemOpen
  \bibfield{author}{%
  \bibinfo {author} {\bibfnamefont{C.}~\bibnamefont{Storm}}, \bibinfo {author}
  {\bibfnamefont{J.~J.}\ \bibnamefont{Pastore}}, \bibinfo {author}
  {\bibfnamefont{F.~C.}\ \bibnamefont{MacKintosh}}, \bibinfo {author}
  {\bibfnamefont{T.~C.}\ \bibnamefont{Lubensky}},\ and\ \bibinfo {author}
  {\bibfnamefont{P.~A.}\ \bibnamefont{Janmey}},\ }%
  \bibfield{journal}{%
  \Doi{10.1038/nature03521}{\bibinfo {journal} {Nature}}\ }%
  \textbf{\bibinfo {volume} {435}},\ \bibinfo {pages} {191} (\bibinfo {year}
  {2005})%
  \bibAnnoteFile{NoStop}{Storm2005}%
\bibitem{Koenderink2009}%
  \BibitemOpen
  \bibfield{author}{%
  \bibinfo {author} {\bibfnamefont{G.~H.}\ \bibnamefont{Koenderink}}, \bibinfo
  {author} {\bibfnamefont{Z.}~\bibnamefont{Dogic}}, \bibinfo {author}
  {\bibfnamefont{F.}~\bibnamefont{Nakamura}}, \bibinfo {author}
  {\bibfnamefont{P.~M.}\ \bibnamefont{Bendix}}, \bibinfo {author}
  {\bibfnamefont{F.~C.}\ \bibnamefont{MacKintosh}}, \bibinfo {author}
  {\bibfnamefont{J.~H.}\ \bibnamefont{Hartwig}}, \bibinfo {author}
  {\bibfnamefont{T.~P.}\ \bibnamefont{Stossel}},\ and\ \bibinfo {author}
  {\bibfnamefont{D.~a.}\ \bibnamefont{Weitz}},\ }%
  \bibfield{journal}{%
  \Doi{10.1073/pnas.0903974106}{\bibinfo {journal} {Proc. Natl. Acad. Sci. U.
  S. A.}}\ }%
  \textbf{\bibinfo {volume} {106}},\ \bibinfo {pages} {15192} (\bibinfo {year}
  {2009})%
  \bibAnnoteFile{NoStop}{Koenderink2009}%
\bibitem{Mofrad2009}%
  \BibitemOpen
  \bibfield{author}{%
  \bibinfo {author} {\bibfnamefont{M.~R.}\ \bibnamefont{Mofrad}},\ }%
  \bibfield{journal}{%
  \Doi{10.1146/annurev.fluid.010908.165236}{\bibinfo {journal} {Annu. Rev.
  Fluid Mech.}}\ }%
  \textbf{\bibinfo {volume} {41}},\ \bibinfo {pages} {433} (\bibinfo {year}
  {2009})%
  \bibAnnoteFile{NoStop}{Mofrad2009}%
\bibitem{NumericalRecipiesBook}%
  \BibitemOpen
  \bibfield{author}{%
  \bibinfo {author} {\bibfnamefont{W.~H.}\ \bibnamefont{Press}}, \bibinfo
  {author} {\bibfnamefont{S.~A.}\ \bibnamefont{Teukolsky}}, \bibinfo {author}
  {\bibfnamefont{W.~T.}\ \bibnamefont{Vetterling}},\ and\ \bibinfo {author}
  {\bibfnamefont{B.~P.}\ \bibnamefont{Flannery}},\ }%
  \emph{\bibinfo {title} {{Numerical Recipes in C: The Art of Scientific
  Computing}}},\ \bibinfo {edition} {2nd}\ ed.\ (\bibinfo {publisher}
  {Cambridge University Press},\ \bibinfo {address} {New York, NY, USA},\
  \bibinfo {year} {1992})%
  \bibAnnoteFile{NoStop}{NumericalRecipiesBook}%
\bibitem{pozrikidis2009fluid}%
  \BibitemOpen
  \bibfield{author}{%
  \bibinfo {author} {\bibfnamefont{C.}~\bibnamefont{Pozrikidis}},\ }%
  \emph{\bibinfo {title} {{Fluid Dynamics: Theory, Computation, and Numerical
  Simulation}}},\ \bibinfo {edition} {2nd}\ ed.\ (\bibinfo {publisher}
  {Springer US},\ \bibinfo {year} {2009})%
  \bibAnnoteFile{NoStop}{pozrikidis2009fluid}%
\bibitem{Marenduzzo2007a}%
  \BibitemOpen
  \bibfield{author}{%
  \bibinfo {author} {\bibfnamefont{D.}~\bibnamefont{Marenduzzo}}, \bibinfo
  {author} {\bibfnamefont{E.}~\bibnamefont{Orlandini}}, \bibinfo {author}
  {\bibfnamefont{M.~E.}\ \bibnamefont{Cates}},\ and\ \bibinfo {author}
  {\bibfnamefont{J.~M.}\ \bibnamefont{Yeomans}},\ }%
  \bibfield{journal}{%
  \Doi{10.1103/PhysRevE.76.031921}{\bibinfo {journal} {Phys. Rev. E}}\ }%
  \textbf{\bibinfo {volume} {76}},\ \bibinfo {pages} {031921} (\bibinfo {year}
  {2007})%
  \bibAnnoteFile{NoStop}{Marenduzzo2007a}%
\bibitem{Strogatz2008Book}%
  \BibitemOpen
  \bibfield{author}{%
  \bibinfo {author} {\bibfnamefont{S.~H.}\ \bibnamefont{Strogatz}},\ }%
  \emph{\bibinfo {title} {{Nonlinear Dynamics and Chaos: With Applications to
  Physics, Biology, Chemistry, and Engineering}}},\ Studies in nonlinearity\
  (\bibinfo {publisher} {Westview Press},\ \bibinfo {year} {2008})\ ISBN
  \bibinfo {isbn} {9780786723959}%
  \bibAnnoteFile{NoStop}{Strogatz2008Book}%
\bibitem{Chaikin2000Book}%
  \BibitemOpen
  \bibfield{author}{%
  \bibinfo {author} {\bibfnamefont{P.~M.}\ \bibnamefont{Chaikin}}\ and\
  \bibinfo {author} {\bibfnamefont{T.~C.}\ \bibnamefont{Lubensky}},\ }%
  \emph{\bibinfo {title} {{Principles of Condensed Matter Physics}}}\ (\bibinfo
  {publisher} {Cambridge University Press},\ \bibinfo {year} {2000})\ ISBN
  \bibinfo {isbn} {9780521794503}%
  \bibAnnoteFile{NoStop}{Chaikin2000Book}%
\bibitem{Giomi2012}%
  \BibitemOpen
  \bibfield{author}{%
  \bibinfo {author} {\bibfnamefont{L.}~\bibnamefont{Giomi}}, \bibinfo {author}
  {\bibfnamefont{L.}~\bibnamefont{Mahadevan}}, \bibinfo {author}
  {\bibfnamefont{B.}~\bibnamefont{Chakraborty}},\ and\ \bibinfo {author}
  {\bibfnamefont{M.~F.}\ \bibnamefont{Hagan}},\ }%
  \bibfield{journal}{%
  \Doi{10.1088/0951-7715/25/8/2245}{\bibinfo {journal} {Nonlinearity}}\ }%
  \textbf{\bibinfo {volume} {25}},\ \bibinfo {pages} {2245} (\bibinfo {year}
  {2012})%
  \bibAnnoteFile{NoStop}{Giomi2012}%
\bibitem{Salbreux2007}%
  \BibitemOpen
  \bibfield{author}{%
  \bibinfo {author} {\bibfnamefont{G.}~\bibnamefont{Salbreux}}, \bibinfo
  {author} {\bibfnamefont{J.-F.}\ \bibnamefont{Joanny}}, \bibinfo {author}
  {\bibfnamefont{J.}~\bibnamefont{Prost}},\ and\ \bibinfo {author}
  {\bibfnamefont{P.}~\bibnamefont{Pullarkat}},\ }%
  \bibfield{journal}{%
  \Doi{10.1088/1478-3975/4/4/004}{\bibinfo {journal} {Phys. Biol.}}\ }%
  \textbf{\bibinfo {volume} {4}},\ \bibinfo {pages} {268} (\bibinfo {year}
  {2007})%
  \bibAnnoteFile{NoStop}{Salbreux2007}%
\bibitem{Dierkes2014}%
  \BibitemOpen
  \bibfield{author}{%
  \bibinfo {author} {\bibfnamefont{K.}~\bibnamefont{Dierkes}}, \bibinfo
  {author} {\bibfnamefont{A.}~\bibnamefont{Sumi}}, \bibinfo {author}
  {\bibfnamefont{J.}~\bibnamefont{Solon}},\ and\ \bibinfo {author}
  {\bibfnamefont{G.}~\bibnamefont{Salbreux}},\ }%
  \bibfield{journal}{%
  \Doi{10.1103/PhysRevLett.113.148102}{\bibinfo {journal} {Phys. Rev. Lett.}}\
  }%
  \textbf{\bibinfo {volume} {113}},\ \bibinfo {pages} {148102} (\bibinfo {year}
  {2014})%
  \bibAnnoteFile{NoStop}{Dierkes2014}%
\bibitem{Aradian2004}%
  \BibitemOpen
  \bibfield{author}{%
  \bibinfo {author} {\bibfnamefont{A.}~\bibnamefont{Aradian}}\ and\ \bibinfo
  {author} {\bibfnamefont{M.~E.}\ \bibnamefont{Cates}},\ }%
  \bibfield{journal}{%
  \Doi{10.1209/epl/i2005-10011-9}{\bibinfo {journal} {Europhys. Lett.}}\ }%
  \textbf{\bibinfo {volume} {70}},\ \bibinfo {pages} {397} (\bibinfo {year}
  {2004})%
  \bibAnnoteFile{NoStop}{Aradian2004}%
\bibitem{Schaller2010}%
  \BibitemOpen
  \bibfield{author}{%
  \bibinfo {author} {\bibfnamefont{V.}~\bibnamefont{Schaller}}, \bibinfo
  {author} {\bibfnamefont{C.}~\bibnamefont{Weber}}, \bibinfo {author}
  {\bibfnamefont{C.}~\bibnamefont{Semmrich}}, \bibinfo {author}
  {\bibfnamefont{E.}~\bibnamefont{Frey}},\ and\ \bibinfo {author}
  {\bibfnamefont{A.~R.}\ \bibnamefont{Bausch}},\ }%
  \bibfield{journal}{%
  \Doi{10.1038/nature09312}{\bibinfo {journal} {Nature}}\ }%
  \textbf{\bibinfo {volume} {467}},\ \bibinfo {pages} {73} (\bibinfo {year}
  {2010})%
  \bibAnnoteFile{NoStop}{Schaller2010}%
\bibitem{Thampi2014b}%
  \BibitemOpen
  \bibfield{author}{%
  \bibinfo {author} {\bibfnamefont{S.~P.}\ \bibnamefont{Thampi}}, \bibinfo
  {author} {\bibfnamefont{R.}~\bibnamefont{Golestanian}},\ and\ \bibinfo
  {author} {\bibfnamefont{J.~M.}\ \bibnamefont{Yeomans}},\ }%
  \bibfield{journal}{%
  \Doi{10.1098/rsta.2013.0366}{\bibinfo {journal} {Philos. Trans. A. Math.
  Phys. Eng. Sci.}}\ }%
  \textbf{\bibinfo {volume} {372}},\ \bibinfo {pages} {14} (\bibinfo {year}
  {2014})%
  \bibAnnoteFile{NoStop}{Thampi2014b}%
\bibitem{Huterer2005}%
  \BibitemOpen
  \bibfield{author}{%
  \bibinfo {author} {\bibfnamefont{D.}~\bibnamefont{Huterer}}\ and\ \bibinfo
  {author} {\bibfnamefont{T.}~\bibnamefont{Vachaspati}},\ }%
  \bibfield{journal}{%
  \Doi{10.1103/PhysRevD.72.043004}{\bibinfo {journal} {Phys. Rev. D}}\ }%
  \textbf{\bibinfo {volume} {72}},\ \bibinfo {pages} {043004} (\bibinfo {year}
  {2005})%
  \bibAnnoteFile{NoStop}{Huterer2005}%
\bibitem{Giomi2015}%
  \BibitemOpen
  \bibfield{author}{%
  \bibinfo {author} {\bibfnamefont{L.}~\bibnamefont{Giomi}},\ }%
  \bibfield{journal}{%
  \Doi{10.1103/PhysRevX.5.031003}{\bibinfo {journal} {Phys. Rev. X}}\ }%
  \textbf{\bibinfo {volume} {5}},\ \bibinfo {pages} {031003} (\bibinfo {year}
  {2015})%
  \bibAnnoteFile{NoStop}{Giomi2015}%
\bibitem{Bozorgi2014}%
  \BibitemOpen
  \bibfield{author}{%
  \bibinfo {author} {\bibfnamefont{Y.}~\bibnamefont{Bozorgi}}\ and\ \bibinfo
  {author} {\bibfnamefont{P.~T.}\ \bibnamefont{Underhill}},\ }%
  \bibfield{journal}{%
  \Doi{10.1016/j.jnnfm.2014.09.016}{\bibinfo {journal} {J. Non-Newtonian Fluid
  Mech.}}\ }%
  \textbf{\bibinfo {volume} {214}},\ \bibinfo {pages} {69} (\bibinfo {year}
  {2014})%
  \bibAnnoteFile{NoStop}{Bozorgi2014}%
\bibitem{Reynolds1883}%
  \BibitemOpen
  \bibfield{author}{%
  \bibinfo {author} {\bibfnamefont{O.}~\bibnamefont{Reynolds}},\ }%
  \bibfield{journal}{%
  \Doi{10.1098/rstl.1883.0029}{\bibinfo {journal} {Philos. Trans. R. Soc.
  London}}\ }%
  \textbf{\bibinfo {volume} {174}},\ \bibinfo {pages} {935} (\bibinfo {year}
  {1883})%
  \bibAnnoteFile{NoStop}{Reynolds1883}%
\bibitem{White2008}%
  \BibitemOpen
  \bibfield{author}{%
  \bibinfo {author} {\bibfnamefont{C.~M.}\ \bibnamefont{White}}\ and\ \bibinfo
  {author} {\bibfnamefont{M.~G.}\ \bibnamefont{Mungal}},\ }%
  \bibfield{journal}{%
  \Doi{10.1146/annurev.fluid.40.111406.102156}{\bibinfo {journal} {Annu. Rev.
  Fluid Mech.}}\ }%
  \textbf{\bibinfo {volume} {40}},\ \bibinfo {pages} {235} (\bibinfo {year}
  {2008})%
  \bibAnnoteFile{NoStop}{White2008}%
\bibitem{Rallison1988}%
  \BibitemOpen
  \bibfield{author}{%
  \bibinfo {author} {\bibfnamefont{J.}~\bibnamefont{Rallison}}\ and\ \bibinfo
  {author} {\bibfnamefont{E.}~\bibnamefont{Hinch}},\ }%
  \bibfield{journal}{%
  \Doi{10.1016/0377-0257(88)85049-3}{\bibinfo {journal} {J. Non-Newtonian Fluid
  Mech.}}\ }%
  \textbf{\bibinfo {volume} {29}},\ \bibinfo {pages} {37} (\bibinfo {year}
  {1988})%
  \bibAnnoteFile{NoStop}{Rallison1988}%
\bibitem{Thampi2014}%
  \BibitemOpen
  \bibfield{author}{%
  \bibinfo {author} {\bibfnamefont{S.~P.}\ \bibnamefont{Thampi}}, \bibinfo
  {author} {\bibfnamefont{R.}~\bibnamefont{Golestanian}},\ and\ \bibinfo
  {author} {\bibfnamefont{J.~M.}\ \bibnamefont{Yeomans}},\ }%
  \bibfield{journal}{%
  \Doi{10.1209/0295-5075/105/18001}{\bibinfo {journal} {Europhys. Lett.}}\ }%
  \textbf{\bibinfo {volume} {105}},\ \bibinfo {pages} {18001} (\bibinfo {year}
  {2014})%
  \bibAnnoteFile{NoStop}{Thampi2014}%
\bibitem{He2010}%
  \BibitemOpen
  \bibfield{author}{%
  \bibinfo {author} {\bibfnamefont{L.}~\bibnamefont{He}}, \bibinfo {author}
  {\bibfnamefont{X.}~\bibnamefont{Wang}}, \bibinfo {author}
  {\bibfnamefont{H.~L.}\ \bibnamefont{Tang}},\ and\ \bibinfo {author}
  {\bibfnamefont{D.~J.}\ \bibnamefont{Montell}},\ }%
  \bibfield{journal}{%
  \Doi{10.1038/ncb2124}{\bibinfo {journal} {Nat. Cell Biol.}}\ }%
  \textbf{\bibinfo {volume} {12}},\ \bibinfo {pages} {1133} (\bibinfo {year}
  {2010})%
  \bibAnnoteFile{NoStop}{He2010}%
\bibitem{Fielding2012}%
  \BibitemOpen
  \bibfield{author}{%
  \bibinfo {author} {\bibfnamefont{S.}~\bibnamefont{Fielding}}, \bibinfo
  {author} {\bibfnamefont{R.}~\bibnamefont{Larson}},\ and\ \bibinfo {author}
  {\bibfnamefont{M.~E.}\ \bibnamefont{Cates}},\ }%
  \bibfield{journal}{%
  \Doi{10.1103/PhysRevLett.108.048301}{\bibinfo {journal} {Phys. Rev. Lett.}}\
  }%
  \textbf{\bibinfo {volume} {108}},\ \bibinfo {pages} {048301} (\bibinfo {year}
  {2012})%
  \bibAnnoteFile{NoStop}{Fielding2012}%
\bibitem{Bray1994}%
  \BibitemOpen
  \bibfield{author}{%
  \bibinfo {author} {\bibfnamefont{A.}~\bibnamefont{Bray}},\ }%
  \bibfield{journal}{%
  \Doi{10.1080/00018739400101505}{\bibinfo {journal} {Adv. Phys.}}\ }%
  \textbf{\bibinfo {volume} {43}},\ \bibinfo {pages} {357} (\bibinfo {year}
  {1994})%
  \bibAnnoteFile{NoStop}{Bray1994}%
\bibitem{Dogic2004}%
  \BibitemOpen
  \bibfield{author}{%
  \bibinfo {author} {\bibfnamefont{Z.}~\bibnamefont{Dogic}}, \bibinfo {author}
  {\bibfnamefont{J.}~\bibnamefont{Zhang}}, \bibinfo {author}
  {\bibfnamefont{A.}~\bibnamefont{Lau}}, \bibinfo {author}
  {\bibfnamefont{H.}~\bibnamefont{Aranda-Espinoza}}, \bibinfo {author}
  {\bibfnamefont{P.}~\bibnamefont{Dalhaimer}}, \bibinfo {author}
  {\bibfnamefont{D.~E.}\ \bibnamefont{Discher}}, \bibinfo {author}
  {\bibfnamefont{P.}~\bibnamefont{Janmey}}, \bibinfo {author}
  {\bibfnamefont{R.}~\bibnamefont{Kamien}}, \bibinfo {author}
  {\bibfnamefont{T.~C.}\ \bibnamefont{Lubensky}},\ and\ \bibinfo {author}
  {\bibfnamefont{A.}~\bibnamefont{Yodh}},\ }%
  \bibfield{journal}{%
  \Doi{10.1103/PhysRevLett.92.125503}{\bibinfo {journal} {Phys. Rev. Lett.}}\
  }%
  \textbf{\bibinfo {volume} {92}},\ \bibinfo {pages} {125503} (\bibinfo {year}
  {2004})%
  \bibAnnoteFile{NoStop}{Dogic2004}%
\bibitem{Note2}%
  \BibitemOpen
  \bibinfo {note} {If $\Delta _C = 0$, the timestep $\Delta t$ needs to be an
  order of magnitude smaller than for $\Delta _C \not =0$, and the
  spatial-stepsize $\Delta x$ must also be at least halved.}%
  \bibAnnoteFile{Stop}{Note2}%
\bibitem{Note3}%
  \BibitemOpen
  \bibinfo {note} {Microscopically, at the heart of the structure, we find a
  pair of $+\protect \frac {1}{2}$ defects though these are close enough that
  the effective director field forms a spiral pattern of topological charge
  $+1$.}%
  \bibAnnoteFile{Stop}{Note3}%
\bibitem{Kruse2004}%
  \BibitemOpen
  \bibfield{author}{%
  \bibinfo {author} {\bibfnamefont{K.}~\bibnamefont{Kruse}}, \bibinfo {author}
  {\bibfnamefont{J.-F.}\ \bibnamefont{Joanny}}, \bibinfo {author}
  {\bibfnamefont{F.}~\bibnamefont{J\"{u}licher}}, \bibinfo {author}
  {\bibfnamefont{J.}~\bibnamefont{Prost}},\ and\ \bibinfo {author}
  {\bibfnamefont{K.}~\bibnamefont{Sekimoto}},\ }%
  \bibfield{journal}{%
  \Doi{10.1103/PhysRevLett.92.078101}{\bibinfo {journal} {Phys. Rev. Lett.}}\
  }%
  \textbf{\bibinfo {volume} {92}},\ \bibinfo {pages} {078101} (\bibinfo {year}
  {2004})%
  \bibAnnoteFile{NoStop}{Kruse2004}%
\bibitem{Elgeti2011}%
  \BibitemOpen
  \bibfield{author}{%
  \bibinfo {author} {\bibfnamefont{J.}~\bibnamefont{Elgeti}}, \bibinfo {author}
  {\bibfnamefont{M.~E.}\ \bibnamefont{Cates}},\ and\ \bibinfo {author}
  {\bibfnamefont{D.}~\bibnamefont{Marenduzzo}},\ }%
  \bibfield{journal}{%
  \Doi{10.1039/c0sm01097a}{\bibinfo {journal} {Soft Matter}}\ }%
  \textbf{\bibinfo {volume} {7}},\ \bibinfo {pages} {3177} (\bibinfo {year}
  {2011})%
  \bibAnnoteFile{NoStop}{Elgeti2011}%
\bibitem{Yang2014a}%
  \BibitemOpen
  \bibfield{author}{%
  \bibinfo {author} {\bibfnamefont{X.}~\bibnamefont{Yang}}, \bibinfo {author}
  {\bibfnamefont{D.}~\bibnamefont{Marenduzzo}},\ and\ \bibinfo {author}
  {\bibfnamefont{M.~C.}\ \bibnamefont{Marchetti}},\ }%
  \bibfield{journal}{%
  \Doi{10.1103/PhysRevE.89.012711}{\bibinfo {journal} {Phys. Rev. E: Stat.
  Nonlinear, Soft Matter Phys.}}\ }%
  \textbf{\bibinfo {volume} {89}},\ \bibinfo {pages} {012711} (\bibinfo {year}
  {2014})%
  \bibAnnoteFile{NoStop}{Yang2014a}%
\bibitem{Shen2011}%
  \BibitemOpen
  \bibfield{author}{%
  \bibinfo {author} {\bibfnamefont{X.~N.}\ \bibnamefont{Shen}}\ and\ \bibinfo
  {author} {\bibfnamefont{P.~E.}\ \bibnamefont{Arratia}},\ }%
  \bibfield{journal}{%
  \Doi{10.1103/PhysRevLett.106.208101}{\bibinfo {journal} {Phys. Rev. Lett.}}\
  }%
  \textbf{\bibinfo {volume} {106}},\ \bibinfo {pages} {208101} (\bibinfo {year}
  {2011})%
  \bibAnnoteFile{NoStop}{Shen2011}%
\bibitem{Gagnon2014}%
  \BibitemOpen
  \bibfield{author}{%
  \bibinfo {author} {\bibfnamefont{D.~A.}\ \bibnamefont{Gagnon}}, \bibinfo
  {author} {\bibfnamefont{N.~C.}\ \bibnamefont{Keim}}, \bibinfo {author}
  {\bibfnamefont{X.}~\bibnamefont{Shen}},\ and\ \bibinfo {author}
  {\bibfnamefont{P.~E.}\ \bibnamefont{Arratia}},\ }%
  \bibfield{journal}{%
  \Doi{10.1063/1.4896598}{\bibinfo {journal} {Phys. Fluids}}\ }%
  \textbf{\bibinfo {volume} {26}},\ \bibinfo {pages} {103101} (\bibinfo {year}
  {2014})%
  \bibAnnoteFile{NoStop}{Gagnon2014}%
\bibitem{Note4}%
  \BibitemOpen
  \bibinfo {note} {There are several differences between our study and
  Ref.~\cite {Bozorgi2013}. In that work (i) the stability analysis is about an
  isotropic base state, (ii) liquid-crystalline stresses are not included,
  (iii) the active particles are self-propelled, and (iv) the concentration
  field can vary in space (we assume a homogeneous concentration).}%
  \bibAnnoteFile{Stop}{Note4}%
\end{thebibliography}%
\end{document}